\documentclass[useAMS,usenatbib,usegraphics]{mn2e}
\usepackage{ifpdf} 
\pdfoutput=1
\usepackage{graphicx,tablefootnote}
\usepackage{latexsym,amssymb}
\usepackage{amsmath}
\usepackage{gensymb}
\usepackage{capt-of}
\usepackage{textpos}
\usepackage{url}
\usepackage{multirow}
\usepackage{color}
\usepackage[T1]{fontenc}
\usepackage{longtable}
\usepackage{hyperref}

\hypersetup{
    bookmarks=true,         % show bookmarks bar?
    unicode=false,          % non-Latin characters in Acrobat�s bookmarks
    pdftoolbar=true,        % show Acrobat�s toolbar?
    pdfmenubar=true,        % show Acrobat�s menu?
    pdffitwindow=false,     % window fit to page when opened
    pdfstartview={FitH},    % fits the width of the page to the window
    pdftitle={My title},    % title
    pdfauthor={Author},     % author
    pdfsubject={Subject},   % subject of the document
    pdfcreator={Creator},   % creator of the document
    pdfproducer={Producer}, % producer of the document
    pdfkeywords={keyword1} {key2} {key3}, % list of keywords
    pdfnewwindow=true,      % links in new window
    colorlinks=true,       % false: boxed links; true: colored links
    linkcolor=black,          % color of internal links (change box color with linkbordercolor)
    citecolor=black,        % color of links to bibliography
    filecolor=magenta,      % color of file links
    urlcolor=black,           % color of external links
    linkbordercolor={1 0 0},
    citebordercolor={0 1 0}
}

\makeatletter
\Hy@AtBeginDocument{%
  \def\@pdfborder{0 0 1}% Overrides border definition set with colorlinks=true
  \def\@pdfborderstyle{/S/U/W 0}% Overrides border style set with colorlinks=true
}
\makeatother

\newcommand{\kms}{km\,s$^{-1}$}

\newcommand{\Msun}{M$_\odot$}

\newcommand{\MLsun}{$\Upsilon_\odot$}

\newcommand{\Reff}{\ensuremath{R_e}}

\newcommand{\MLstar}{\ensuremath{\Upsilon_\star}}

\newcommand{\Mbh}{\ensuremath{M_\mathrm{\bullet}}}
\newcommand{\Mstar}{\ensuremath{M_{\star}}}
\newcommand{\hst}{\textit{HST}}                     % HST
\newcommand{\ppak}{\textit{PPAK}}                     % PPAK
\newcommand{\sersic}{S\'{e}rsic}		%S�rsic

\title[The structural and dynamical properties of compact elliptical galaxies]{The structural and dynamical properties of compact elliptical galaxies}
\author[A. Y{\i}ld{\i}r{\i}m et al.]
{\parbox{\textwidth}{Ak{\i}n Y{\i}ld{\i}r{\i}m$^{1,2}$\thanks{E-mail: yildirim@mpia.de},
Remco C. E. van den Bosch$^{1}$,
Glenn van de Ven$^{1}$,
Ignacio Mart\'{i}n-Navarro$^{3}$,
Jonelle L. Walsh$^{4}$,
Bernd Husemann$^{5}$,
Kayhan G\"ultekin$^{6}$
and Karl Gebhardt$^{7}$}\vspace{0.4cm}\\
\parbox{\textwidth}{
$^{1}$Max Planck Institute for Astronomy, K\"onigstuhl 17, 69117 Heidelberg, Germany\\
$^{2}$Max Planck Institute for Astrophysics, Karl-Schwarzschild-Str. 1, 85741 Garching, Germany\\
$^{3}$University of California Observatories, 1156 High Street, Santa Cruz, CA 95064, USA\\
$^{4}$George P. and Cynthia Woods Mitchell Institute for Fundamental Physics and Astronomy, Department of Physics and Astronomy, Texas A\&M University, College Station, TX 77843, USA\\
$^{5}$European Southern Observatory, Karl-Schwarzschild-Str. 2, 85748 Garching, Germany\\
$^{6}$Department of Astronomy, University of Michigan, Ann Arbor, MI 48109, USA\\
$^{7}$Department of Astronomy, The University of Texas at Austin, Austin, TX 78712, USA\\}}
\begin{document}

\date{Accepted for publication in MNRAS}
%\date{Accepted 1988 December 15. Received 1988 December 14; in original form 1988 October 11}

%\pagerange{\pageref{firstpage}--\pageref{lastpage}} \pubyear{2002}

\def\LaTeX{L\kern-.36em\raise.3ex\hbox{a}\kern-.15em
    T\kern-.1667em\lower.7ex\hbox{E}\kern-.125emX}

\maketitle

\label{firstpage}

\begin{abstract}
Dedicated photometric and spectroscopic surveys have provided unambiguous evidence for a strong stellar mass-size evolution of galaxies within the last 10\,Gyr. The likely progenitors of today's most massive galaxies are remarkably small, disky, passive and have already assembled much of their stellar mass at redshift z=2. An in-depth analysis of these objects, however, is currently not feasible due to the lack of high-quality, spatially-resolved photometric and spectroscopic data. In this paper, we present a sample of nearby compact elliptical galaxies (CEGs), which bear resemblance to the massive and quiescent galaxy population at earlier times. Hubble Space Telescope (\hst) and wide-field integral field unit (IFU) data have been obtained, and are used to constrain orbit-based dynamical models and stellar population synthesis (SPS) fits, to unravel their structural and dynamical properties. We first show that our galaxies are outliers in the present-day stellar mass-size relation. They are, however, consistent with the mass-size relation of compact, massive and quiescent galaxies at redshift z=2. The compact sizes of our nearby galaxies imply high central stellar mass surface densities, which are also in agreement with the massive galaxy population at higher redshift, hinting at strong dissipational processes during their formation. Corroborating evidence for a largely passive evolution within the last 10\,Gyr is provided by their orbital distribution as well as their stellar populations, which are difficult to reconcile with a very active (major) merging history. This all supports that we can use nearby CEGs as local analogues of the high-redshift, massive and quiescent galaxy population, thus providing additional constraints for models of galaxy formation and evolution. 
\end{abstract}


%============================= Section 1 =============================
\section{Introduction}
\label{sec:introduction}
%=====================================================================

The structural and dynamical properties of early-type galaxies (ETGs) are key to our understanding of their formation and evolution. The homogenous properties of the most massive ellipticals, including their evolved stellar populations and high metallicities \citep[e.g.][]{1992MNRAS.254..601B,2006MNRAS.370.1106G,2007ApJ...669..947J,2009ApJ...693..486G,2009ApJ...698.1590G,2010MNRAS.408...97K}, boxy isophotes and high \sersic\ indices \citep[e.g.][]{1996ApJ...464L.119K,2009ApJS..182..216K}, isotropic velocity distributions and non-rotating, pressure supported velocity profiles \citep[e.g.][]{1978MNRAS.183..501B,1983ApJ...266...41D,1991MNRAS.253..710V,2007MNRAS.379..401E,2007MNRAS.379..418C,2011MNRAS.414..888E}, are considered as cornerstones, which a successful theory of galaxy formation and evolution has to be able to reproduce. Understanding the formation and evolution of ETGs in general - i.e. of the entire population of disky fast-rotators and boxy slow-rotators -  becomes even more important when considering that more than half of the stellar mass in the universe is confined to this class of objects and the spheroidal components of disk galaxies, which closely resemble elliptical galaxies of similar luminosity  \citep{1998ApJ...503..518F,2002AJ....124..646H,2003ApJS..149..289B,2004ApJ...600..681B}.\\

Monolithic collapse \citep{1962ApJ...136..748E} has long been considered the principal formation mechanism for galaxies in general and the population of ETGs in particular. However, the advent of deeper imaging and the possibility to obtain $K$-band number counts \citep{1998MNRAS.297L..23K} as well as galaxy luminosity functions \citep{2007ApJ...665..265F} at high redshift ($z \sim 1$) has effectively ruled out this simple formation scenario. An alternative to the monolithic collapse theory for the formation and evolution of ETGs is provided by the merging paradigm \citep{1972ApJ...178..623T,1977egsp.conf..401T} which, within a cosmological framework \citep{1978MNRAS.183..341W}, predicts a hierarchical build-up of galaxies through successive minor and major merging events. Considerable efforts have therefore been made in developing high-resolution $N$-body simulations of dissipational and dissipationless (un-)equal mass mergers of (disk-)galaxies \citep{1988ApJ...331..699B,1989Natur.338..123B,1996ApJ...471..115B,1992ApJ...400..460H,1993ApJ...409..548H,2003ApJ...597..893N,2005MNRAS.360.1185J,2006ApJ...641...21R,2006ApJ...650..791C}. While these simulations have been remarkably successful in reproducing the global photometric and kinematic properties of disky fast-rotating and boxy slow-rotating ellipticals, discrepancies remain in recovering e.g. the detailed dynamics of the most massive, slow-rotating ellipticals \citep{2008ApJ...685..897B} as well as their chemical abundance ratios \citep{2009ApJ...690.1452N}. 

The issue of reproducing massive ETGs in the local universe is even more severe if their structural and morphological transformation through cosmic time is taken into account. By tracing back a population of galaxies at constant number density \citep{2010ApJ...709.1018V} their detailed mass and size evolution can be recovered, indicating that the likely progenitors of today's red, massive galaxy population (a.k.a. "red nuggets") have been considerably smaller and denser at earlier times \citep[][]{2005ApJ...626..680D,2006ApJ...650...18T,2007ApJ...656...66Z,2008ApJ...688...48V,2008ApJ...677L...5V,2010ApJ...714L.244S,2012ApJ...749..121S,2014ApJ...788...28V}. Star formation was already quenched for a large fraction of this galaxy population and hence is unable to explain their drastic growth in size and mass since $z=2$ \citep[\Reff\ $\propto$ $\Mstar^{2}$;][]{2006ApJ...649L..71K,2007ApJ...671..285T,2008A&A...482...21C,2008ApJ...682..896K,2009ApJ...700..221K,2009ApJ...691.1879W,2010ApJ...709.1018V}. Moreover, these galaxies are disky \citep{2005ApJ...624L...9T,2006ApJ...650...18T,2011ApJ...730...38V,2013ApJ...762...83C}, with small \sersic\ indices, and have high stellar velocity dispersions \citep{2009Natur.460..717V,2011ApJ...737L..31B,2012ApJ...754....3T,2013ApJ...771...85V}. Here again, monolithic collapse fails to understand their evolution, as their present-day descendants would be too small and too low in number density \citep{2008ApJ...682..896K,2009ApJ...697.1290B}. On the other hand, binary mergers of equal mass (disk-)galaxies result in a comparable growth in size and mass \citep[$M \propto \Reff$;][]{2012MNRAS.425.3119H,2013MNRAS.429.2924H} and the required level of mergers would lead to an overestimation of the local galaxy mass function at the upper end \citep{2013MNRAS.428.1088M}.

Focus has therefore shifted to a two-phase growth, in which the assembly history of massive galaxies is dominated by an initial dissipative stage, where stars are formed in-situ, followed by an inside-out growth through few major and numerous minor merging events \citep{2009ApJ...699L.178N,2009MNRAS.398..898H,2010ApJ...709.1018V,2010ApJ...725.2312O,2012ApJ...744...63O,2012MNRAS.425..641L,2012MNRAS.425.3119H,2013MNRAS.429.2924H,2013ApJ...764L...1P,2013ApJ...766...15P,2013MNRAS.431..767B,2016MNRAS.456.1030W}.

The theoretical efforts are motivated by snapshots in the evolution of ETGs, provided by objects in the local universe and observational progress in obtaining photometric and spectroscopic data of galaxies at higher redshift ($z \le 2$). These, however, are notoriously difficult and expensive with current observational facilities. More importantly though, accurate measurements of stellar masses for galaxies at $z=2$, for instance, rely on fits to their spectral energy distribution (SED), which are liable to the stellar population synthesis (SPS) models \citep{2009ApJ...699..486C}. Similarly, stellar dynamical masses - the gold standard, which is generally employed to countercheck the aforementioned stellar mass estimates - are currently based on the measurement of the central velocity dispersions only \citep{2009ApJ...700..221K,2009Natur.460..717V,2011ApJ...736L...9V,2012ApJ...754....3T,2013ApJ...764L...8B,2013ApJ...771...85V} and virial mass estimators which are calibrated for the population of nearby ellipticals \citep{2006MNRAS.366.1126C}. Hence our understanding of the evolution of ETGs suffers from both systematic uncertainties in deriving accurate measurements as well as modelling assumptions due to the lack of spatially resolved data.

Optimally, one would like to study the progenitors of today's massive galaxy population in the local universe where, according to the stochastic nature of the merging mechanism, a non-negligible number is expected to survive unaltered \citep{2013ApJ...773L...8Q}. High quality photometric and spectroscopic data would allow a more detailed investigation and yield a more complete picture of their sizes and masses, their luminous and dark matter content, their stellar angular momentum profiles and pristine stellar populations, which in turn contain information regarding their origin and their (non-)violent growth mechanisms. Even though some individual objects have already been found and are claimed to be the relics of the early universe \citep[e.g.][]{2012Natur.491..729V,2014ApJ...780L..20T,2015ApJ...808...79F,2015MNRAS.452.1792Y}, no study yet has aimed to inspect the global properties of this leftover galaxy population by accumulating a large, homogenous and complementary set of high-quality data while employing state-of-the-art modelling techniques.\\

In this paper, we present a sample of 16 nearby ($\le 112$\,Mpc) compact elliptical galaxies (CEGs). These objects have been discovered by the Hobby-Eberly Telescope Massive Galaxy Survey \citep[HETMGS;][]{2015ApJS..218...10V}. The goal of the HETMGS was to obtain spatially resolved spectroscopic data of a large number of galaxies in order to assess their suitability for follow-up observations, which might be able to ultimately resolve their black hole sphere-of-influence (SOI) and thus enable a dynamical measurement of their supermassive black hole (SMBH) mass. The CEGs have been picked-up by the HETMGS based on the SOI argument, which naturally targets galaxies with the highest stellar velocity dispersions, and a diversity sampling strategy to cover a wider range of host galaxy properties with respect to the objects that already populate the black hole scaling relations \citep[e.g.][]{2013ARA&A..51..511K}. Detailed investigations on a case-to-case study, based on high-spatial resolution imaging and spectroscopy as well as wide-field IFU data, revealed interesting results with respect to their SMBHs and dark matter halos \citep[e.g.][]{2015ApJ...808..183W,2016MNRAS.456..538Y}. Here, however, we will gain insight into the sample's global structural and dynamical properties and aim to provide a tight observational link to the likely progenitors of today's massive galaxy population at redshift $z\sim2$. Given the body of evidence that is presented throughout this paper, we will postulate that these objects are indeed (largely) passively evolved analogues of the compact, massive galaxy population at $z \sim 2$. Consequently, their characteristics provide a benchmark of the ETG population of $\sim10$\,Gyr ago and can be used as constraints for formation models which try to recover the evolutionary path of massive ETGs within a cosmological context.\\

The paper proceeds as follows: We briefly revisit the data acquisition and reduction pipelines in Section \hyperref[sec:data]{\ref{sec:data}}. The wealth of photometric and spectroscopic information is exploited by means of our orbit-based dynamical models and stellar population synthesis fits, with the details of the modelling machineries highlighted in Sec. \hyperref[sec:analysis]{\ref{sec:analysis}}. We derive accurate stellar and dark masses, stellar mass surface density profiles, total mass density slopes and spatially resolved stellar ages, metallicities and abundance ratios, which are all presented in Sec. \hyperref[sec:results]{\ref{sec:results}}, and draw a comparison to galaxies at high and low redshift. We qualitatively discuss the impact of minor and major mergers - the currently adopted and prominent formation channel for ETGs since $z\sim2$ - and assess if these mechanisms are capable of evolving CEGs into and reconciling their properties with the massive and nearby ETG population. Finally, we summarise our findings in Sec. \hyperref[sec:summary]{\ref{sec:summary}}.

Throughout this paper we adopt 5th year results of the Wilkinson Microwave Anisotropy Probe \citep[\textit{WMAP};][]{2009ApJS..180..225H}, with a Hubble constant of $H_0=70.5$\,\kms\ Mpc$^{-1}$, a matter density of $\Omega_{M}=0.27$ and a cosmological constant of $\Omega_{\lambda}=0.73$.

%============================= Section 3 =============================
\section{Data}
\label{sec:data}
%=====================================================================

%---------------------------------------------------------------------
\subsection{HST imaging}
\label{sec:imaging}
%---------------------------------------------------------------------

We covered the photometric pipeline in detail in \cite{2015MNRAS.452.1792Y}. For the sake of completeness, we encapsulate the most relevant aspects.\\

Single orbit imaging of 15 CEGs have been obtained with the \hst\ WFC3 in \textit{I-} (F814W) and \textit{H}-band (F160W), as part of program GO: 13050 (PI: van den Bosch). The \textit{I-}band data comprises three dithered full-array exposures with a total integration time of 500\,s, whereas the \textit{H-}band data consists of three dithered full-array and four sub-array exposures with a total integration time of $\sim$ 1400\,s for each galaxy. The \textit{H-}band full-arrays are 450\,s long exposures, covering a nominal field of view (FOV) of 136\,\arcsec\ $\times$ 123\,\arcsec. The 1.7\,s short 16\,\arcsec\ $\times$ 16\,\arcsec\ sub-array exposures serve the purpose of mitigating possible saturation of the high surface brightness nucleus in the long full-array exposures.

For the analysis presented throughout this paper, we solely make use of the deeper \textit{H-}band photometry. Observations in the near-infrared (NIR) F160W filter have been chosen to minimise the effect of dust extinction and line blanketing while being aware that the NIR is also a better tracer of the stellar mass. Contamination from more recent starbursting events is minimal at longer wavelengths and the stellar mass-to-light ratio (\MLstar) thus becomes a weaker function of the underlying stellar populations \citep{2001ApJ...550..212B,2001MNRAS.326..255C}.

The reduction and combination of the exposures is carried out via \textsc{Astrodrizzle} \citep{2012drzp.book.....G}. Here, flat-field calibrated images are corrected for geometric distortions and drizzled onto a reference frame, where they are aligned according to the pointing information provided in the header of the respective files. The aligned images are then combined; the final output image is weighted by the exposure times of the individual input images. This final output image is then drizzled back onto the original, geometrically distorted but flat-field calibrated images, to identify cosmic rays and bad pixels. The sky level in the individual exposures is determined by iterative sigma-clipping of uniformly distributed pixels. However, both the deep full-array and sub-array exposures are dominated by galaxy light, either from the nucleus or the extended stellar halo of these objects. We therefore determine the sky level manually in less contaminated regions of the full-array exposures, while the background flux of the sub-array exposures is derived by measuring the flux difference between the sky subtracted full-arrays and the non-sky subtracted individual sub-arrays.

\begin{figure*}
		\begin{center}
		\includegraphics[width=.95\textwidth]{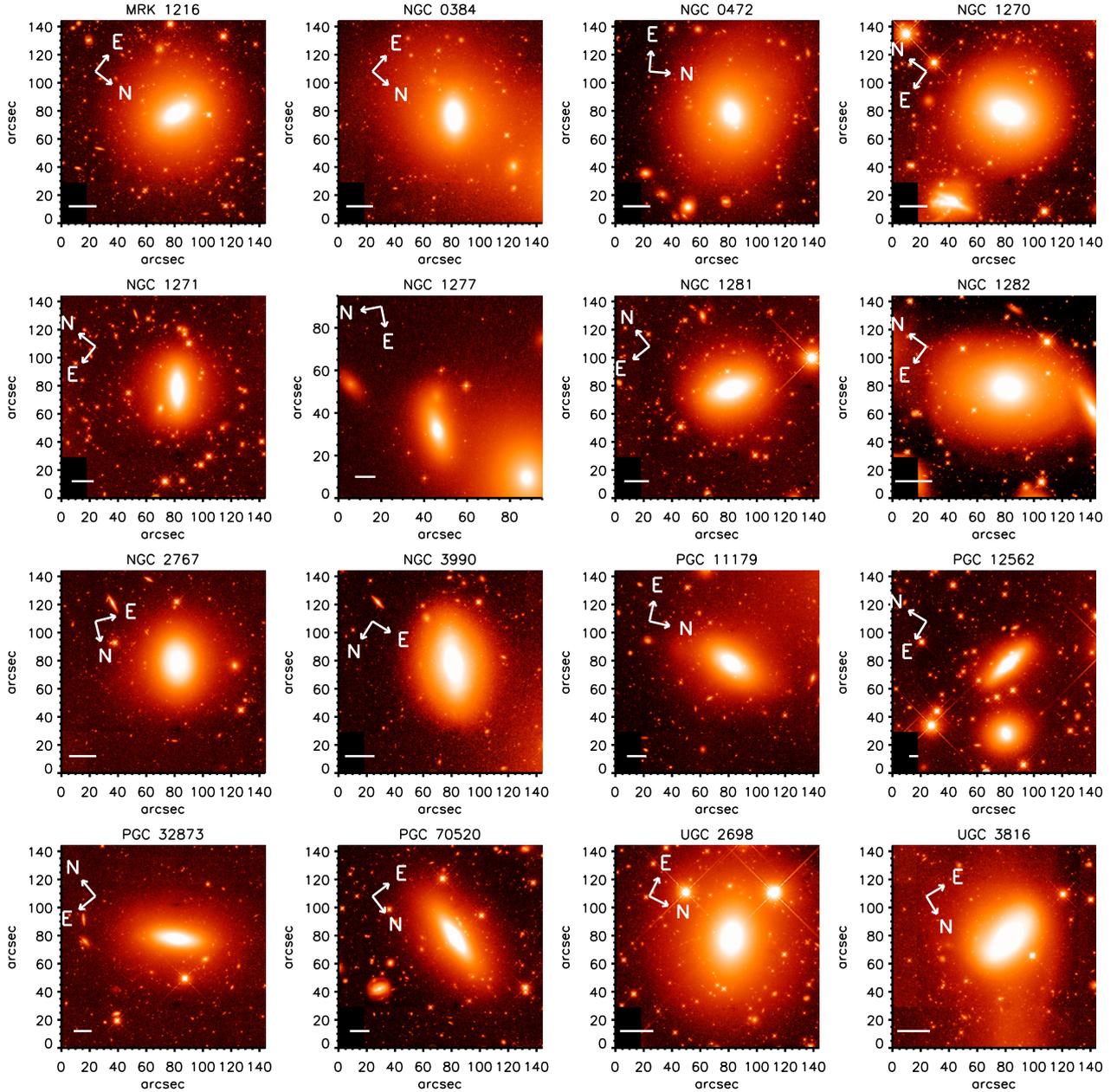}
		\end{center}
	\caption{\hst\ \textit{H}-band photometry (with the exception of NGC\,1277, where we rely on \textit{V}-band imaging) of the local CEGs, with a FOV of $\sim$ 150\,$\square$\,\arcsec\ (i.e. $\ge$ 50\,kpc$^{2}$, given the distance of the closest object) and a final scale of 0.06\,\arcsec/pixel. The horizontal bar on the bottom of each panel represents 3\,\Reff\ and thus illustrates the deep coverage of the \hst\ imaging data, with the compass showing the imaging orientation.}
	\label{fig:hst}
\end{figure*}

We present the final supersampled output of the photometric pipeline in Fig. \hyperref[fig:hst]{\ref{fig:hst}}, where we illustrate the \hst\ \textit{H}-band images of 15 compact ellipticals with a FOV of $\sim$ 150\,$\square$\,\arcsec\ at a resolution of 0.06\,\arcsec\ per pixel. To this sample we add the compact elliptical galaxy NGC\,1277. \textit{H-}band photometry of NGC\,1277 is missing and we therefore rely on \hst\ archival data (F550M) for this object, which has been obtained as part of program GO: 10546 (PI:Fabian). We refer the reader to \cite{2012Natur.491..729V} for the details, but provide an illustration of the shallower \hst\ \textit{V-}band imaging along with the other galaxies in Fig. \hyperref[fig:hst]{\ref{fig:hst}}. The specifics of NGC\,1277's photometric data are also discussed in \cite{2013MNRAS.433.1862E}, \cite{2015MNRAS.452.1792Y},\cite{2016ApJ...817....2W} and \cite{2016ApJ...819...43G}.

The point spread function (PSF) has been adopted from the Cosmic Assembly Near-Infrared Deep Extragalactic Survey \citep[CANDELS;][]{2012ApJS..203...24V}. To this end, a model PSF has been created via \textsc{TinyTim} in the F160W filter and is centred on the WFC3 detector, where distortion is minimal. The synthetic PSFs are generated for different dither positions and are then drizzled onto a common frame, reproducing a PSF at the same scale as our final science image with a PSF size of 0.17\,\arcsec\ FWHM.\\

In Sec. \hyperref[sec:analysis]{\ref{sec:analysis}}, we derive accurate stellar masses via our orbit-based dynamical models which, in turn, need a stellar mass model within which a representative library of orbits can be calculated. The stellar mass model is obtained from a deprojection of the surface brightness (SB) distribution for a given set of viewing angles. We parametrise the SB with a set of multiple, two-dimensional Gaussian functions \citep[MGE;][]{1992A&A...253..366M,1994A&A...285..723E}. The Gaussians do not form a complete set and the deprojection is not unique even in the case of an axisymmetric stellar system \citep{1987IAUS..127..397R}, but the SB distributions of realistic multi-component galaxies are commonly well reproduced \citep{2002MNRAS.333..400C} and the MGE method is convenient, since the convolution with the PSF as well as the deprojection can be handled analytically. All MGEs, except two, have been obtained with a constant position angle (PA) and a common centre for each galaxy. This is a necessary though insufficient condition for the deprojection in an oblate axisymmetric case, but reasonable given their small variation in the PA ($\Delta$PA $\le 5$\,\degree) and the acceptable reproduction of their SB profiles. For NGC\,2767 and PGC\,11179 we observe a PA twist in the centre, hinting at the presence of a bar, a dust disk and/or spiral arms and therefore stick to the assumption of an oblate disk model with axial symmetry. NGC\,0472 and NGC\,1282, however, show a significant variation of the PA from the innermost to the outermost regions, indicating that these two galaxies are in fact rather triaxial. The details of all MGEs are listed in Appendix \hyperref[sec:appendix_mge]{\ref{sec:appendix_mge}}. Magnitude measurements have been corrected for galactic extinction \citep{2011ApJ...737..103S} and the luminosity density has been obtained by adopting an absolute magnitude of 3.32 for the sun in \textit{H}-band \citep{1998gaas.book.....B}.

\begin{table*}
	\caption{Photometric properties of 16 CEGs and details of their imaging observations. The columns depict the galaxy (1), the corresponding \hst\ filter with which the observations have been acquired (2), the effective major axis radius from a single \sersic\ fit (3), the flattening of the single \sersic\ (4), the effective radius measured from a circular aperture that contains half of the light (5), the effective radius measured along the major axis of a best-fitting ellipse that contains half of the light, which we adopt as their trues sizes throughout this paper (6), the total apparent magnitude (extinction corrected) in \textit{H}-band (except of NGC\,1277, for which we rely on \textit{V}-band imaging) (7), and the adopted distance (8).}
	\begin{center}
	\centerline{
	\begin{tabular}{ c  c  c  c  c  c  c  c  c  c  c  c  c}
		\hline
		Galaxy & \hst\ & $R_{e,ser}$ & $b/a$ & $R_{e,circ}$ & $R_{e,ell}$ & mag & D \\
		& & [kpc] & & [kpc] & [kpc] & [H,Vega] & [Mpc] \\
		(1) & (2) & (3) & (4) & (5) & (6) & (7) & (8) \\
		\hline
		MRK\,1216	& F160W	&	2.8 $\pm$ 0.1 	& 0.58	& 2.3 $\pm$ 0.1	& 3.0 $\pm$ 0.1 	& 10.41	& 94 $\pm$ 2		\\
		NGC\,0384	& F160W	&	2.0 $\pm$ 0.1 	& 0.68	& 1.5 $\pm$ 0.1	& 1.8 $\pm$ 0.1 	& 10.32	& 59 $\pm$ 1		\\
		NGC\,0472	& F160W	&	3.0 $\pm$ 0.1 	& 0.72	& 2.0 $\pm$ 0.1	& 2.4 $\pm$ 0.1	& 10.55	& 74 $\pm$ 1		\\
		NGC\,1270	& F160W	&	2.1 $\pm$ 0.1	& 0.68	& 1.9 $\pm$ 0.1	& 2.2 $\pm$ 0.1	& 9.79	& 69 $\pm$ 1		\\
		NGC\,1271	& F160W	&	2.1 $\pm$ 0.1	& 0.43	& 1.4 $\pm$ 0.1	& 2.0 $\pm$ 0.1	& 10.72	& 80 $\pm$ 2		\\
		NGC\,1277	& F550M	&	1.3 $\pm$ 0.1	& 0.52	& 1.2 $\pm$ 0.1	& 1.3 $\pm$ 0.1	&  13.80	& 71 $\pm$ 1		\\
		NGC\,1281	& F160W	& 2.0 $\pm$ 0.1	& 0.64	&	1.3 $\pm$ 0.1	& 1.6 $\pm$ 0.1	& 10.46	& 60 $\pm$ 1		\\
		NGC\,1282	& F160W	&	1.8 $\pm$ 0.1 	& 0.82	& 1.2 $\pm$ 0.1	& 1.3 $\pm$ 0.1	& 10.02	& 31 $\pm$ 2		\\
		NGC\,2767	& F160W	& 2.8 $\pm$ 0.1	& 0.75	&	1.9 $\pm$ 0.1	& 2.4 $\pm$ 0.1	& 10.64	& 74 $\pm$ 1		\\
		NGC\,3990	& F160W	&	0.6 $\pm$ 0.1 	& 0.50	& 0.4 $\pm$ 0.1	& 0.6 $\pm$ 0.1	& 9.94	& 15 $\pm$ 1		\\
		PGC\,11179	& F160W	&	2.1 $\pm$ 0.1 	& 0.66	& 1.8 $\pm$ 0.1	& 2.1 $\pm$ 0.1	& 10.74	& 94 $\pm$ 2		\\
		PGC\,12562	& F160W	&	1.0 $\pm$ 0.1 	& 0.53	& 0.7 $\pm$ 0.1	& 0.7 $\pm$ 0.1	& 11.08	& 67 $\pm$ 1		\\
		PGC\,32873	& F160W	&	2.3 $\pm$ 0.1 	& 0.53	& 1.9 $\pm$ 0.1	& 2.3 $\pm$ 0.1	& 11.02	& 112 $\pm$2		\\
		PGC\,70520	& F160W	&	1.8 $\pm$ 0.1 	& 0.49	& 1.2 $\pm$ 0.1	& 1.6 $\pm$ 0.1	& 10.62	& 72 $\pm$ 1		\\
		UGC\,2698	& F160W	& 4.1 $\pm$ 0.1	& 0.73	&	3.1 $\pm$ 0.1	& 3.7 $\pm$ 0.1	& 9.93	& 89 $\pm$ 2		\\
		UGC\,3816	& F160W	&	2.9 $\pm$ 0.1 	& 0.69	& 1.8 $\pm$ 0.1	& 2.1 $\pm$ 0.1	& 9.66	& 51 $\pm$ 1		\\
		\hline
	\end{tabular}
	}
%	\vspace{2ex}
	\label{tab:phot_all}
	\end{center}
\end{table*}

We further analyse the photometry by carrying out single \sersic\ fits, but derive accurate sizes also in terms of the major axis radius of a best-fitting circular and elliptical isophote that contains half of the light. We adopt the elliptical effective radii as their true sizes throughout this paper, unless mentioned otherwise, and adopt errors of 0.1\,kpc in order to account for uncertainties in the distance. As pointed out by \cite{2010MNRAS.401.1099H} and illustrated in \cite{2013MNRAS.432.1709C}, this measurement of the effective radius is less prone to inclination effects. Given the high quality of our photometric observations, and in contrast to observations of the red and massive galaxy population at higher redshifts \citep[e.g.][]{2014ApJ...788...28V}, we can follow the low SB wings down to more than 10 magnitudes below the central SB \citep[see][]{2015MNRAS.452.1792Y} and therefore do not need to rely on parametrised fits to derive accurate sizes. We compile the photometric properties of our sample in Table \hyperref[tab:phot_all]{\ref{tab:phot_all}}.

%---------------------------------------------------------------------
\subsection{PPAK kinematics}
\label{sec:kinematics}
%---------------------------------------------------------------------

We present the data acquisition and reduction pipeline for our sample of 16 CEGs. This yields the wide-field IFU stellar kinematics which are used for the kinematic analysis and as input for our orbit-based dynamical models. For a more detailed overview, though, we refer the reader to \cite{2015MNRAS.452.1792Y} where the kinematic pipeline is also discussed at length.\\

Spectroscopic data of all galaxies have been obtained at the 3.5m telescope at Calar Alto, during multiple runs between December 2011 and October 2014. Mounted on the telescope is the \textit{Potsdam Multi Aperture Spectrograph} \citep[PMAS;][]{2005PASP..117..620R} and the \ppak\ fibre module \citep{2004AN....325..151V,2006PASP..118..129K}. The \ppak\ module consists of 382 fibres, each with a diameter of 2.7\,\arcsec\ projected on the sky, which are bundled to a hexagonal shape, covering a FOV of $\sim$ 1.3 $\square$\,\arcmin. A total of 36 fibres, bundled in pairs of 6 fibres each, are located 72\,\arcsec\ away from the centre and dedicated to sample the sky while 15 fibres are used for calibration purposes. The remaining fibres are science fibres with an inter-fibre distance of 3.2\,\arcsec\ and hence provide a filling factor of 60 per cent across the entire FOV, but a 100 per cent filling factor as well as an increase in spatial resolution is ensured when used with a three-point dither pattern.

Two setups, consisting of a medium-resolution V1200 and a low-resolution V500 grating, are available. Observations of MRK\,1216 and NGC\,1277 have been obtained with the former and were already presented and discussed in \cite{2015MNRAS.452.1792Y}. In brief, the V1200 grating has a resolving power of \textit{R} $=$ 1650 at 4000 \AA\ with a spectral resolution of 2.3 \AA\ FWHM across the nominal spectral range of 3400 - 4840 \AA, based on line-width measurements in the arc lamp exposures. This corresponds to an instrumental velocity dispersion of $\sigma =$ 85\,\kms. Both galaxies have been observed for a total of 1.5h on-source, with two science exposures of 900\,s in each of the three dither pointings. On the other hand, observations for the remaining galaxies have been carried out with the lower resolution V500 grating which has a resolving power of \textit{R} $=$ 850 at 5400 \AA\ and a spectral resolution of 6.3 \AA\ FWHM across the nominal spectral range of 3745 - 7500 \AA. This configuration corresponds to an instrumental velocity dispersion of $\sigma =$ 150\,\kms. For each of these 14 objects, we obtained data with a total of 6 hours on-source integration, split into three complete runs with three dither pointings of 2400\,s each, divided into two frames \`{a} 1200\,s. 

The data reduction follows a dedicated pipeline, which has been developed for the Calar Alto Legacy Integral Field Spectroscopy Area Survey (\textsc{CALIFA}) \citep{2012A&A...538A...8S,2013A&A...549A..87H} and includes bias and stray light subtraction, flat-fielding, cosmic ray cleaning \citep{2012A&A...545A.137H}, sky subtraction by means of the 30 faintest sky fibres, wavelength calibration, flexure correction and flux calibration using spectrophotometric standard stars. Spectra from the three pointings were then combined and resampled into a data cube with a 1\,\arcsec\ sampling, using a distance-weighted interpolation algorithm, followed by a correction for differential atmospheric refraction.

\begin{figure}
		\begin{center}
		\includegraphics[width=.47\textwidth]{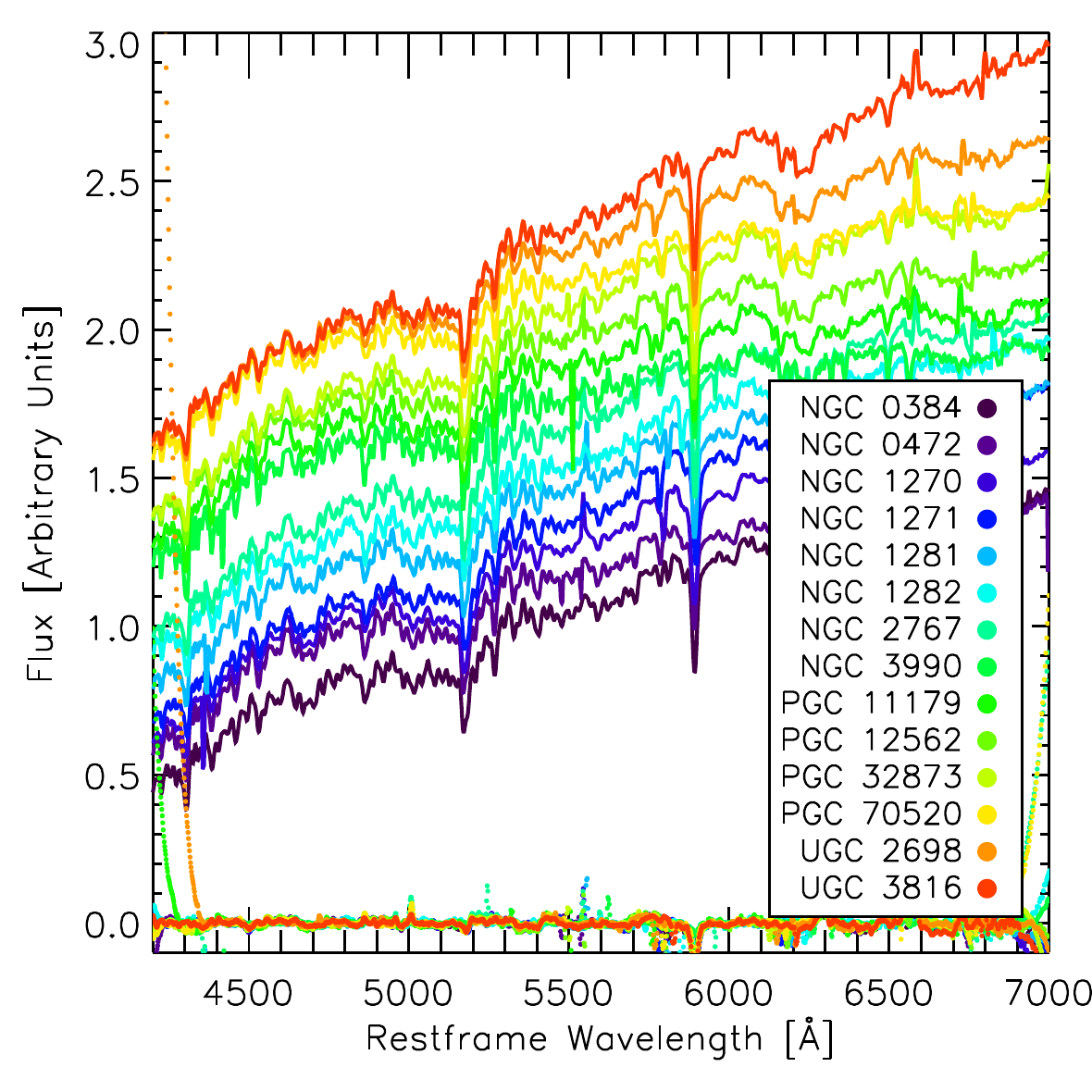}
		\end{center}
	\caption{Central \ppak\ spectrum of 14 CEGs, for which observations with the V500 grating have been acquired. MRK\,1216 and NGC\,1277 are excluded from the figure given their different observational setup, resulting in a narrower and more blue-shifted spectral coverage of 3400 - 4840 \AA. The spectra have been fitted with the Indo-U.S. stellar templates, convolved with the best-fitting LOSVD, and the corresponding model residuals are shown in the bottom. Fluxes have been shifted by an arbitrary amount for the sake of clarity. Sky and emission line features are generously masked and excluded from the fits, explaining the high residuals at e.g. $\sim$ 4300 and 5500 \AA.}
	\label{fig:ppak_spec}
\end{figure}

The line-of-sight velocity distribution (LOSVD) has been extracted from the reduced data by fitting the spatially binned spectra \citep{2003MNRAS.342..345C} with a non-negative linear combination of stellar templates \citep{2004PASP..116..138C}, in the vignetting and sensitivity limited useful spectral range of 4200 - 7000 \AA. This spectral range covers important absorption features, including H$\beta$, Mg$b$ and Fe 5015. The spectra were binned to reach a minimum signal-to-noise ratio (S/N) of 40 in each Voronoi zone. This target S/N is deemed sufficient to reliably extract the line-of-sight velocity $v$, velocity dispersion $\sigma$ as well as the higher order Gauss-Hermite moments $h_3$ and $h_4$, which quantify the asymmetric and symmetric deviations from a Gaussian LOSVD \citep{1993ApJ...407..525V,1994MNRAS.269..785B}. Likewise, a conservative S/N threshold of 10 for each spaxel has been chosen to achieve a trade-off between spatial resolution and coverage of the kinematic data, which leads to up to 600 spatially binned measurements of the LOSVD (for the highest quality data cube). For the stars, we made use of the Indo-U.S. stellar library with 328 spectral templates and a nominal spectral resolution of 1.2 \AA\ FWHM. Sky and emission line features in each data cube have been identified and masked, spatial masks have been employed to exclude any contamination by fore- and background objects and additive Legendre polynomials of 15th order have been used to correct the template continuum shapes during the fitting process. The central spectrum of each galaxy is shown in Fig. \hyperref[fig:ppak_spec]{\ref{fig:ppak_spec}}, along with the model residuals, showcasing the high quality of the \ppak\ observations and of the corresponding fits. As a result, we obtained large scale measurements of the mean line-of-sight velocity $v$, velocity dispersion $\sigma$ and higher order Gauss-Hermite moments $h_3$ and $h_4$, which map the 2D stellar kinematics out to several effective radii. 

The measurement errors are determined via Monte Carlo simulations by adding random Gaussian noise to the spectrum based upon the PPXF model residuals. Within the effective radius, the measurement errors are in general less than 10\,\kms\ for $v$ and $\sigma$, but can reach values of up to 50\,\kms\ for some of the outermost bins well beyond 3\,\Reff, as those bins cannot accumulate enough spaxels to reach the target S/N. Moreover, the robustness of the measurements has been tested by employing the MILES stellar library instead with a subset of 117 spectral templates \citep{2006MNRAS.371..703S,2011A&A...532A..95F}, by including multiplicative Legendre polynomials and by varying the width of the spectral masks. The measurements are largely consistent within the 2$\sigma$ measurement errors and the mean deviations amount to 5 and 10\,\kms\ for $v$ and $\sigma$ respectively and to roughly 0.02 for $h_3$ and $h_4$. The largely consistent measurements of the LOSVD encourage us to stick to the fiducial kinematics, but we take into account the systematic measurement uncertainties between the individual runs by adding them in quadrature to the formal 1$\sigma$ fitting uncertainties.

\begin{table*}
	\caption{Kinematic properties of 16 CEGs and details of their spectroscopic observations. The columns depict the galaxy (1), the corresponding \ppak\ grating with which the observations have been taken (2), the central stellar velocity dispersion (3), the stellar velocity dispersion within the best-fitting circular aperture that contains half of the light (4), the specific stellar angular momentum at one (5) and three elliptic effective radii (6), and the reconstructed PSF, expanded by an inner (7) and outer round Gaussian (8) with its corresponding weight and dispersion.}
	\begin{center}
	\centerline{
	\begin{tabular}{ c  c  c  c  c  c  c  c  c  c}
		\hline
		Galaxy & \ppak & $\sigma_c$ & $\sigma_{e,circ}$ & $\lambda_{1\,\Reff,ell}$ & $\lambda_{3\,\Reff,ell}$ & 1st Gaussian & 2nd Gaussian \\
		& & [\kms] & [\kms] & &  & weight - dispersion [\arcsec] & weight - dispersion [\arcsec] & \\
		(1) & (2) & (3) & (4) & (5) & (6) & (7) & (8) & \\
		\hline
		MRK\,1216	& V1200	& 335$\pm$6	& 308$\pm$7	& 	0.36	&	0.40	&	0.767 - 1.339	&	0.233 - 3.719\\
		NGC\,0384	& V500	& 240$\pm$5	& 221$\pm$6	&  	0.49	&	0.53	&	0.793 - 1.637	&	0.207 - 6.000\\
		NGC\,0472	& V500	& 252$\pm$7	& 217$\pm$8	&	0.23	&	0.19	&	0.837 - 1.063	&	0.163 - 4.544\\
		NGC\,1270	& V500	& 376$\pm$9	& 327$\pm$8	&	0.30	&	0.37	&	0.857 - 1.202	&	0.143 - 5.953\\
		NGC\,1271	& V500	& 302$\pm$8	& 295$\pm$6	&	0.50	&	0.62	&	0.822 - 1.478	&	0.178 - 4.289\\
		NGC\,1277	& V1200	& 355$\pm$5	& 317$\pm$5	&	0.45	&	0.63	&	0.751 - 1.208	&	0.249 - 2.440\\
		NGC\,1281	& V500	& 263$\pm$6	& 240$\pm$7	&	0.37	&	0.54	&	1.000 - 1.637	&	0.000 - 0.000\\
		NGC\,1282	& V500	& 204$\pm$6	& 201$\pm$4	&	0.31	&	0.45	&	0.802 - 1.847	&	0.198 - 5.492\\
		NGC\,2767	& V500	& 247$\pm$9	& 223$\pm$9	&	0.36	&	0.57	&	0.769 - 1.177	&	0.231 - 5.486\\
		NGC\,3990	& V500	& 108$\pm$18	& 93$\pm$9		&	0.22	&	0.48	&	0.844 - 1.462	&	0.156 - 6.000\\
		PGC\,11179	& V500	& 292$\pm$7	& 266$\pm$9	&	0.41	&	0.57	&	0.643 - 1.411	&	0.357 - 5.279\\
		PGC\,12562 & V500	& 260$\pm$7	& 256$\pm$9 	&	0.13	&	0.41	&	0.345 - 0.877	&	0.655 - 1.587\\
		PGC\,32873	& V500	& 308$\pm$9	& 304$\pm$8	&	0.44	&	0.62	&	0.877 - 1.353	&	0.123 - 5.775\\
		PGC\,70520 & V500	& 259$\pm$8	& 248$\pm$8	&	0.40	&	0.68	&	0.893 - 1.500	&	0.107 - 6.000\\
		UGC\,2698	& V500	& 351$\pm$8	& 304$\pm$6	&	0.19	&	0.21	&	0.576 - 1.368	&	0.424 - 6.000\\
		UGC\,3816	& V500	& 251$\pm$7	& 224$\pm$6	&	0.43	&	0.66	&	0.683 - 1.840	&	0.317 - 6.000\\
		\hline
	\end{tabular}
	}
%	\vspace{2ex}
	\label{tab:kin_all}
	\end{center}
\end{table*}

For illustration purposes, we display the bi-symmetrised stellar kinematic maps (along with the best-fitting dynamical model predictions, which will be discussed in more detail in Sec. \hyperref[sec:schwarzschild]{\ref{sec:schwarzschild}}) in Appendix \hyperref[sec:data_model]{\ref{sec:data_model}}. To visualise the extent of the kinematic data and to facilitate a direct comparison with the photometry, we overplot contours of constant SB at 1 and 3\,\Reff, given the measurements from the high-spatial resolution imaging. All objects show fast and regular rotation around the short axis with velocities of up to 280\,\kms. The kinematics reveal a strong anti-correlation between $v$ and $h_3$ (see also \citealt{2015MNRAS.452.1792Y,2016MNRAS.456..538Y,2015ApJ...808..183W,2016ApJ...817....2W}), with many galaxies also exhibiting exceptional central stellar velocity dispersion peaks of up to 380\,\kms. We also point out that the velocity dispersions of almost all galaxies are well above the \ppak\ instrumental resolution within the effective radius. Only in some individual cases do we observe a drop below this threshold for the outermost bins. Still, the bulk of our measurements are not affected by the resolution limit except for NGC\,3990, where the measurements commonly fall below 150\,\kms\ and thus question the reliability of this data set.

Finally, the PSFs of the spectroscopic observations have been recovered by convolving the deconvolved MGEs with two round Gaussians in order to match the collapsed \ppak\ data cubes. The details of the spectroscopic observations and the kinematic properties of our sample are summarised in Table \hyperref[tab:kin_all]{\ref{tab:kin_all}}.

%============================= section 4 =============================
\section{Analysis}
\label{sec:analysis}
%=====================================================================

%---------------------------------------------------------------------
\subsection{Schwarzschild models}
\label{sec:schwarzschild}
%---------------------------------------------------------------------

We construct orbit-based dynamical models for all galaxies in our sample. To this end, we make use of the triaxial realisation of Schwarzschild's orbit superposition method \citep{2008MNRAS.385..647V}. The working principles, details and application of this method to CEGs are also discussed in \cite{2015MNRAS.452.1792Y,2016MNRAS.456..538Y} and \cite{2015ApJ...808..183W,2016ApJ...817....2W}. Here, we confine ourselves to a brief description of the main steps.\\

We start with a mass model within which a representative orbit library is being calculated. The orbit library comprises 7776 orbits, sampled along 32 logarithmically spaced equipotential shells. Whereas the innermost shell is at a fixed location of 0.003\,\arcsec\ away from the centre, the location of the outermost shell is adjusted for each galaxy to ensure that the gravitational potential is well sampled out to at least four times the size of the largest Gaussian in the MGE. Every shell is then used as a starting point for 9 orbits. We employ the (x,z) start space twice, to account for pro- and retrograde orbits, whereas the ($\theta,\phi$) start space is populated only once and comprises triaxial orbit families, given the triaxial nature of the modelling machinery that can be run in the axisymmetric limit. The orbits are numerically integrated 200 times the period of a closed elliptical orbit of equal energy and their projected and deprojected quantities are stored and PSF convolved for comparison with the data. We find a non-negative linear superposition of the orbits that best matches the binned LOSVD in a $\chi^2$-sense, with the intrinsic and aperture masses provided by the MGE being used as additional constraints which have to be recovered with an accuracy of 2 per cent. The contributions of the individual gravitational constituents are then varied, and the aforementioned steps are reiterated, in order to assess the confidence intervals for the parameters of interest.

Our mass model includes the stellar mass component \Mstar, which is the deprojected intrinsic luminosity density times the constant stellar mass-to-light ratio \MLstar/\MLsun\ in \textit{H-} and \textit{V-}band respectively (the latter only for NGC\,1277), the mass of a supermassive black hole \Mbh\ and a dark matter component which is parametrised by a spherically symmetric NFW profile \citep{1996ApJ...462..563N,1997ApJ...490..493N} with concentration $c_\mathrm{DM}$ and total virial mass $M_\mathrm{DM}=M_{200}$. The concentration is usually not well constrained by our orbit-based dynamical models \cite[e.g][]{2015MNRAS.452.1792Y,2015ApJ...808..183W}, which is why we fix the concentration to $c_\mathrm{DM} =10$. Moreover, the \ppak\ data with its 2.7\,\arcsec\ wide fibres and a PSF of $\sim$ 3\,\arcsec\ FWHM is usually not sufficient to resolve the black hole sphere of influence. We therefore fix the black hole mass according to the black hole mass - stellar velocity dispersion relation \citep[$\Mbh-\sigma$;][]{2016ApJ...831..134V}, adopting the \ppak\ central velocity dispersion in each galaxy as a conservative estimate for $\sigma$. Even in the case of our CEGs with pronounced stellar velocity dispersion peaks in the centre, hinting at the presence of very massive SMBHs, the $\Mbh-\sigma$ relation still provides a decent proxy of the central SMBH in contrast to the relation between black hole mass and bulge luminosity. This is particularly true for NGC\,1271, NGC\,1277 and MRK\,1216, where our black hole mass estimate agrees well within a factor of two with the results from orbit-based dynamical models of adaptive optics assisted IFU observations \citep{2015ApJ...808..183W,2016ApJ...817....2W,2017ApJ...835..208W}. Our models thus probe the two parameter space in \MLstar\ $\in$ [0.5,3] and log($M_\mathrm{DM}/\Mstar$) $\in$ [-2,5], in steps of 0.05 and 0.5 respectively.

We assume a close to oblate axisymmetric shape (with an intermediate to long-axis ratio of $q=0.99$) for all but two CEGs. Modelling galaxies in the axisymmetric limit enables us to recover their LOSVD while benefiting from additional (i.e. triaxial) orbit families that are vital only for the support of triaxial mass configurations. Axial symmetry is a well justified assumption for the bulk of our sample, considering their fast and regular rotation around the short axis, the anti-correlation between $v$ and $h_3$, the negligible PA variation, the agreement of the photometric and kinematic PA and results from shape inversions of a large sample of fast-rotating galaxies \citep{2014MNRAS.444.3340W}. We note, however, that two galaxies - namely NGC\,0472 and NGC\,1282 - exhibit strong isophotal twists ($\Delta$PA $\ge 10$\,\degree), which cannot be explained by the presence of a bar and/or spiral features in the centre where the oblate disk assumption would still hold. In these instances, we pick a viable viewing orientation $\theta, \phi, \psi$ (which are directly linked to the intrinsic shape parameters $p$, $q$ and $u$) for which the MGE can still be deprojected. For NGC\,0472 and NGC\,1282 this is (54.114\,\degree, -20.994\,\degree, 89.998\,\degree) and (62.137\,\degree, 50.075\,\degree, 90.002\,\degree) respectively, which translates to a close to oblate axisymmetric system in the centre that becomes mildly triaxial towards the outer regions\footnote{Regarding the viewing orientations and intrinsic shape parameters, we follow the notation in \cite{2008MNRAS.385..647V}.}.

Under the assumption of oblate axial symmetry, the inclination $i$ is the only viewing parameter that is needed to pin down the intrinsic shape of the galaxy. Usually, the inclination is treated as another fitting parameter in the models, but difficult to constrain unless distinct kinematic features exist \citep{2005MNRAS.357.1113K,2009MNRAS.398.1117V}. Models with different inclinations can therefore reproduce the LOSVD equally well, but the inclination will only have a significant role for the recovery of the stellar mass-to-light ratio (and hence also for the recovery of the black hole and dark halo mass) if highly face-on projections are allowed \citep{2006MNRAS.366.1126C} which, however, is not the case for our sample. The minimum angle for the deprojection in an axisymmetric case is given by the flattest Gaussian in the respective MGE and the range of possible inclinations is $50\,\degree \le i \le 90\,\degree$ (with $90\,\degree$ being edge-on) even for the roundest object in our sample. In each instance we therefore choose to employ the midpoint value for the inclination, unless a central dust disk is present from which we can directly infer the inclination assuming that the disk is intrinsically flat and traces the PA of its host.

We would like to emphasise that no regularisation has been employed during the fitting process since the required level of regularisation is not known a priori and the fact that it can lead to an artificial narrowing of the $\chi^2$ contours. Moreover, we bi- and point-symmetrise the kinematics beforehand, depending on whether the galaxy is assumed to be axisymmetric or triaxial, in order to reduce noise and systematic effects in the data which helps in particular for the recovery of the higher order Gauss-Hermite moments.

%---------------------------------------------------------------------
\subsection{Stellar population synthesis}
\label{sec:stellar_pops}
%---------------------------------------------------------------------

For the stellar population analysis we made use of the most recent and extended version of the MILES stellar population models \citep{2010MNRAS.404.1639V,2015MNRAS.449.1177V}. The models are fed with the solar-scaled and $\alpha$-enhanced BasTI isochrones \citep{2004ApJ...612..168P,2006ApJ...642..797P} and the MILES stellar library, consisting of 925 spectral templates. This set of models has a wide range in metallicity, which extends up to $+0.40$ dex, thus ideal for our sample of galaxies composed mainly of massive and therefore metal rich objects. We cover a cosmologically motivated range in age from 0.03 to 14\,Gyr, assuming a single power-law stellar initial mass function (IMF), which corresponds to $\Gamma_\mathrm{b} = 1.3$ (Kroupa-like) and $\Gamma_\mathrm{b} = 2.3$ (Salpeter-like) in the MILES notation, and apply a lower and upper mass-cutoff of 0.1 and 100 \Msun\ respectively.

To derive the radial variations of the stellar population properties - namely their age, metallicity and $
\alpha$-abundance profiles - we first binned up the IFU data. For this purpose, elliptical apertures with a fixed ellipticity and position angle, as inferred from the high-resolution imaging, have been employed and a minimum S/N of 80 in each bin has been imposed during the binning process. Given the wide spectral coverage of the \ppak\ observations, ages and metallicities were then derived by fitting the age and metallicity sensitive spectral indices H$_{\beta o}$ and [MgFe]$'$ in this range, using a standard H$_{\beta o}$ \citep{2009MNRAS.392..691C} vs [MgFe]$'$ \citep{2003MNRAS.339..897T} index-index grid. Contamination of H$_{\beta o}$ from nebular emission has been taken care of by following the approach outlined in \cite{2013MNRAS.433.3017L}. In order to maximise the information extracted from the spectra, the comparisons between line-strength model predictions and observations were done at the resolution of each radial bin, i.e., that of the model (2.51 \AA) plus the local velocity dispersion of the galaxy.

%============================= section 5 =============================
\section{Results \& Discussion}
\label{sec:results}
%=====================================================================

%---------------------------------------------------------------------
\subsection{Modelling constraints}
\label{sec:modelling_constraints}
%---------------------------------------------------------------------

\begin{figure*}
		\begin{center}
		\includegraphics[width=.95\textwidth]{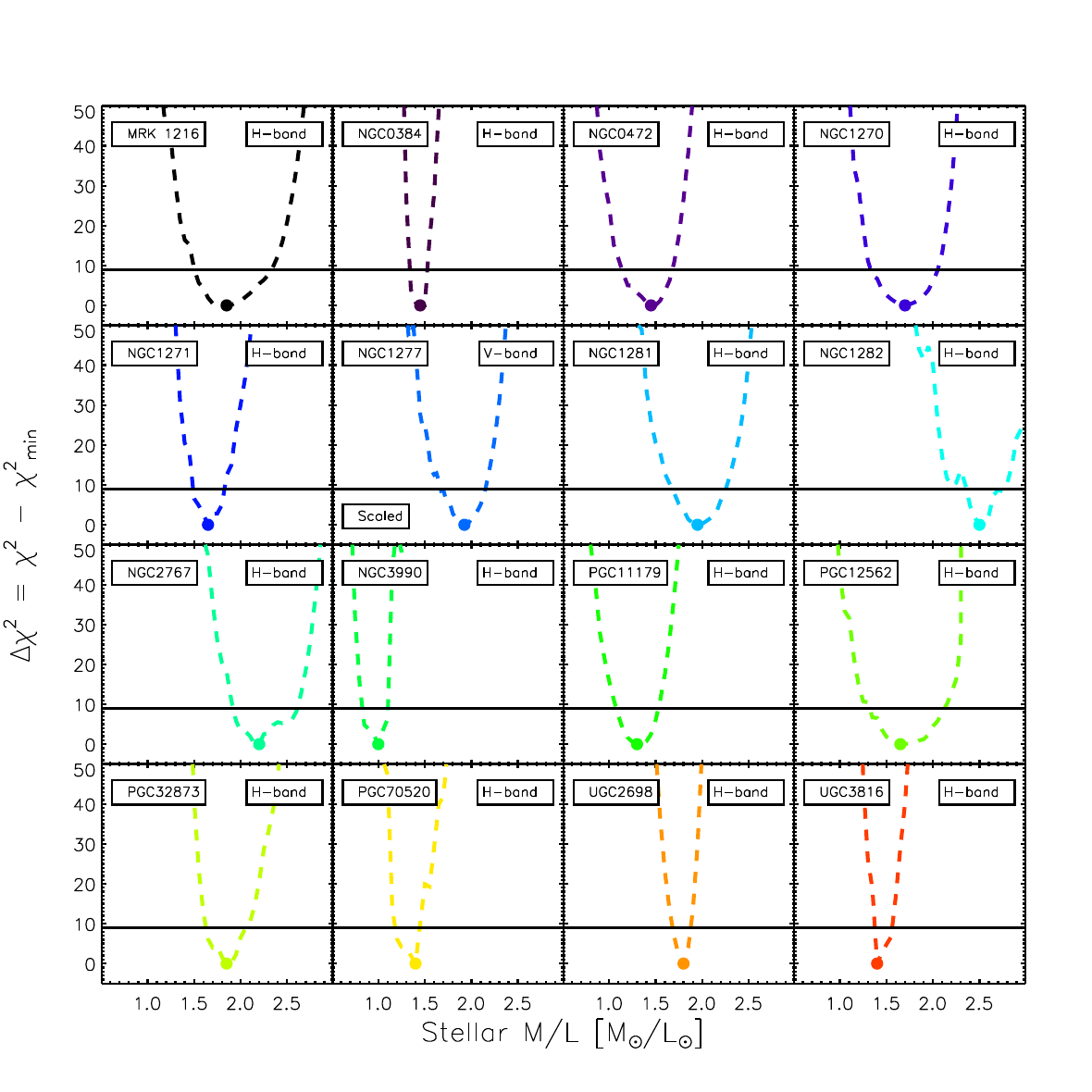}
		\end{center}
	\caption{Stellar $M/L$ constraints for each galaxy in our sample, marginalised over the dark halo virial mass $M_\mathrm{DM}$. The plots correspond to the modelling results with a fixed halo concentration of $c_\mathrm{DM}=10$, a fixed inclination angle (obtained either from a dust disk or by roughly adopting a midpoint value that is constrained by the MGE) and a fixed black hole mass according to the $\Mbh-\sigma$ relation. The coloured dots mark the best-fitting values. The horizontal lines denote a $\Delta\chi^2$ difference of 9, which corresponds to statistical 3$\sigma$ uncertainties for one degree of freedom. In the case of NGC\,1277, the $V$-band stellar $M/L$ has been scaled down by a factor of 4 for illustration purposes.}
	\label{fig:ml_constraints}
\end{figure*}

\begin{figure*}
		\begin{center}
		\includegraphics[width=.95\textwidth]{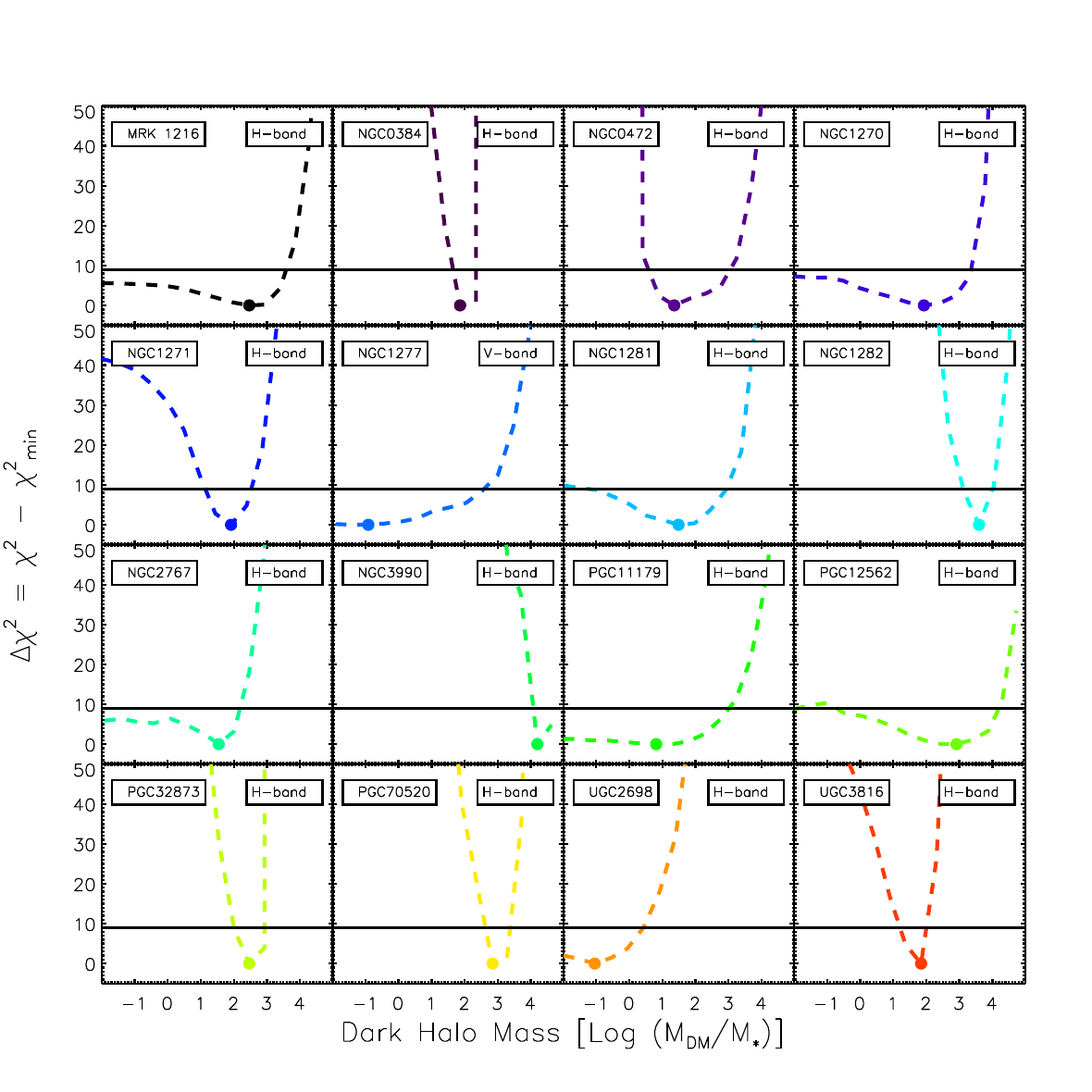}
		\end{center}
	\caption{Dark halo virial mass constraints for each galaxy in our sample, marginalised over the stellar $M/L$. The plots correspond to the modelling results with a fixed halo concentration of $c_\mathrm{DM}=10$, a fixed inclination angle (obtained either from a dust disk or by roughly adopting a midpoint value that is constrained by the MGE) and a fixed black hole mass according to the $\Mbh-\sigma$ relation. The coloured dots mark the best-fitting values. The horizontal lines denote a $\Delta\chi^2$ difference of 9, which corresponds to statistical 3$\sigma$ uncertainties for one degree of freedom.}
	\label{fig:dm_constraints}
\end{figure*}

Prior to the comparison of CEGs and the population of massive and passive galaxies at $z\sim2$, we first discuss the results from our orbit-based dynamical models and present the parameter constraints for the stellar mass-to-light ratio and dark halo virial mass for each of our 16 objects in Fig. \hyperref[fig:ml_constraints]{\ref{fig:ml_constraints}} and \hyperref[fig:dm_constraints]{\ref{fig:dm_constraints}}. These figures show the statistical 3$\sigma$ uncertainties for models with a fixed inclination angle, which was either obtained by the presence of a central dust disk or by simply adopting the midpoint value from the range of inclinations that are allowed by the deprojection of the MGE, a fixed black hole mass according to the $\Mbh-\sigma$ relation and a fixed halo concentration of $c_\mathrm{DM}=10$.

We have already taken the systematic uncertainties in the extraction of the stellar kinematics into account, by adding the mean offsets from the individual PPXF runs in quadrature to the fiducial formal fitting uncertainties. To assess the systematic uncertainties that are associated with our assumptions in the dynamical models, we also explore a more face-on and edge-on viewing orientation for each galaxy, perform fits to the unsymmetrised kinematic data and  change the central black hole mass by a factor of 2 as well as employ a different solver, which relaxes the mass constraints and thus fits the kinematic moments only. Here again, we add the mean differences in the best-fitting values in quadrature to the formal fitting errors, with the final parameter constraints summarised in Table \hyperref[tab:results]{\ref{tab:results}}.

\begin{table*}
	\caption{Parameter constraints from the orbit-based dynamical models and the stellar population synthesis fits. The columns depict each galaxy (1), its stellar dynamical $M/L$ in \textit{H}-band (with the exception of NGC\,1277, for which we quote the \textit{V}-band value) (2), its stellar dynamical mass (3), its black hole mass (which has been fixed according to the $\Mbh-\sigma$ relation) (4), its dark halo virial mass (with the dark halo being parametrised by a spherically symmetric NFW profile with a fixed halo concentration of $c_\mathrm{DM}=10$) (5) and the adopted inclination (6). The corresponding stellar age (7), metallicity (8) and $\alpha$-abundance measurements (9) have been obtained by fitting the integrated spectra within a $\sim$ 3\,\Reff\ wide aperture. A stellar M/L prediction in \textit{H}-band is also provided for comparison, based on the best-fitting stellar population parameters and under the assumption of a Kroupa-like stellar IMF (10). Regarding the inclination; an asterisk highlights those galaxies for which a dust disk is present and thus an estimate of the inclination is possible. The adopted inclination is roughly the midpoint value that is allowed by the deprojection of the MGE, if no dust disk is available. The measurement uncertainties represent the 3$\sigma$ confidence intervals in each fitting parameter in the dynamical analysis, with the systematic errors added in quadrature to the formal fitting errors, and the formal 1$\sigma$ confidence intervals in each fitting parameter for the stellar population analysis.}
	\begin{center}
	\begin{tabular}{ c c c c c c c c c c }
		\hline
		Galaxy &  \MLstar/\MLsun  & log(\Mstar/\Msun) & log(\Mbh/\Msun) & log($M_\mathrm{DM}$/\Mstar) & Inc. &  Age  & Metallicity & $\alpha$/Fe & \MLstar/\MLsun\\
			 &	&	&	&	& [deg]	& Gyr	& dex	& dex	&  \\
		(1)	 & (2)	& (3)	& (4)	& (5)	& (6) & (7) & (8) & (9) & (10) \\
		\hline
		MRK 1216	& 1.85$^{+0.52}_{-0.40}$	&	11.34$^{+0.11}_{-0.10}$	&9.50 &	2.47$^{+2.68}_{-5.11}$	&	70 	& - & - & - & - \\
		NGC 0384	& 1.45$^{+0.14}_{-0.17}$	&	10.96$^{+0.05}_{-0.05}$	&8.70 &	1.86$^{+0.84}_{-0.68}$	&	75 	& 13.7$\pm$0.36 &  0.05$\pm$0.05 &  0.27$\pm$0.02 & 1.30\\
		NGC 0472	& 1.45$^{+0.46}_{-0.51}$	&	11.07$^{+0.06}_{-0.11}$	&8.90 &	1.36$^{+1.35}_{-0.51}$	&	70 	& 13.4$\pm$0.31 &  0.03$\pm$0.03 &  0.22$\pm$0.02 & 1.29\\
		NGC 1270	& 1.70$^{+0.41}_{-0.41}$	&	11.31$^{+0.10}_{-0.12}$	&9.80 &	1.93$^{+1.49}_{-3.96}$	&	80*	& 14.0$\pm$0.50 &  0.34$\pm$0.02 &  0.23$\pm$0.02 & 1.33\\
		NGC 1271	& 1.65$^{+0.25}_{-0.25}$	&	11.06$^{+0.07}_{-0.07}$	&9.30 &	1.92$^{+0.65}_{-0.66}$	&	83 	& 14.0$\pm$0.50 &  0.13$\pm$0.02 &  0.31$\pm$0.02 & 1.32\\
		NGC 1277	& 7.70$^{+1.14}_{-1.22}$	&	11.13$^{+0.06}_{-0.07}$	&9.70 &	-0.91$^{+3.58}_{-1.44}$	&	75*	& - & - & - & - \\
		NGC 1281	& 1.95$^{+0.35}_{-0.31}$	&	11.00$^{+0.08}_{-0.08}$	&8.90 &	1.49$^{+1.57}_{-2.54}$	&	70*	& 14.0$\pm$0.50 &  0.21$\pm$0.03 &  0.27$\pm$0.02 & 1.33\\
		NGC 1282	& 2.50$^{+0.56}_{-0.51}$	&	10.77$^{+0.09}_{-0.09}$	&8.40 &	3.60$^{+0.61}_{-0.77}$	&	70 	& 9.64$\pm$0.15 &  0.00$\pm$0.02 &  0.20$\pm$0.02 & 1.08\\
		NGC 2767	& 2.20$^{+0.46}_{-0.34}$	&	11.12$^{+0.09}_{-0.08}$	&8.80 &	1.54$^{+0.83}_{-3.52}$	&	70 	& 14.0$\pm$0.50 &  0.11$\pm$0.02 &  0.29$\pm$0.02 & 1.32\\
		NGC 3990	& 1.00$^{+0.23}_{-0.25}$	&	9.68$^{+0.12}_{-0.13}$	&6.90 &	4.20$^{+0.76}_{-0.62}$	&	80 	& 12.9$\pm$0.17 & -0.15$\pm$0.03 &  0.20$\pm$0.01 & 1.23\\
		PGC 11179	& 1.30$^{+0.21}_{-0.21}$	&	11.16$^{+0.06}_{-0.08}$	&9.20 &	0.81$^{+1.97}_{-3.00}$	&	75 	& 14.0$\pm$0.50 &  0.11$\pm$0.03 &  0.23$\pm$0.02 & 1.32\\
		PGC 12562	& 1.65$^{+0.46}_{-0.32}$	&	10.74$^{+0.10}_{-0.09}$	&8.90 &	2.92$^{+1.08}_{-4.93}$	&	80 	& 14.0$\pm$0.50 &  0.23$\pm$0.03 &  0.26$\pm$0.03 & 1.33\\
		PGC 32873	& 1.85$^{+0.22}_{-0.22}$	&	11.28$^{+0.04}_{-0.04}$	&9.30 &	2.47$^{+0.48}_{-0.46}$	&	84*	& 14.0$\pm$0.50 &  0.27$\pm$0.02 &  0.28$\pm$0.02 & 1.33\\
		PGC 70520	& 1.40$^{+0.35}_{-0.40}$	&	10.95$^{+0.10}_{-0.12}$	&8.90 &	2.85$^{+0.87}_{-0.75}$	&	82 	& 14.0$\pm$0.50 &  0.09$\pm$0.02 &  0.22$\pm$0.01 & 1.32\\
		UGC 2698	& 1.80$^{+0.10}_{-0.10}$	&	11.58$^{+0.01}_{-0.03}$	&9.60 &	-1.05$^{+1.32}_{-1.32}$	&	67*	& 14.0$\pm$0.50 &  0.20$\pm$0.02 &  0.27$\pm$0.02 & 1.33\\
		UGC 3816	& 1.40$^{+0.21}_{-0.15}$	&	10.96$^{+0.06}_{-0.04}$	&8.80 &	1.85$^{+1.46}_{-0.47}$	&	75 	& 14.0$\pm$0.50 &  0.08$\pm$0.02 &  0.29$\pm$0.02 & 1.32\\
	\end{tabular}
%	\vspace{2ex}
	\label{tab:results}
	\end{center}
\end{table*}

In general, the models reproduce the features of the LOSVD in each galaxy very well (see Appendix \hyperref[sec:data_model]{\ref{sec:data_model}}). Only in a few cases do the models struggle to fit the detailed kinematic moments. In NGC\,1282, for instance, this can be linked to the choice of viewing orientation, which we omitted to explore further as the search in three additional parameters would be computationally expensive, and the strong residuals in the parametrisation of its stellar light distribution by the MGE. In NGC\,3990, on the other hand, this is due to difficulties in the measurement of the LOSVD, and no satisfying fit can be obtained when fitting all four kinematic moments simultaneously. The models fail in particular with the recovery of the high $h_3$ and $h_4$ values, which is why (for the time being) we constrain the fit to $v$ and $\sigma$, which can be reproduced sufficiently well. This issue, however, will be alleviated in the future, thanks to higher-spectral resolution observations that have been obtained with the \textit{VIRUS-W} spectrograph in the meantime.

The reduced $\chi^2$ values of the fits range from 0.07 to 0.25. Yet, keep in mind that these are most likely underestimated and do not represent the actual quality of the fit. Very low $\chi^2$ values usually hint at an overestimation of the measurement errors, but (in our case) are also driven by the symmetrisation of the kinematics as well as the fact that we do not have 4$\times N$ (with $N$ being the number of bins) independent measurements of the LOSVD, due to the S/N correlation in the 3-point dithered, spatially binned \ppak\ spectra. In fact, the $\chi^2$ values approach unity when e.g. fits to the unsymmetrised data are performed, with negligible changes in the parameter constraints that are presented in Table \hyperref[tab:results]{\ref{tab:results}}.\\

When it comes to the stellar population analysis, systematic uncertainties may arise from a wide variety of sources, starting from our assumption of a single stellar population-like star formation history to differences in the populations synthesis models' ingredients. The latter effect can be partially tested within the MILES SSP models by comparing how line-strength predictions depend on the adopted set of isochrones. We found that for old (10\,Gyr) and metal-rich ([M/H] = +0.22) populations, there is a systematic offset of $\sim 0.05$ dex in the \textit{H}-band mass-to-light ratio between PADOVA-\citep{2000A&AS..141..371G} and BaSTI-based \citep{2004ApJ...612..168P,2006ApJ...642..797P} predictions. Note that, in contrast to e.g. differences in the star formation histories, this systematic does not affect the comparison between galaxies once a certain stellar populations model is assumed. In Table \hyperref[tab:results]{\ref{tab:results}} we list the stellar population constraints from fits to the integrated spectra within a 3\,\Reff\ wide aperture and complement these with predictions for the \textit{H}-band stellar M/L, based on the best-fitting values and under the assumption of a Kroupa-like IMF. Note also, that the stellar population synthesis fits have only been carried out for those galaxies for which data in the V500 setup are available (see also Sec. \hyperref[sec:stellar_populations]{\ref{sec:stellar_populations}}).

%---------------------------------------------------------------------
\subsection{Stellar mass-size relation}
\label{sec:stellar_mass_size}
%---------------------------------------------------------------------

Massive ETGs have grown significantly in stellar mass and half-light size since redshift $z=2$ \citep[e.g.][]{2006ApJ...650...18T,2008ApJ...688..770F,2008ApJ...677L...5V,2010ApJ...709.1018V,2008ApJ...688...48V,2014ApJ...788...28V}. The progenitors of today's massive galaxy population are also found to be more flattened and disky \citep[e.g.][]{2005ApJ...624L...9T,2006ApJ...650...18T,2011ApJ...730...38V,2013ApJ...762...83C}, very compact \citep[e.g.][]{2007ApJ...656...66Z,2008ApJ...687L..61B,2008ApJ...677L...5V,2010ApJ...714L.244S} with little to no ongoing star formation \citep[e.g.][]{2009ApJ...691.1879W} and exceptional central velocity dispersion peaks \citep[e.g.][]{2009Natur.460..717V}.

Descendants of the compact, red and massive galaxy population are rare in the local universe \citep{2009ApJ...692L.118T,2010ApJ...720..723T}. The agreement, however, between the photometric and kinematic properties of our compact elliptical galaxy sample - namely their high central velocity dispersions, compact sizes, rapid and regular rotation and disky SB profiles -  and the compact and massive ellipticals at $z \sim 2$ is remarkable and suggests that these objects are actually passively evolved analogues. Corroborating evidence has been provided by the investigation of their dynamical structures in \cite{2015MNRAS.452.1792Y}, indicating that these galaxies have not undergone a recent, active phase of a few major and numerous minor mergers, which is assumed to be the main driver of the mass and size evolution of massive ellipticals since $z=2$ \citep[][]{2009ApJ...699L.178N,2010ApJ...725.2312O,2012MNRAS.425..641L,2012ApJ...744...63O,2012MNRAS.425.3119H,2013MNRAS.429.2924H,2013MNRAS.431..767B}. Furthermore, the stellar populations of NGC\,1277 \citep{2014ApJ...780L..20T} and NGC\,1281 \citep{2015MNRAS.452.1792Y} have been investigated in detail and stellar age estimates and star formation histories have also been derived for some galaxies in our sample in \cite{2015ApJ...808...79F}, based on SDSS spectroscopic data, showing that they are comprised of a uniformly old stellar population (> 10\,Gyr) without a recent (i.e. < 10\,Gyr) star formation event that might have been triggered by e.g. gas-rich (i.e. "wet") mergers.
 
\begin{figure}
		\begin{center}
		\includegraphics[width=.47\textwidth]{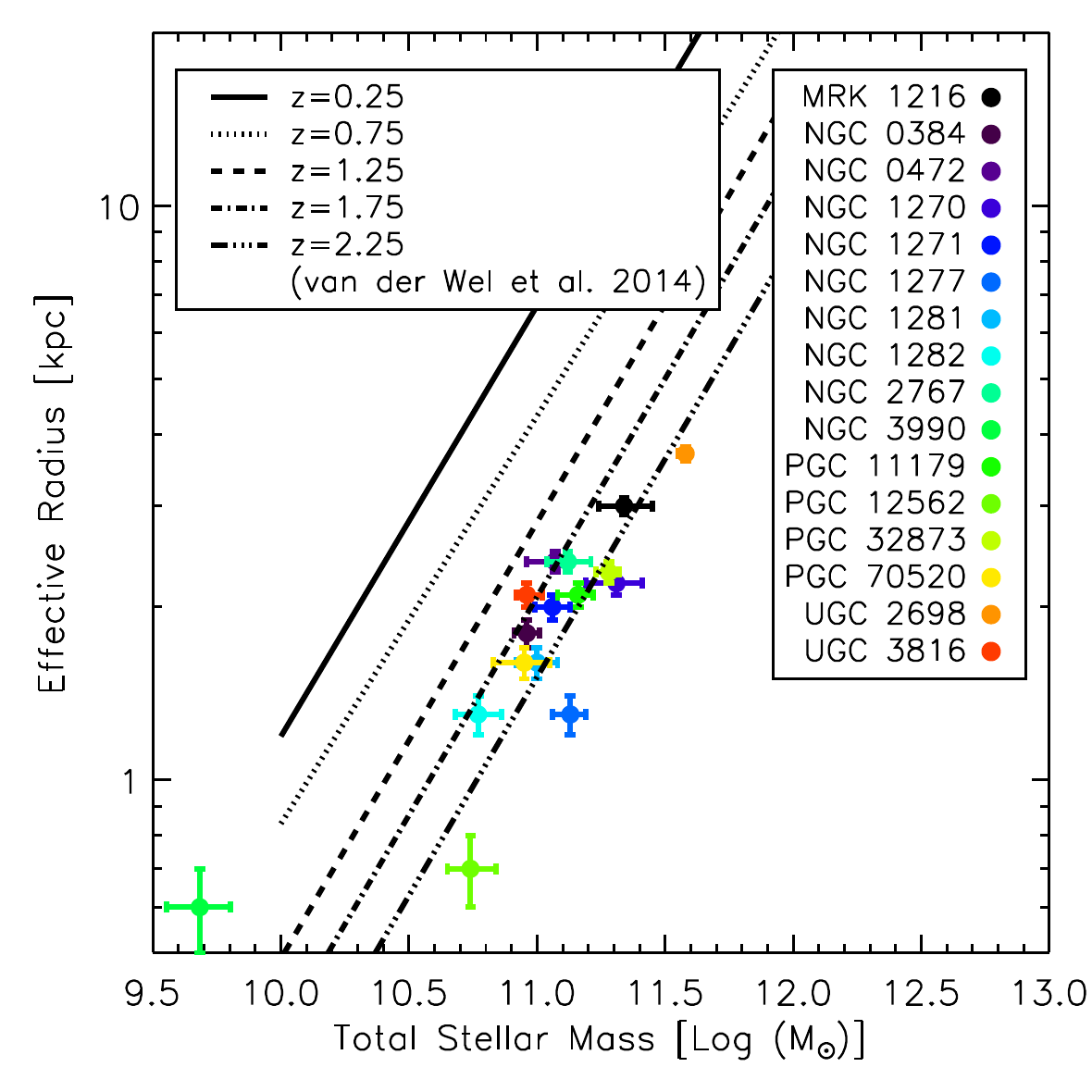}
		\end{center}
	\caption{Stellar mass-size relation of early-type galaxies at different redshifts (black), derived from the \textsc{CANDELS/3D-HST} survey \citep{2014ApJ...788...28V}. The sizes of the CEGs have been measured from the deep \hst\ \textit{H}-band images (with the exception of NGC\,1277, for which we rely on archival \textit{V}-band imaging), based on a best-fitting elliptical isophote that contains half of the light. The total stellar masses have been inferred from our orbit-based dynamical models. All galaxies are outliers from the local mass-size relation with an intrinsic scatter of $\sigma$(log \Reff) $=0.10$, but consistent with the relation at $z \sim 2$, except for the (most likely) tidally stripped object NGC\,3990.}
	\label{fig:mass_size_relation}
\end{figure}

By means of the orbit-based dynamical models, we can derive accurate total stellar masses and are now in a position to constrain their location in the mass-size relation, both of which are viewed as basic parameters in theories of galaxy formation and evolution. Fig. \hyperref[fig:mass_size_relation:]{\ref{fig:mass_size_relation}} exhibits our measurements of all 16 compact galaxies in our sample, with the associated uncertainties in mass and size.  Accurate measurements of stellar masses and sizes of a large sample of galaxies through cosmic time have been obtained as part of the \textsc{3D-HST+CANDELS} survey \citep{2014ApJ...788...28V}, where ETGs are distinguished as non-actively forming stars through colour-colour selections. We overplot their mass-size relations in the redshift range $0 \le\ z \le\ 3$, obtained assuming a log-normal distribution with scatter $\sigma$(log \Reff), intercept A and slope $\alpha$, and the relation being parametrised by \Reff\ $=$ A (\Mstar\ / 5 $\times$ 10$^{10}$ $M_{\scriptscriptstyle \odot}$)$^{\alpha}$. The figure convincingly demonstrates the affiliation of our compact objects with the sample at $z \sim 2$. All galaxies are outliers from the mass-size relation at $z \sim 0$, which has an intrinsic scatter in size of $\sigma$(log \Reff) $ = 0.10$,  and Kolmogorov-Smirnov tests on the distributions of mass and size rules out the null hypothesis that the compact galaxies have been drawn from the same parent distribution of local ETGs at better than 95 per cent confidence. NGC\,3990 is the only outlier from our sample, being consistent with the mass-size relation at $z \le 0.75$. However, this galaxy does not share the same characteristics as the rest of our sample, being roughly a magnitude smaller in stellar mass and size and devoid of a pronounced stellar velocity dispersion peak. On the contrary, being a companion of NGC\,3998, a close-by lenticular galaxy which has drawn attention due to a disparity of its gas and stellar dynamical black hole measurement as well as due to the lack of clear evidence for the presence of a dark halo \citep{2006A&A...460..439D,2012ApJ...753...79W,2016MNRAS.460.3029B}, we suspect that NGC\,3990 is stripped. Fig. \hyperref[fig:mass_size_relation:]{\ref{fig:mass_size_relation}} (in conjunction with the stellar mass surface density profiles, which are discussed in Sec. \hyperref[sec:stellar_mass_surface]{\ref{sec:stellar_mass_surface}}) therefore points out that the bulk of our CEG sample cannot simply be the leftovers of tidal interactions, but are in fact passively evolved analogues of the "red nuggets".

It is worth noting here that an overestimation of the stellar masses, as a consequence of e.g an overestimation of the stellar $M/L$ in the dynamical models, will affect our conclusions only marginally. Decreasing the stellar $M/L$ by 25 per cent - the mean deviation between the dynamically inferred stellar $M/L$ and those that are expected from their old stellar populations based on a Kroupa rather than a Salpeter IMF - moves the galaxies closer to the stellar mass-size relation at 1.25 $\le z \le$ 1.75, but the sample still remains an outlier from the present-day mass-size relation.

%---------------------------------------------------------------------
\subsection{Stellar mass surface density}
\label{sec:stellar_mass_surface}
%---------------------------------------------------------------------

\begin{figure}
		\begin{center}
		\includegraphics[width=.47\textwidth]{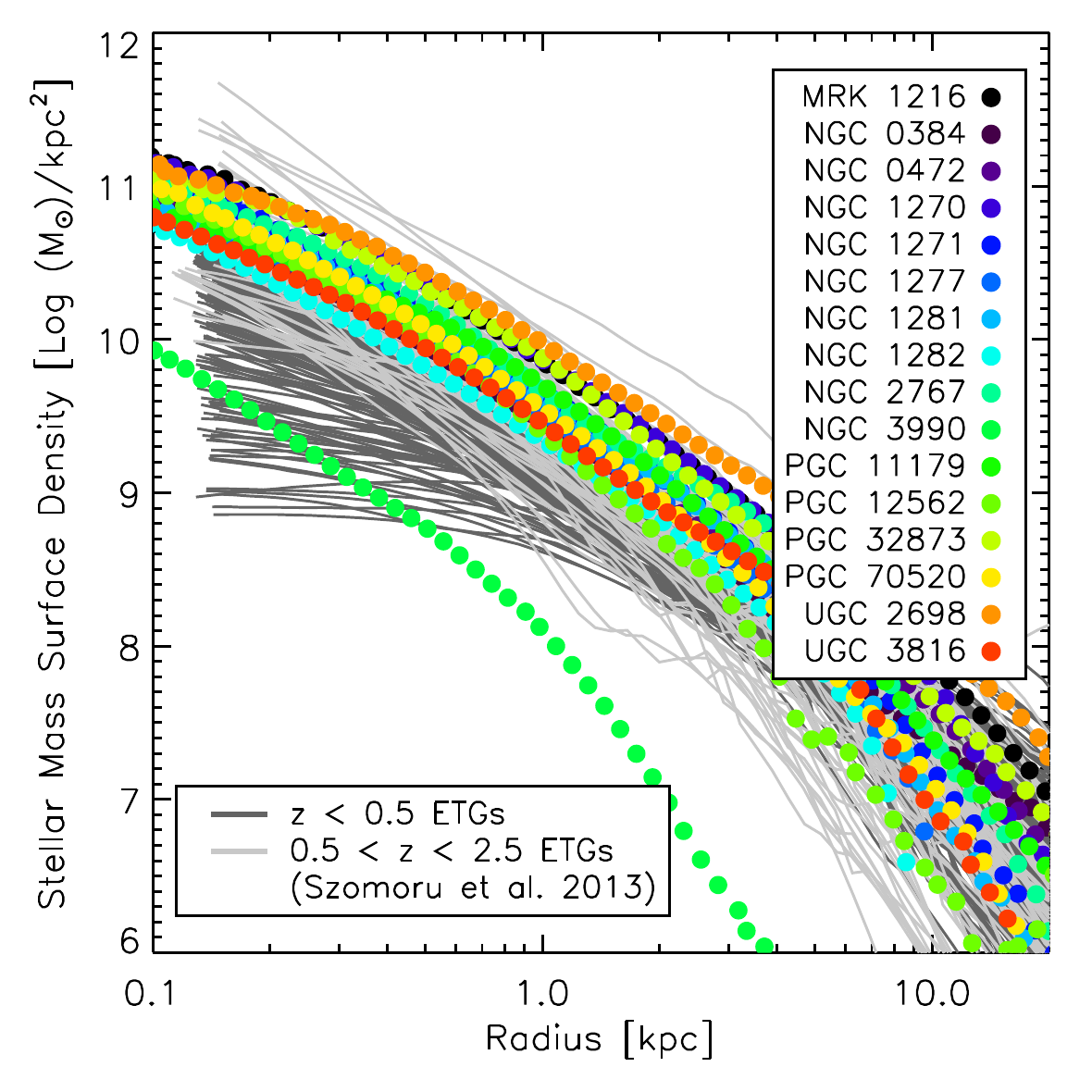}
		\end{center}
	\caption{Stellar mass surface density profiles of massive ETGs at high (light grey) and low (dark grey) redshifts  \citep{2013ApJ...763...73S}. Overplotted are the stellar mass surface density profiles of the CEGs, based on the deep \hst\ photometry and the stellar $M/L$ estimates from our orbit-based dynamical models of the wide-field IFU stellar kinematics. Apart from NGC\,3990, the stellar mass surface density profiles of our sample agree with those of compact, massive and passive galaxies at high redshift. For the largest CEGs, we observe a disproportional growth in mass at large radii, where the surface mass density profiles start to resemble nearby elliptical galaxies.}
	\label{fig:surface_mass_densities}
\end{figure}

By virtue of the deep, high-spatial resolution \hst\ photometry, we can obtain accurate surface density profiles out to large radii. Having constrained the stellar $M/L$ dynamically, we can now convert the SB profiles into surface mass density profiles, which are presented in Fig. \hyperref[fig:surface_mass_densities]{\ref{fig:surface_mass_densities}}.

The central stellar mass surface densities of the red, massive galaxy population at high redshift (light grey) are 2-3 times higher than for local ETGs (dark grey) \citep{2009ApJ...697.1290B,2012ApJ...749..121S,2013ApJ...763...73S}. This is a result of their remarkable compactness. The high stellar surface mass densities are assumed to be triggered by gas-rich major mergers at even higher redshift \citep{2007ApJ...658..710N,2010ApJ...722.1666W} of presumably even more compact, starbursting submillimeter galaxies (SMGs) \citep{2014ApJ...782...68T} or by dynamical instabilities which feed galaxies with cold gas from the intergalactic medium (IGM) \citep{2003MNRAS.345..349B,2005MNRAS.363....2K,2009Natur.457..451D,2014MNRAS.438.1870D}. Alternatively, the progenitors of the compact, massive and quiescent galaxy population at $z \sim 2$ could be heavily obscured and slightly larger star-forming galaxies (SFGs), which decrease their sizes and thus increase their central stellar mass surface densities as a result of centrally located, intense star formation rates \citep{2013ApJ...765..104B,2014ApJ...791...52B,2016ApJ...827L..32B,2015ApJ...813...23V}

In any case, the central stellar surface mass densities of our sample are in agreement with the values provided by the bulk of galaxies at high redshift, except for the (most likely) tidally stripped object NGC\,3990. Considering that our compact galaxies are outliers in the local mass-size relation (Fig. \hyperref[fig:mass_size_relation]{\ref{fig:mass_size_relation}}), this is expected. Self-evidently, the stellar mass surface density profiles are also steeper, with a sharp fall-off at large radii, reflecting their compactness. Remarkably, several galaxies in our sample, e.g. MRK\,1216 and UGC\,2698, have a more extended profile, which is consistent with the shallower density profile of local ETGs in the remote regions. These objects are at the same time the largest and most massive galaxies in our sample, with sizes and masses more than twice as large as for instance NGC\,1277. This indicates that they have already grown considerably with respect to the rest of the sample and might have already entered the path of becoming a regular ETG, also promoted by our orbital analysis of MRK\,1216 which closely follows the relationship between $\beta_z$ and $\delta$ of a sample of nearby ETGs \citep{2015MNRAS.452.1792Y}. Moreover, the extended surface mass density profile of MRK\,1216 and UGC\,2698 suggests that the growth in mass (and consequently in size) has predominantly been deposited in the outer parts. Indeed, integrating the density profile shows that the mass content in the outer regions has grown disproportionally, with $\sim$ 40 per cent of the total stellar mass already being located beyond 5\,kpc in MRK\,1216 and UGC\,2698, in contrast to e.g. NGC\,1277 which only harbours 10 per cent of the total stellar mass at these radii.

The small sizes, density profiles and the disproportional growth in mass in the remote regions of the largest galaxies in our sample endorse the theory in which our galaxies are i) (largely) passively evolved analogues of the "red nuggets" and ii) expected to form the cores of massive present-day ellipticals. As suggested by the density profiles, the evolution into massive spheroids is closely related to the accretion of low surface density material in the outer parts while at the same time decreasing the central densities. Whether this evolution can solely be ascribed to minor and/or major merging and is capable of reconciling their detailed properties with the massive, present-day ETG population, though, requires a more thorough inspection of their total mass density profiles (Sec. \hyperref[sec:total_mass_density]{\ref{sec:total_mass_density}}), orbital distributions (Sec. \hyperref[sec:stellar_angular_momentum]{\ref{sec:stellar_angular_momentum}}) as well as of their stellar populations (Sec. \hyperref[sec:stellar_populations]{\ref{sec:stellar_populations}}).

Regarding the peak stellar mass surface densities; simple virial arguments \citep{2009ApJ...697.1290B} as well as numerical simulations \citep{2013MNRAS.429.2924H} show that major merging cannot reduce the central surface mass densities. Minor merging may account for this fact and is highly necessary to explain the stellar mass build-up in the outer parts. However, if minor mergers are the dominant growth channel for CEGs we would expect a decrease of the central surface mass densities, which is not evident from the profiles of the most massive and largest objects in our sample. Also, subsequent minor merging is expected to increase the scatter in ETG scaling relations - i.e. the scatter in the relation between stellar mass, size and velocity dispersion - \citep{2012MNRAS.422.1714N}, which is not observed, and the required rate of minor mergers has been challenged, too \citep{2012ApJ...746..162N}. Additional processes such as adiabatic expansion, because of mass loss during active galactic nuclei (AGN) feedback \citep{2008ApJ...689L.101F,2010ApJ...718.1460F}, may therefore be necessary.

%---------------------------------------------------------------------
\subsection{Total mass density slope, dark matter fraction and mass-to-light ratio}
\label{sec:total_mass_density}
%---------------------------------------------------------------------

\begin{figure}
		\begin{center}
		\includegraphics[width=.47\textwidth]{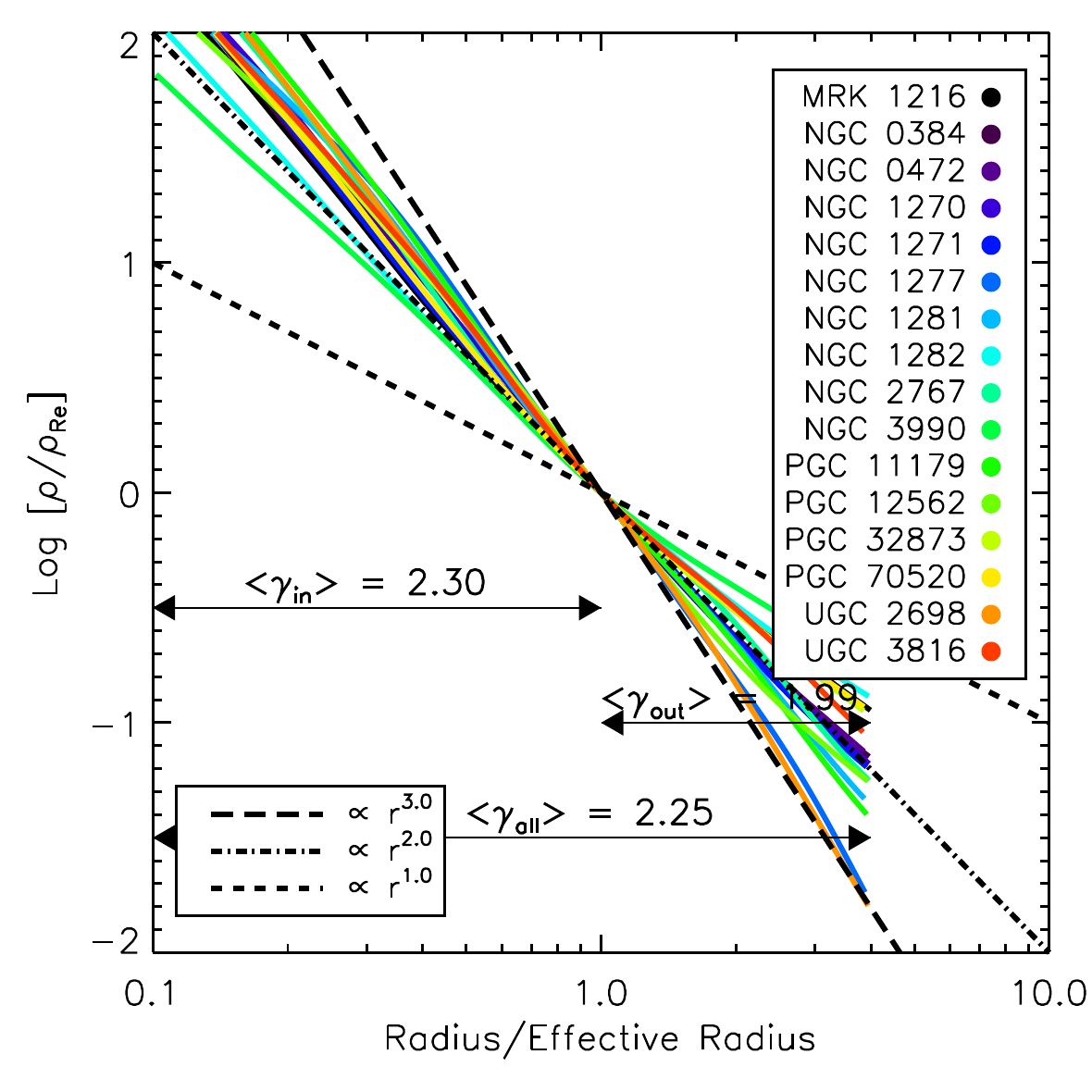}
		\end{center}
	\caption{Total mass density slopes of our CEG sample from our best-fitting dynamical models. The density profiles fall off steeper than the isothermal density profile ($\rho \propto r^{2.0}$) within the effective radius, with a mean slope of 2.30, but are on average almost perfectly isothermal beyond. The deviation of the central total mass density slopes from an isothermal profile is related to the high stellar mass densities, which point to highly dissipative formation scenarios, and the close connection with the massive ETG population at $z \sim 2$ calls for an evolution of the slope with redshift ($\gamma \propto z$).}
	\label{fig:density_slopes}
\end{figure}

Rather than dissecting the contribution of luminous and dark matter to the total mass budget, we can investigate the total mass profile, which is assumed to hold additional clues regarding the formation and evolution of galaxies \citep[e.g.][]{2013ApJ...766...71R,2013MNRAS.432.2496D,2017MNRAS.464.3742R}. To this end, we illustrate the total mass density profiles, consisting of the stellar and dark matter density profile, as a function of effective radius in Fig. \hyperref[fig:density_slopes]{\ref{fig:density_slopes}}. The profile of each galaxy has been derived from the respective best-fitting dynamical model and the corresponding total mass density slope $\gamma$ from a least squares power-law fit to this profile, within the range of 0.1\,\Reff\ $\le$ r $\le$ 1.0\,\Reff\ ($\gamma_{in}$), 1.0\,\Reff\ $\le$ r $\le$ 4.0\,\Reff\ ($\gamma_{out}$) and 0.1\,\Reff\ $\le$ r $\le$ 4.0\,\Reff\ ($\gamma_{all}$). The black dashed and dash-dotted lines depict density profiles with slopes of $\gamma=1$, 2 and $\gamma=3$, with $\gamma=2$ corresponding to an isothermal density profile that is commonly observed in ETGs from both lensing and dynamics \citep[][]{2006ApJ...649..599K,2007ApJ...667..176G,2009ApJ...703L..51K,2010ApJ...724..511A,2009MNRAS.399...21B,2011MNRAS.415.2215B}. As evident from Fig. \hyperref[fig:density_slopes]{\ref{fig:density_slopes}}, the total mass density profiles of our compact galaxy sample are in general steeper than the isothermal density profile within the effective radius, reflecting the dominance of the baryonic mass accumulation in this region, and reach peak values of $\gamma_{in} = 2.54$ in individual cases. The mean slope of $<\gamma_{in}> = 2.30$ with an rms scatter of $\sigma_{\gamma}=0.17$ is also noticeably steeper than the average total mass density slope of regular ETGs from the \textsc{SLUGGS} and \textsc{ATLAS$^{3D}$} sample, measured within the same radial extent \citep{2015ApJ...804L..21C}. On the other hand, the density profiles are almost perfectly isothermal on average, with a slope of $<\gamma_{out}> =1.99$, beyond 1\,\Reff\ and close to the average value of 2.19$\pm$0.03 reported in \cite{2015ApJ...804L..21C} when fitting the profile between 0.3 and 4\,\Reff\ ($<\gamma_{all}> =2.25$).

Given the association of CEGs with the red and massive galaxy population at $z \sim 2$ (Fig. \hyperref[fig:mass_size_relation]{\ref{fig:mass_size_relation}} and \hyperref[fig:surface_mass_densities]{\ref{fig:surface_mass_densities}}) and the comparison with present-day ETGs, the total mass density slopes within the effective radius ($\gamma_{in}$) would imply a change of the slope over cosmic time. Even if our sample size is currently too small to be statistically significant, there is a clear trend towards higher total mass density slopes with increasing redshift. This is at odds with the claim of no evolution of the slope from the \textsc{SLACS} survey \citep{2009ApJ...703L..51K} and also disagrees with the tentative evidence presented in \cite{2011ApJ...727...96R} and \cite{2012ApJ...757...82B}, where a steepening of $\gamma$ with decreasing redshift is observed.

Interestingly, \cite{2017MNRAS.464.3742R} analyse the total mass density slopes of a suite of cosmological (zoom) simulations, for galaxies in the redshift range $0 \le\ z \le\ 2$. In these simulations, a correlation between the total mass density slope and the dark matter fraction within 1\,\Reff\ is presented ($\gamma \propto f_\mathrm{DM}^{-1}$), as well as an evolution of the density slope with redshift ($\gamma \propto z$). After postulating that the CEGs in this study are indeed passively evolved analogues of the "red nuggets", we can confirm the trend towards higher values with increasing redshift, as promoted by the simulations. In the simulations, higher density slopes imply higher fractions of stars formed in-situ due to e.g. gas-rich (binary) mergers. While purely "wet" binary mergers are an unlikely formation scenario for present-day ellipticals \citep[see also][]{2008ApJ...685..897B}, they are qualitatively a plausible formation mechanism of compact and massive ellipticals at high redshift and thus for our sample. Actually, gas-rich mergers are more frequent at high redshift and necessary to explain the old stellar populations and short formation timescales of the massive, red galaxy population at $z=2$. SMGs at $3 \le z \le 6$, for instance, are poster children of gas-rich, compact and disky systems. \cite{2014ApJ...782...68T} show that their co-moving number densities match the number densities of the compact and massive ellipticals at $z=2$, if their duty cycles are of the order of 42\,Myr. The gas-rich major mergers would also be able to explain the positive deviation of the central density slopes of our compact galaxies from $\gamma \sim 2.1$. Because of its dissipative nature, gas condenses in the centre of galaxies and gives rise to massive starbursts which increase the central stellar mass densities and hence the total mass density slope in these regions while keeping the galaxies compact. Similarly, \cite{2013MNRAS.432.2496D} perform an investigation of the total mass density slopes for models within the $\Lambda$CDM framework. Although their definition of the total mass density slope slightly differs from ours, a clear correlation between $\gamma$ and the stellar mass surface density as well as between $\gamma$ and the effective radius is found, with smaller and denser galaxies having overall higher total mass density slopes. Taking into account the fact of smaller half-light radii and higher densities for both the CEG sample as well as for the population of massive and passive galaxies at $z \sim 2$, this would advocate our findings of a positive correlation of the slope with increasing redshift.

We note, though, that the mean total mass density slope of our CEGs within 4\,\Reff\ ($<\gamma_{all}>$) is in conflict with the predicted slope of $\gamma = 2.45 $ for galaxies at $z \sim 2$ from \cite{2017MNRAS.464.3742R}. This can be traced back to the influence of the dark halo. The median dark matter fraction within 1\,\Reff\ is only 11 per cent for our sample, but the dark halo can become quite dominant beyond this distance and thus flattens the total mass density slope $\gamma_{all}$ significantly. Considering that \cite{2017MNRAS.464.3742R} obtain a dark matter fraction of $f_\mathrm{DM}=10\pm5$ per cent within the effective radius for their objects at $z=2$, one would expect a comparable influence of the dark halo beyond the effective radius and thus shallower total mass density slopes in their simulations, unless their central stellar mass density slopes differ even stronger from an isothermal profile than the values presented here.\\

Next, we compare the dark matter fraction of our sample to literature values of dynamically inferred dark matter fractions. From axisymmetric Jeans models of a statistically significant sample of 260 ETGs in the local volume, \cite{2013MNRAS.432.1709C} derive a median dark matter fraction of 11 per cent within 1\,\Reff. However, the total stellar masses of the \textsc{ATLAS$^{3D}$} sample span a range of $9.5 \le$ log(\Mstar) $\le\ 12$, whereas our sample has a median total stellar mass of log$(\Mstar)=11.08$, with NGC\,3990 being the only strong outlier from this value with a total stellar mass of only log$(\Mstar)=9.63$. Given this median total stellar mass estimate from our dynamical models, the expected dark matter fraction of our sample would be 19 per cent, according to \cite{2013MNRAS.432.1709C}. We attribute the lower dark matter fraction to i) the smaller effective sizes of our galaxies and ii) a systematic variation of the stellar $M/L$ and hence of the IMF\footnote{An in-depth investigation of an IMF variation in these objects and its physical drivers is beyond the scope of this paper, but will be addressed in a future publication.}. Assuming that the dark halo profile in our objects can be parametrised by a spherically symmetric NFW profile, the smaller effective radius encompasses less of the dark volume and therefore yields lower dark matter fractions within the same radial extent. Whereas the dark matter fraction is almost 50 per cent lower than expected from local ETGs, it is in line with dynamical dark matter estimates of compact ($\le$ 3\,kpc) and massive (log$(\Mstar) \ge\ 11$) galaxies at $z \ge 1.5$ where, based on the central velocity dispersion measurement and a virial mass estimator, \cite{2013ApJ...771...85V} report a dark matter fraction of roughly 10 per cent\footnote{The dark matter estimates in \cite{2013ApJ...771...85V} have been derived under the assumption of a Chabrier IMF and are thus prone to the aforementioned systematic variation of the IMF.}.

Keep in mind that both the increase of the dark matter fraction within the effective radius as well as the decrease of the total mass density slopes as a function of cosmic time are qualitatively in agreement with an evolution that is dominated by minor mergers. As pointed out above, the radical size growth within the last 10\,Gyr leads to an increase of the dark volume and, since the stellar mass density profile falls off steeper than the dark halo density profile, this entails a flattening of the total mass density slopes towards an isothermal density profile as well as an increase of the enclosed dark matter fraction to the total mass content.\\

The stellar M/L and hence the dark matter fractions that are employed throughout this paper are constrained via our orbit-based dynamical models and are thus insensitive to any assumptions about the IMF. They depend, however, on the assumption of a radially constant M/L and a dark halo shape, which can be parametrised by a spherically symmetric NFW profile. Given the uniformly old ages (see Sec. \hyperref[sec:stellar_populations]{\ref{sec:stellar_populations}}), the former assumption appears to be a reasonable choice. Yet, recent claims of a radially dependent IMF \citep[e.g.][]{2015MNRAS.447.1033M} would imply a radially varying M/L and hence induce a deviation from the sample's aforementioned median dark matter fraction. To test this scenario, we have limited the dynamical models to fit the kinematic data within one effective radius only. A radially varying IMF should manifest itself in vastly different M/L constraints at different radii (unless the break is well within the effective radius, which would not be resolved by our data). Fits to the data within \Reff, however, do not dramatically change the derived M/L and thus lead to insignificant variations in the inferred dark matter fractions. We observe changes in the M/L of maximally 18 per cent when compared to the fiducial fits, but there is no trend apparent; the variations in the M/L and dark matter fraction for the whole sample cancel out such that the median dark matter fraction within one effective radius remains roughly 11 per cent.

When it comes to the stellar M/L, we notice slightly higher values (on average) from the dynamics than inferred from the stellar populations, while assuming a canonical Kroupa-like IMF for the latter. The dynamical stellar M/L estimates are, however, lower than expected from a Salpeter-like IMF. In some cases a Salpeter IMF can even be excluded, as the SPS stellar mass estimate would overshoot the total mass estimate from the dynamics, which also includes dark matter. We will provide a more detailed discussion of possible IMF constraints in the future, but briefly comment here that we cannot support the rather simplified claim of a very bottom-heavy IMF in high velocity dispersion galaxies. Even if the dynamical M/L constraints point towards an IMF that is "heavier" than a canonical Kroupa IMF, a more complex IMF shape (such as a double power-law profile) might be necessary to bring both the dynamical and stellar population estimates into agreement \citep[e.g.][]{2016MNRAS.463.3220L}.

%---------------------------------------------------------------------
\subsection{Specific stellar angular momentum}
\label{sec:stellar_angular_momentum}
%---------------------------------------------------------------------

Efforts to classify the population of ETGs have so far used the shape of their isophotes \citep{1996ApJ...464L.119K}, the deficit/excess of light in their centre \citep{2009ApJS..182..216K} or the amount of ordered versus random motion \citep{1983ApJ...266...41D}. But, while the latter attempt has been shown to be susceptible to projection effects \citep{2005MNRAS.363..597B}, the former tries to encompass the dynamical state and evolutionary history of galaxies purely based on information encoded in their photometric profiles. The \textsc{SAURON} and \textsc{ATLAS$^{3D}$} surveys have conducted a large photometric and spectroscopic analysis of a volume-limited, representative sample of ellipticals in the nearby universe. They define a new parameter $\lambda_r$ -  the specific stellar angular momentum - which is effective in discriminating between the two classes of slow and fast-rotating ETGs \citep{2007MNRAS.379..401E}. This classification scheme does not only provide a good estimate of the amount of large-scale rotation in galaxies, but more importantly is tightly related to the orbital configuration of galaxies \citep{2007MNRAS.379..418C} and hence is assumed to contain information regarding their mass assembly \citep[e.g.][]{2008ApJ...685..897B,2014MNRAS.444.3357N}. For this purpose, we show the specific angular momentum profile out to 3\,\Reff. All galaxies exhibit a rising angular momentum profile except NGC\,0384 and NGC\,0472, which are peaked at $\sim 1$\,\Reff\ and have a minimal dip beyond that radius. Naturally, all galaxies could be considered as fast rotators, according to the revised classification scheme in \cite{2011MNRAS.414..888E}. However, inspecting the line-of-sight velocity moments in more detail, we will show that the most massive galaxy UGC\,2698 is only at the edge of being a fast rotator and has probably been slowed down due to a slightly more complex merging history.\\

\begin{figure}
		\begin{center}
		\includegraphics[width=.47\textwidth]{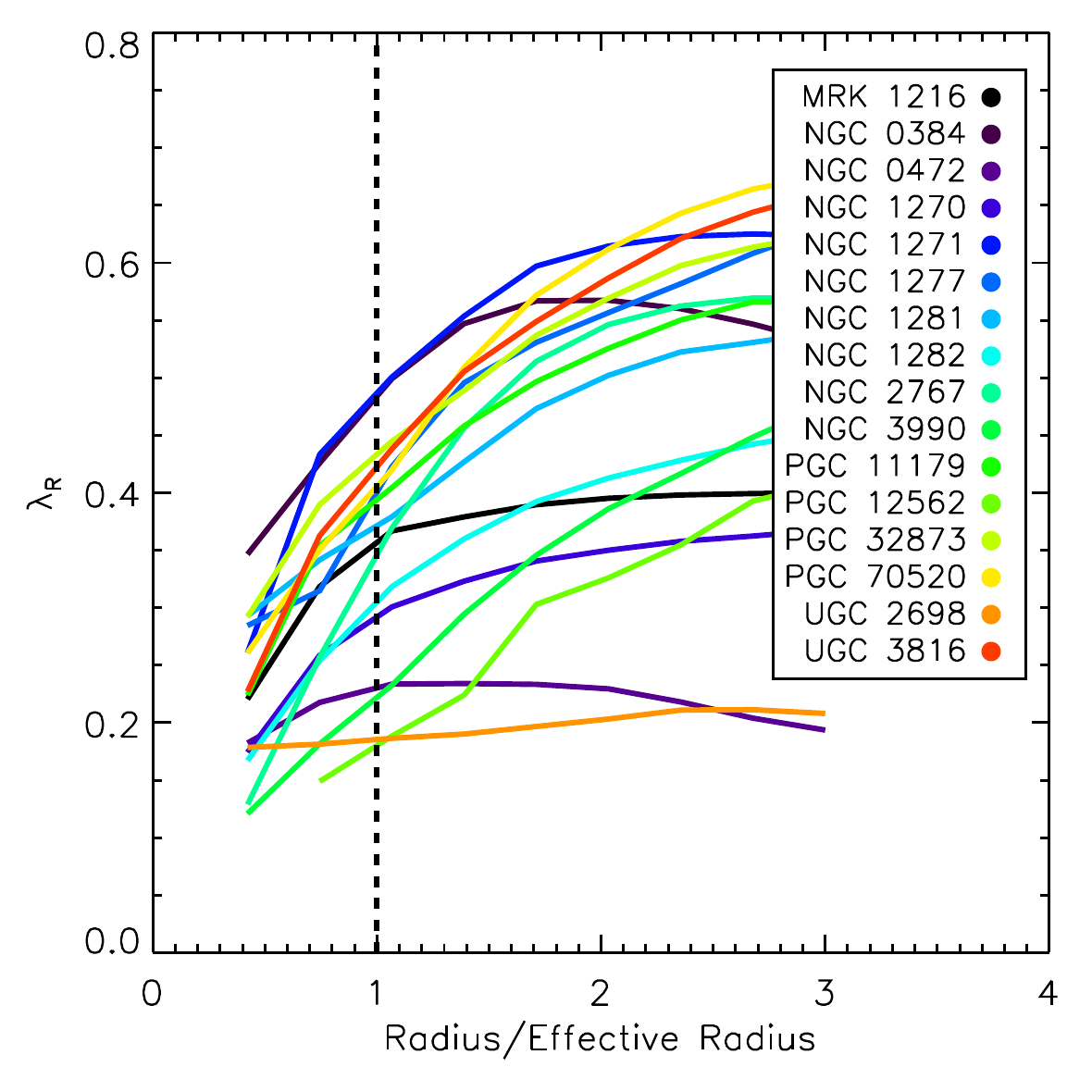}
		\end{center}
	\caption{Measurement of the specific stellar angular momentum of the CEG sample as a function of effective radius, with typical uncertainties of $\Delta\lambda_r \sim 0.03$ at 1\,\Reff. The angular momentum profiles are constantly rising, except for NGC\,0384 and NGC\,0472. All galaxies are fast rotators, as is expected from their rapid and regular rotation. When compared with hydrodynamical simulations, the $\lambda_r$ profiles promote the theory in which these objects might be the products of gas-rich major mergers.}
	\label{fig:lambda_r}
\end{figure}

\begin{figure}
		\begin{center}
		\includegraphics[width=.47\textwidth]{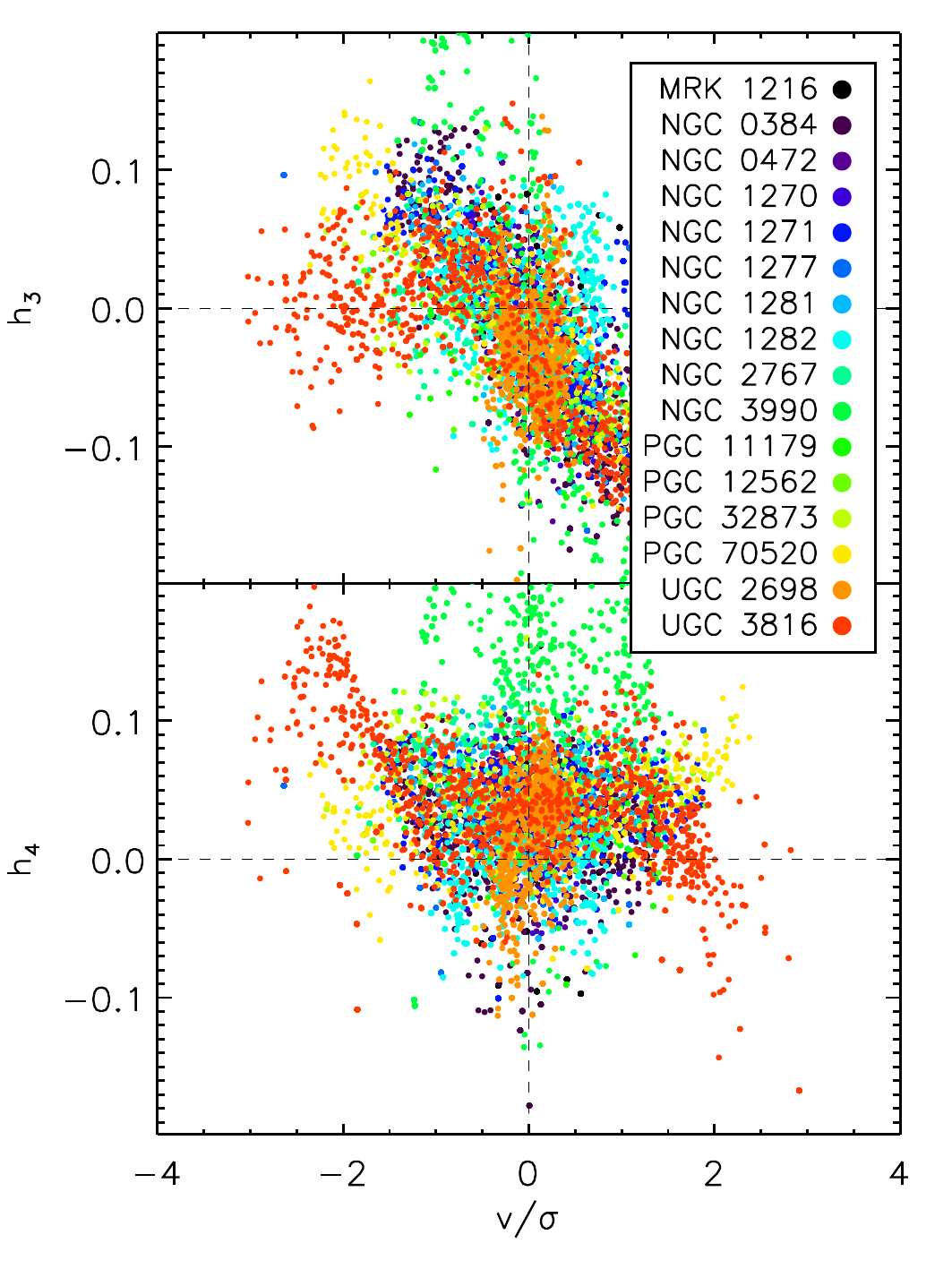}
		\end{center}
	\caption{$v/\sigma$ vs. $h_3$ (top) and $v/\sigma$ vs. $h_4$ (bottom) distribution for all CEGs. The points correspond to the measurements of the line-of-sight velocity moments in the spatially binned \ppak\ kinematics, in each of the 16 data cubes. The figure illustrates the strong anti-correlation between $v/\sigma$ and $h_3$ for all galaxies but UGC\,2698 and NGC\,1282. The anti-correlation is a signature of rotationally supported components in the orbital distribution and, according to simulations, linked to high gas fractions involved during their formation. Moreover, minor merging cannot redistribute the occupation fractions of different orbital families. This implies that the largest galaxy UGC\,2698 must have also experienced a significant "dry" (major-)merger in the past, whereas the bulk of the sample has retained much of its rotation.}
	\label{fig:vsigma_h3h4}
\end{figure}

Analogous to Sec. \hyperref[sec:total_mass_density]{\ref{sec:total_mass_density}}, we draw a comparison to cosmological hydrodynamical simulations. Even if these simulations are usually tailored to recover the cosmological formation paths of present-day ellipticals, much can be inferred with respect to the mass assembly history of our objects and hence of the population of compact ellipticals at higher redshift. In \cite{2014MNRAS.444.3357N}, $\lambda_r$ profiles such as those presented in Fig. \hyperref[fig:lambda_r]{\ref{fig:lambda_r}} are ascribed to fast rotators of class A or B. These objects distinguish themselves via a gas-rich major merging event which has led to a spin-up of the merger remnant and to massive amounts of in-situ star formation. In the simulations, this spin-up yields the highest rotational velocities of the order of $\sim 200$\,\kms\ and a strong anti-correlation between $v$ and $h_3$, indicative of the presence of a large, rotationally supported component. Both features, along with a centrally peaked velocity dispersion profile, are commonly found in our sample. The imprint of (non-)dissipational processes in the orbital distribution of galaxies can also be visualised by inspecting the correlation of the line-of-sight velocity moments \citep{2009ApJ...705..920H,2010ApJ...723..818H}. This is shown in Fig. \hyperref[fig:vsigma_h3h4]{\ref{fig:vsigma_h3h4}}, where we display the $v/\sigma$ vs. $h_3$ and $v/\sigma$ vs. $h_4$ relation of all CEGs in our sample. A strong anti-correlation between $v/\sigma$ and $h_3$ is found in 14 out of the 16 galaxies in our sample and related to the dominance of orbits with a high net angular momentum \citep[see also][for the orbital decomposition of individual objects in our sample]{2015MNRAS.452.1792Y,2015ApJ...808..183W}. This effect is well studied in idealised "wet" merger simulations. A fraction of the cold gas looses its angular momentum to strong torques, which arise as a consequence of the non-axisymmetric perturbations during the merger; gas settles in the centre and the gradient in the gravitational potential well steepens. Stars on box orbits are then more difficult to maintain due to scattering processes \citep[e.g.][]{1996ApJ...471..115B,2007MNRAS.376..997J}. In addition, some gas retains its angular momentum during the merger and rebuilds an embedded stellar disk afterwards \citep[][and references therein]{2009ApJ...691.1168H}. On top of the surviving disk, the stars formed in the embedded disk are preferentially found in short-axis tube orbits, thus contributing to the net rotation and the skewed velocity profile.

If taken at face value, these correlations provide corroborating evidence for a theory in which these objects are the products of highly dissipational (merger) events at $z \ge 2$. The fast rotator UGC\,2698, however, harbours a non-negligible amount of box orbits in our models and shows no anti-correlation between $v/\sigma$ and $h_3$. UGC\,2698 is the most massive and largest galaxy in our sample. As highlighted in Sec. \hyperref[sec:stellar_mass_surface]{\ref{sec:stellar_mass_surface}}, the tail of the surface mass density profile endorses a picture where this object has grown predominantly via mass accumulation in the outer regions. Fig. \hyperref[fig:vsigma_h3h4]{\ref{fig:vsigma_h3h4}}, on the other hand, implies that minor merging alone cannot account for the change in the orbital distribution of UGC\,2698. If this would be the case, we would expect a similar impact on the orbital distribution in its slightly smaller siblings (i.e. no $v/\sigma$ vs. $h_3$ anti-correlation, prevalence of box orbits, diminished $\lambda_r$ values etc.), which have also grown in mass and size, but which is not evident in e.g. MRK\,1216. Major merging, on the other hand, is known to be able to redistribute the occupation fractions of the different orbital families more effectively. Yet, late ($z \le 2$) "wet" major mergers are expected to rejuvenate the stellar content, which is not apparent from our SPS fits (see Sec. \hyperref[sec:stellar_populations]{\ref{sec:stellar_populations}}). Similarly, late ($z \le 2$) "dry" (i.e. gas-poor) major mergers can largely be ruled out for the bulk of our sample, as collisionless major mergers tend to wash out any (anti-)correlation between $v/\sigma$ and $h_3$ \citep{2006MNRAS.372..839N} because of thermalisation of the LOSVD during violent relaxation \citep{1967MNRAS.136..101L}, but an additional "dry" (major or at least intermediate mass) merger appears to be a likely scenario for UGC\,2698.\\

Unfortunately, spatially resolved measurements of the LOSVD for galaxies beyond $z \ge 2$ are almost non-existent, owing to their small angular sizes and the difficulties in obtaining high S/N of the continuum. As a consequence, a direct comparison between our CEG sample and their analogues at higher redshift is currently not possible, except for a gravitationally lensed compact and quiescent galaxy at $z\sim2.6$ \citep{2015ApJ...813L...7N}. While higher-order velocity moments could not be extracted in that particular case, \cite{2015ApJ...813L...7N} report a high degree of rotation, despite the galaxy's rather low ellipticity, and thus support our findings of rotationally supported orbital configurations. Clearly, this accordance can only be regarded as anecdotal evidence at the moment and the lensed compact galaxy is not necessarily representative of the "red nuggets", but the high-quality observations for the local CEGs can serve as a benchmark until more measurements at high-$z$ become available.\\

The classification of CEGs as fast rotators is not surprising given their disky SB profiles, which are merely a representation of their rotationally supported orbital configurations. More important, though, is the opportunity of reverse engineering their photometric and kinematic properties in order to unravel their early ($z\ge2$) formation paths, given their almost passive evolution afterwards, in contrast to their local descendants where the (violent) growth mechanisms might have diluted much of this information. Once the detailed properties of the ETG population at $z\sim2$ (or of their local analogues) have been pinned down, they can be employed as points of reference to put our currently favoured formation scenarios to the test. For instance, if CEGs are indeed the progenitors of the most massive ellipticals today and constitute their cores, then some of our disky fast-rotating objects must also evolve into rounder slow-rotating ellipticals. The observed build-up of the low surface mass density wings (Sec. \hyperref[sec:stellar_mass_surface]{\ref{sec:stellar_mass_surface}}) shows that minor merging is effective in increasing their sizes and a plausible mechanism to explain the evolution of the total mass density slopes and dark matter fractions (Sec. \hyperref[sec:total_mass_density]{\ref{sec:total_mass_density}}), but (as pointed out previously) can neither reduce the central surface mass densities nor radically change the stellar angular momentum profile and redistribute the orbital composition. In addition to the drastic size growth implied by minor merging, the necessity of major mergers can therefore not be underestimated for the recovery of the diverse kinematic and photometric properties of massive ETGs that are commonly observed in the local universe. This is also in line with cosmological hydrodynamical simulations. In \cite{2016MNRAS.456.1030W}, the evolutionary paths of compact, massive and quiescent galaxies at high redshift are followed. Although the majority of these objects survives, by inhabiting the central parts of their more massive descendants, the diverse evolution of this galaxy population as a whole underscores the necessity of more than just one principal channel for the formation of massive present-day ellipticals.

%---------------------------------------------------------------------
\subsection{Stellar ages, metallicities and abundance ratios}
\label{sec:stellar_populations}
%---------------------------------------------------------------------

\begin{figure}
		\begin{center}
		\includegraphics[width=.47\textwidth]{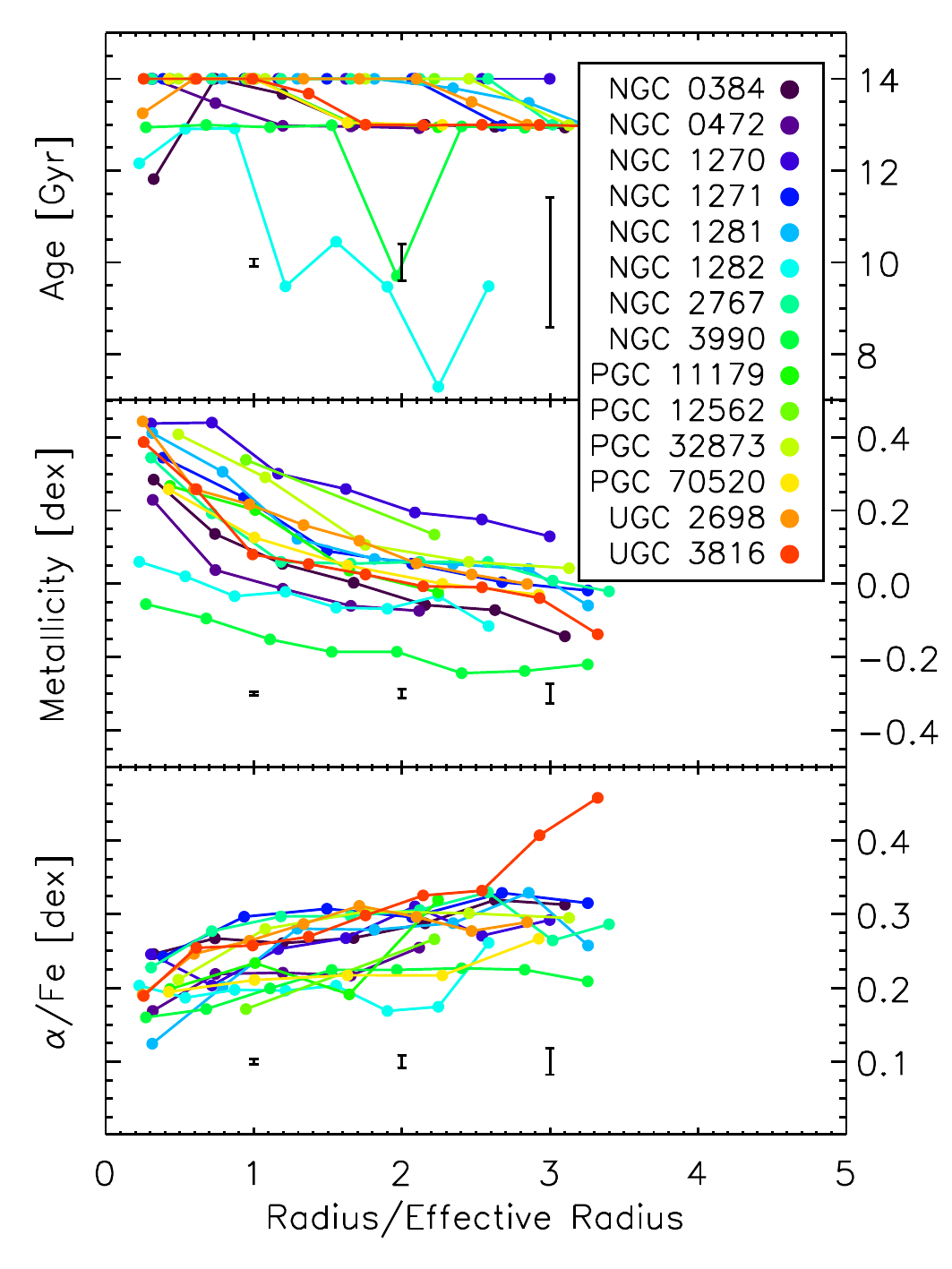}
		\end{center}
	\caption{Stellar population properties (coloured) and associated formal 1-$\sigma$ errors (black bars) of 14 CEGs, for which \ppak\ data in the V500 setup is available. Stellar ages (top), metallicities (middle) and abundance ratios (bottom) are plotted as a function of effective radius. The uniformly old stellar ages out to 3\,\Reff\ support the theory that these objects have been assembled more than 10\,Gyr ago, with little to no star formation that might have been triggered by more recent "wet" mergers. The metallicities are super-solar within the effective radius and decrease rapidly outwards, whereas the $\alpha$-abundances show a mild increase.}
	\label{fig:stellar_pops}
\end{figure}

So far, the resemblance to the "red nuggets" and any conclusion with respect to their assembly history has been established from a purely dynamical and photometric point of view. The stellar populations present an independent method to assess the robustness of this tight observational link and the feasibility of the aforementioned early ($z \ge 2$) and late ($z \le 2$) formation and evolution channels. In Fig. \hyperref[fig:stellar_pops]{\ref{fig:stellar_pops}}, we show the stellar age, metallicity and $\alpha$-abundance measurements as a function of effective radius, in 14 out of our 16 galaxies for which data in the V500 setup is available. This excludes MRK\,1216 and NGC\,1277 from the analysis (which are missing relevant absorption line features, owing to the narrower coverage of the V1200 setup), but we refer the reader to \cite{2014ApJ...780L..20T}, \cite{2015MNRAS.451.1081M} and \cite{2017arXiv170105197F} for the spatially resolved stellar populations of NGC\,1277 and MRK\,1216 based on long-slit spectroscopic data.

As illustrated in Fig. \hyperref[fig:stellar_pops]{\ref{fig:stellar_pops}}, the stellar ages of our sample are uniformly old, with little to no variation out to 3\,\Reff. Even after taking into account the formal fitting uncertainties, the stars cannot be younger than 10\,Gyr for most objects. Only NGC\,1282 and NGC\,3990 contain a significantly younger stellar population beyond the effective radius. This, however, is expected, based on ample evidence in both their photometric and kinematic properties for recent merger/stripping interactions. Thus, the stellar content of CEGs must have been assembled beyond redshift $z=2$, similar to the central stellar populations of nearby and massive ETGs \citep[e.g.][]{2005ApJ...621..673T}, followed by a passive (i.e. non-star forming) evolution thereafter. Turning to the metallicity and $\alpha$-abundance measurements, we observe well-known radial trends. The metallicity decreases rapidly as a function of effective radius, dropping from super-solar values of 0.2 $\le$ Fe/H $\le$ 0.4 to roughly solar metallicities at the outermost bins. Likewise, super-solar metallicities in this range are usually observed in the central parts of the most massive and oldest ETGs \citep[i.e. CEGs are positive outliers from the mass-metallicity relation;][]{2005MNRAS.362...41G,2008MNRAS.391.1117P,2014ApJ...791L..16G}, demonstrating that these measurements can only be reconciled with the locally established relations if these objects either end up in the cores of today's massive ETG population or if their central metallicities somehow decrease at a fixed stellar mass. While the former scenario is supported by the gradual build-up of a stellar envelope (Fig. \hyperref[fig:surface_mass_densities]{\ref{fig:surface_mass_densities}}), the latter is very unlikely as any substantial change in the stellar metallicities is mainly driven by (major) mergers and accompanied by a non-negligible stellar mass increase in a universe where structures grow hierarchically \citep{1980MNRAS.191P...1W,2004MNRAS.347..740K,2009A&A...499..427D,2010ApJ...721L..48K,2015MNRAS.449..528H}. Furthermore, the chemical abundance ratios are also super-solar and show a mild rise with increasing distance from the centre, hinting at highly efficient and fast star formation time scales and indicating that significant late minor accretion is highly unlikely for the bulk of our sample as those stellar populations would result in lower abundances due to their formation in shallower gravitational potential wells \citep{2015A&A...582A..46W}. From a photometric point of view, the observed trends also imply modest colour gradients \citep{2016MNRAS.456..538Y}, mainly driven by the metallicity variation. This stands in contrast to the steep gradients reported at high and intermediate redshifts \citep{2011ApJ...735...18G,2012MNRAS.425.2698G,2016MNRAS.457.2845T,2017MNRAS.465.3185O}. Note, however, that minor differences in the stellar population ages will further drive the colour differences in the former (i.e. at high redshifts, age variations will be an important contributor to the colour gradients), whereas some evolution (presumably due to minor merging) might explain the enhanced gradients in the latter.\\

\begin{figure}
		\begin{center}
		\includegraphics[width=.47\textwidth]{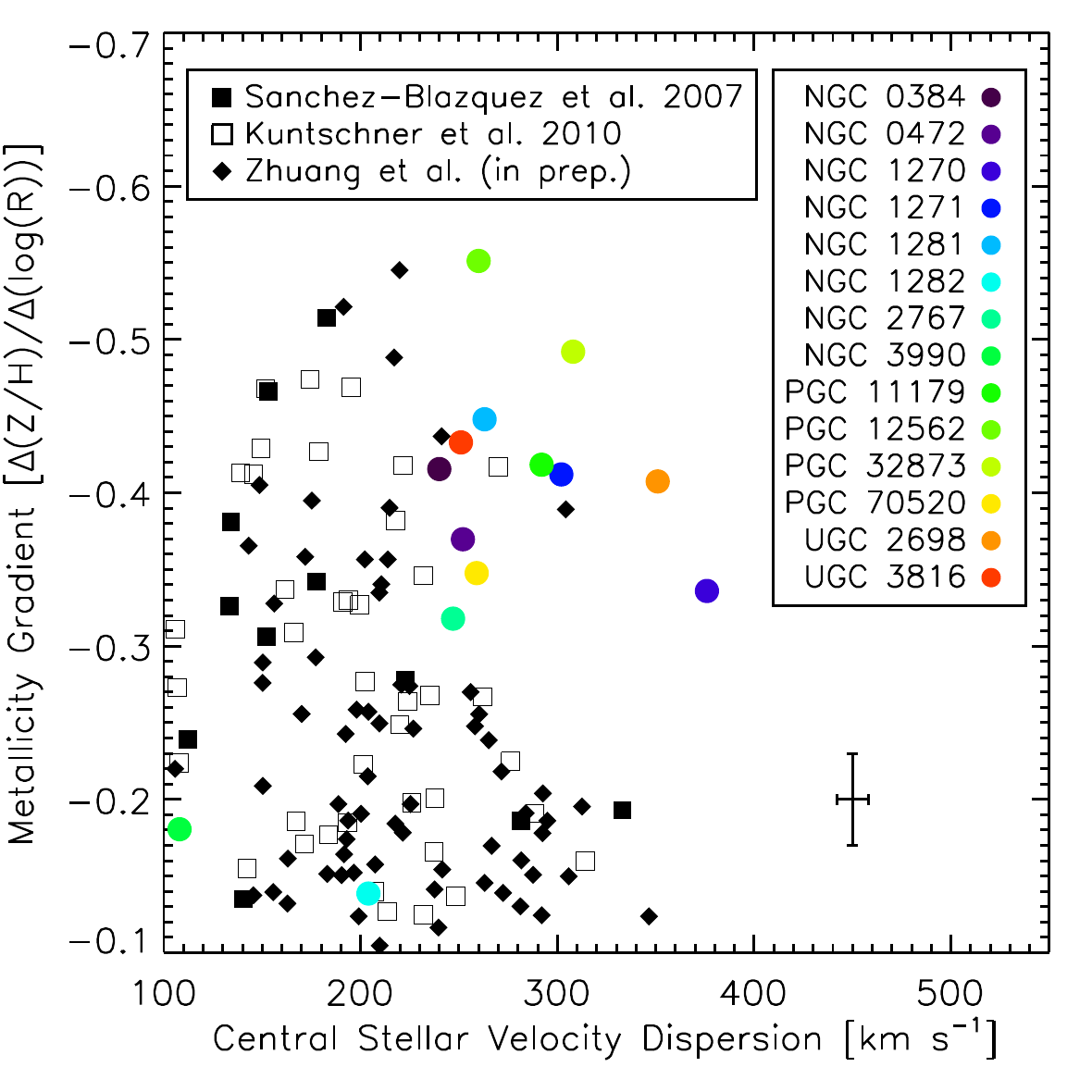}
		\end{center}
	\caption{Metallicity gradient vs. central stellar velocity dispersion of 14 CEGs. The black error bar highlights the typical measurement uncertainties. In comparison, we show the metallicity gradients of regular ETGs from dedicated long-slit \citep{2007MNRAS.377..759S} and IFU observations, including the \textsc{SAURON} \citep{2010MNRAS.408...97K} and \textsc{CALIFA} (Zhuang et al. in prep.) survey. Besides the tidally stripped galaxy NGC\,3990 and the recent merger remnant NGC\,1282, CEGs have very steep metallicity gradients and populate a region which is generally avoided by ordinary ETGs in the local universe.}
	\label{fig:metallicity_gradients}
\end{figure}

Dedicated long-slit and IFU observations have also obtained spatially resolved metallicity and chemical abundance ratios of elliptical galaxies in the nearby universe. In Fig. \hyperref[fig:metallicity_gradients]{\ref{fig:metallicity_gradients}}, we present a small compilation of literature measurements \citep{2007MNRAS.377..759S}, including the results of the \textsc{SAURON} \citep{2010MNRAS.408...97K} and \textsc{CALIFA} (Zhuang et al. in prep.) survey. The figure displays the metallicity slope $\Delta$(Z/H)/$\Delta$(log(R) (hereafter $\Delta$\,Z/H) as a function of central stellar velocity dispersion $\sigma_c$, with the slopes of the CEGs highlighted in colour. Whereas both the \textsc{SAURON} and \textsc{CALIFA} galaxies sample a large range in $\Delta$\,Z/H and $\sigma_c$, the CEGs are highly clustered in $\sigma_c$ (apart from the two objects NGC\,1282 and NGC\,3990). More interestingly, however, is the occupation of a region which is generally avoided by local ETGs. Given their small half-light radii and the large \ppak\ spaxels (with a 1\,\arcsec\ sampling), the central stellar velocity dispersion measurements of our CEGs cover a larger radius (in a relative sense) than the literature values (which are generally measured within \Reff/8). Furthermore, the steep central gradients are not resolved in the \ppak\ data, with its $\sim3$\,\arcsec\ wide PSF. As a result, the central stellar velocity dispersion measurements as well as the metallicity gradients presented here will be underestimated and the discrepancy in the coverage of the $\Delta$\,Z/H $-$ $\sigma_c$ plot will further increase. Nonetheless, and irrespective of these shortcomings, the mean slope of our sample is -0.41$\pm$0.07 and hence already higher than anticipated for regular old ($\ge$ 8\,Gyr) ellipticals within e.g. the \textsc{SAURON} survey (-0.25$\pm$0.11). Only the highest velocity dispersion galaxies in \cite{2010MNRAS.408..272S} exhibit steep metallicity gradients comparable to the ones observed here. Yet, those are generally brightest cluster galaxies (BCGs) where the gradients might originate from a completely different mechanism, where e.g. cold gas infall and subsequent star formation in the centre of the cluster potential well increases the slope very late in their evolution (but see also \citealt{2007MNRAS.375....2D} for a semi-analytical discussion of the gas and stellar content of BCGs).\\

Steep metallicity gradients are generally predicted by classical models of dissipative collapse \citep{1974MNRAS.166..585L,1984ApJ...286..403C} and their more sophisticated variants - which try to account for radiative cooling, supernovae feedback and chemical enrichment \citep[e.g.][]{2002MNRAS.335..335C,2010MNRAS.407.1347P} - as well as by models within the $\Lambda$CDM framework, where highly dissipative mergers at high redshifts are frequent \citep[e.g.][]{2004MNRAS.347..740K,2009ApJS..181..135H}. In both models, steep gradients arise as a consequence of metal enriched gas that is continuously funnelled towards the centre. Based on the observations, it is currently not possible to clearly distinguish between an early (and strictly) dissipative collapse and a gas-rich merger for the formation of CEGs. The two scenarios are capable of reproducing metallicity gradients in the observed range. Dissipative collapse, for instance, is known to preserve the initial gradients, which can be as high as $\Delta$\,Z/H $=-1$. But, as the stellar populations of the CEGs have gradually and passively aged, the differences in their stellar populations must have faded in the meantime, suggesting that $\sim$ 10\,Gyr ago the metallicity gradients could well agree with the estimates of models of dissipative collapse. Note that some minor merging, as indicated by the surface mass density profiles in Fig \hyperref[fig:surface_mass_densities]{\ref{fig:surface_mass_densities}}, might have countered the fading by depositing metal poor stars in the outer parts \citep{2015MNRAS.449..528H}. Similarly, \cite{2009ApJS..181..135H} demonstrate that steep gradients ($\Delta$\,Z/H $\le -0.8$) will be observed in early ($z\ge2$) galaxy-galaxy encounters, where pre-merger gas fractions of 40 per cent and more are involved and which have been detected in the SMG population \citep{2010ApJ...714.1407C,2011ApJ...739L..31R,2014ApJ...782...68T}.

Of course, dissipative collapse and gas-rich mergers are not mutually exclusive. On the contrary, high densities, the abundance of cold gas and increasing merger rates are attributes of the early universe, and the galaxies presented here might have formed through either one of these formation channels or maybe even via a combination of both. In fact, \cite{2015MNRAS.449..361W} show that dissipative collapse as well as gas-rich mergers can equally contribute to the population of massive and passive galaxies at $z\sim2$. Within their simulations they find observable differences between the two formation channels, such as the dependence of the stellar ages as a function of radius, with dissipative collapse yielding younger stellar ages at larger radii. Even though we observe a slight trend of younger stellar populations with increasing distance from the centre, the measurements are fully consistent with a uniformly old population within the margin of error. Keep also in mind that mergers, too, are predicted to give rise to slightly younger stellar ages in the outer parts. Those younger stars could be a sign of embedded disks that form out of cold gas after the merger, adding to the strong anti-correlation in the line-of-sight velocity moments (Fig. \hyperref[fig:vsigma_h3h4]{\ref{fig:vsigma_h3h4}}), so that (at this point) we cannot employ the information encoded in the stellar ages as a key discriminator between both.\\

Whereas the currently available data stops us from pinning down the high-redshift formation history of these objects, the stellar populations provide constraints for their evolution since $z\sim2$. As mentioned previously, "wet" major mergers are very unlikely for these objects after $z\sim2$. In principle, they can be reconciled with the high stellar mass density and the rotationally supported orbital configuration of our CEGs. Yet, the rejuvenation of the stellar content after a gas-rich merger within the last 10\,Gyr should leave imprints in the stellar age estimates. The impact of such an event is evident in NGC\,1282 for instance, but clearly missing for the majority of our sample. In comparison, "dry" major mergers of galaxies with comparably old stellar populations will preserve the consistently old stellar age estimates. However, violent relaxation during the merger will decrease their metallicity and chemical abundance gradients by as much as a factor of two \citep{1980MNRAS.191P...1W,2009A&A...499..427D,2015MNRAS.449..528H} which, after taking into account the fading effect and the underestimation of the central gradients, is difficult to reconcile with the measurements presented here as this would entail gradients at the time of their formation that most likely overshoot even the most optimistic estimates from simulations.

Concerning the possible evolution of CEGs into regular ETGs; as they grow in half-light radius because of minor merging, this will result in a lower measurement of the central stellar velocity dispersion, thus moving CEGs closer to the population of regular, nearby ellipticals in Fig. \hyperref[fig:metallicity_gradients]{\ref{fig:metallicity_gradients}}. Also, the stellar mass increase after successive minor mergers will be able to align their mass-metallicity values with those of their more massive descendants. However, when it comes to the reproduction of the metallicity gradients of the most massive ETGs in the local neighbourhood, major merging is a much more efficient way of decreasing the slopes of CEGs, indicating that - even if many CEGs might end up in the cores of present-day ellipticals as a result of minor mergers - some CEGs must experience a major merger in oder to recover the shallowest metallicity gradients at the highest stellar velocity dispersions. This is also supported by the most recent analysis of the stellar populations within the Illustris simulations \citep{2014MNRAS.444.1518V} where the stellar metallicity gradients depend on the amount of ex-situ accretion, with the shallowest gradients usually found in those galaxies with the strongest accretion histories \citep{2016ApJ...833..158C}.\\

%============================= Section 6 =============================
\section{Summary \& Outlook}
\label{sec:summary}
%=====================================================================

In this paper, we presented an investigation of a sample of 16 compact elliptical galaxies (CEGs) in the nearby universe ($\le$ 112\,Mpc). Based on deep \hst\ photometric and wide-field \ppak\ IFU spectroscopic observations, we have shown that all objects are small (\Reff\ $\le$ 4\,kpc), disky ($b/a \le\ 0.73$) and fast-rotating ($v \le 280$\,\kms), with exceptional central stellar velocity dispersions ($\sigma_c \le 380$\,\kms). These objects closely resemble the population of massive and passive galaxies at $z=$ 2 and therefore present the opportunity to obtain an in-depth glimpse into the progenitors of today's most massive galaxies.\\

With this suite of data, we constructed orbit-based dynamical models to constrain their stellar and dark matter content, their stellar mass surface density profiles as well as their total mass density slopes. In addition, we investigated their stellar angular momentum profiles and spatially resolved stellar populations. The analysis yielded following conclusions:
\begin{itemize}
\item CEGs in the local universe are outliers from the present-day stellar mass-size relation, but in formidable agreement with the stellar mass-size relation at $z \sim 2$.
\item The stellar mass surface density profiles of CEGs follow the profiles of massive and quiescent galaxies at higher redshift and therefore endorse the theory that these objects are indeed passively evolved analogues of the "red nuggets". Moreover, the spread in their structural properties, such as their stellar masses and sizes, allows us to constrain their subsequent evolution channels into regular and massive ellipticals. In particular the largest galaxies in our sample show a build-up of stellar mass, which is predominantly deposited in the remote regions ($\ge$ 5\,kpc) and indicative of a stellar mass assembly that (thus far) is dominated by minor mergers.
\item The central ($\le 1\,\Reff$) total mass density slopes of CEGs exceed the density slope of an isothermal profile. This is driven by the high stellar mass concentration and the negligible contribution of dark matter within one half-light radius. Given the unambiguous affiliation with the population of massive and passive galaxies at $z\sim2$, this implies an evolution of the total mass density slopes with redshift ($\gamma \propto z$), as predicted by e.g. highly dissipational merger simulations. Moreover, the progenitors of our galaxies (and consequently of the compact and massive galaxy population at earlier times) must have been disky and gas rich systems that have either merged or collapsed directly, in order to provide the strong deviations from an isothermal density profile.

Adopting an NFW profile, we also constrained their dark halo contribution to the total mass budget. Within one effective radius, CEGs are highly dominated by the stellar component, with a mean stellar-to-dynamical mass ratio of 0.89. The high stellar-to-dynamical mass ratios are driven by the low amount of dark matter due to the decrease in effective radius which, then again, encompasses less of the dark volume. Our dynamical results therefore suggest an evolution of the dark matter fraction with increasing redshift ($f_\mathrm{DM} \propto z$).
\item All CEGs are fast rotators. Their specific stellar angular momentum profiles are in line with those that are observed within hydrodynamical simulations, where early ($z \ge 2$) "wet" mergers yield a spin-up of the merger remnant. "Dry" major mergers can largely be ruled out for the bulk of our sample as violent relaxation is expected to wash out their rotationally supported orbital configurations, which manifest themselves in a clear anti-correlation of the line-of-sight velocity moments $v$ and $h_3$.
\item The spatially resolved stellar populations of all galaxies are uniformly old ($\ge$ 10\,Gyr), except for two objects which show clear signs of recent merger/stripping interactions. The stellar mass assembly must therefore have happened beyond redshift $z\sim2$, followed by a passive evolution thereafter. This provides corroborating evidence for a largely passive evolution within the last 10\,Gyr and the close resemblance to the "red nuggets".

Their central stellar metallicities are super solar, ranging from 0.2 $\le$ Fe/H $\le$ 0.4, making CEGs positive outliers in the mass-metallicity relation. Their metallicities, however, are consistent with the central metallicities of the oldest and most massive ellipticals in the nearby universe, showing that both measurements can be brought into agreement if CEGs form the cores of today's massive ETG population.

The stellar metallicity and chemical abundance gradients are considerably higher than those found in local elliptical galaxies, with a mean of ($\Delta$(Z/H)/$\Delta$(log(R))$=-0.41\pm0.07$. Given their passive evolution since $z\sim2$, the differences in their light-weighted stellar populations are expected to decrease such that their gradients at the time of formation must have been even stronger. Whereas gradients of this magnitude are predicted by both monolithic collapse and dissipative mergers and therefore cannot be used to strictly pin down their formation paths at this stage, they (in combination with the stellar age estimates and the dynamical characteristics) again raise a "wet" and/or "dry" major merger scenario within the last 10\,Gyr into question.
\end{itemize}

The body of evidence presented throughout this paper is unambiguous and confirms that we have found mostly passively evolved analogues of compact, massive and passive galaxies at $z\sim2$. These objects therefore not only offer an unprecedented view of the structural, dynamical and chemical properties of high-redshift galaxies, which is currently not possible for their analogues at earlier times, but can also serve as tight observational constraints for models of galaxy formation and evolution.

Their largely pristine characteristics hide important information about their (violent) formation histories and their subsequent evolution channels. But, despite the intriguing possibilities of exploiting this wealth of information, the analysis presented here also poses some challenges for the current two-phase paradigm within which the strong structural evolution of massive and passive galaxies since $z\sim2$ is thought to be dictated by numerous minor mergers. If minor merging is indeed the dominant evolutionary path after $z\sim2$, is it capable of i) reconciling the metallicity gradient measurements with those of their more massive counterparts in our vicinity, ii) redistributing the orbital configuration in such a way that slow-rotating galaxies, too, can be formed, iii) explaining the evolution of the total mass density slope over cosmic time and iv) decreasing the peak surface mass densities in order to match those of their local descendants?

It is certainly beyond of the scope of this paper to provide a definitive answer to these questions, but we have tried to qualitatively assess the impact of minor merging on the aforementioned issues. Minor merging will affect i) and iii); by increasing the half-light radius, the central velocity dispersion will decrease and move the CEG sample closer to the population of local ellipticals in the $\Delta$\,Z/H $-$ $\sigma_c$ diagram. Whether the absolute values of the metallicity gradients decrease, however, strongly depends on the chemical composition of satellite galaxies, but it is clear that some evolution is necessary in order to align the measurements of the mean metallicity slopes of both nearby ETGs and CEGs, in particular at the high mass end. Likewise, the growth in size will increase the dark matter contribution to the total mass content within the effective radius, which will yield shallower total mass density slopes that are more consistent with the locally observed isothermal density profile. In contrast, minor merging alone will struggle to explain ii) and iv); it is currently not clear how minor mergers can decrease the stellar angular momentum profile by redistributing the stellar orbits effectively, since satellite material is mainly stored beyond a few effective radii. Furthermore, we do not observe a decrease in the peak stellar mass surface densities even for those objects in our sample which have already grown considerably by building up a stellar envelope. We therefore conclude that, while minor merging is key for the strong size evolution of massive and passive galaxies since $z\sim2$, minor merging alone is insufficient to reconcile the detailed structural properties of CEGs with their present-day descendants and cannot be regarded as the single channel for the evolution of CEGs into massive present-day ellipticals. 

Processes, such as AGN feedback, have been put forward as an explanation of the strong stellar mass-size evolution since $z\sim2$ and as a possible resort for the discrepancy in the central stellar mass surface densities of massive and passive ETGs at high and low-redshift \citep{2008ApJ...689L.101F}. Yet, those processes are expected to dramatically decrease the central stellar velocity dispersions \citep{2010ApJ...718.1460F} and therefore disagree with observational evidence which hints at a modest evolution \citep{2009ApJ...696L..43C,2014ApJ...789...92B}. Similarly, \cite{2013MNRAS.431.2350I} propose AGN-driven star formation in the outskirts to partially account for the evolution of the "red nuggets" at higher redshifts. However, this is at odds with the uniformly old stellar age estimates of our sample, indicating that theses scenarios can only assist in the evolution of the "red nuggets" rather than being considered as the main underlying physical drivers.\\

It is also worth noting here that, despite the many differences in the structural and dynamical properties of CEGs and local ETGs, CEGs still line up in several of the continuous sequences of galaxy properties as observed e.g. within the \textsc{ATLAS$^{3D}$} survey \citep{2016ARA&A..54..597C}. For instance, the ages, velocity dispersions and mass density slopes of our CEGs can roughly be estimated by extrapolating the \textsc{ATLAS$^{3D}$} values towards the most compact and densest fast rotators. That is, the compact elliptical galaxy sample in our study basically straddles the zone of exclusion (ZOE) relation and thus represents the most extreme objects at a given stellar mass. Nonetheless, important differences remain; the stellar M/L are generally lower than expected from the \textsc{ATLAS$^{3D}$} sample and the presence of a prominent bulge component is more than dubios. The CEGs in this work are commonly fast-rotating and rather flat, with low \sersic\ indices, and even if a bulge component is detected via an ambiguous photometric two-component decomposition, the decomposition itself remains inconclusive as it contradicts the dynamical decomposition \citep{2015MNRAS.452.1792Y}. We therefore argue that CEGs do not necessarily follow the same evolutionary path of regular fast-rotating ETGs in the mass-size diagram, as put forward by the \textsc{ATLAS$^{3D}$} survey \citep{2013MNRAS.432.1862C}. The lack of a bulge component also casts doubt on the efficiency of morphological quenching and its importance for the shutoff of star formation \citep{2009ApJ...707..250M} in CEGs in general. If anything, the presence of high central densities alone seems to be sufficient to quench galaxies rapidly \citep{2015ApJ...813...23V}. Interestingly, \cite{2016ApJ...832L..11M} finds a correlation between a galaxy's SMBH mass and its stellar populations, with galaxies harbouring overmassive black holes also showing the oldest and most $\alpha$-enhanced stellar populations. So far, three black hole mass constraints are available for our CEG sample. With all three being positive outliers from the black hole scaling relations \citep{2015ApJ...808..183W,2016ApJ...817....2W,2017ApJ...835..208W}, this suggests that black hole growth and feedback might have played an equally decisive role in the evolution of CEGs and thus for the most massive ETGs; a conclusion which was also reached by \cite{2013ApJ...765..104B}, where AGN feedback was frequently observed for the likely progenitors of the compact and quiescent galaxy population at $z\sim2$.\\

Finally, we emphasise that our sample size here is merely a result of our strict sample selection in the HETMGS (with a hard cut in SOI $\ge 0.05$\,\arcsec\ and \Reff\ $\le 2$\,kpc), yielding a lower limit for the number density of these objects of $\sim 2.5\times10^{-6}$\,Mpc$^{3}$ in the local universe (i.e. within a distance of 112\,Mpc and excluding NGC\,3990 from the CEG sample). However, more CEGs can be found by relaxing the selection criteria, thus increasing the sample size to a statistically more significant number \citep[e.g.][]{2015A&A...578A.134S}.

%=====================================================================

\section*{Acknowledgements}
A. Y{\i}ld{\i}r{\i}m acknowledges support from the Max Planck Institute for Astronomy and Hans-Walter Rix in particular, and thanks Arjen van der Wel and Thorsten Naab for helpful discussions as well as Yulong Zhuang and Harald Kuntschner for sharing their SPS data. The data presented here is based on observations collected at the Centro Astronómico Hispano Alemán (CAHA) at Calar Alto, operated jointly by the Max Planck Institute for Astronomy and the Instituto de Astrofísica de Andalucía (CSIC). This research also based on observations made with the NASA/ESA \textit{Hubble Space Telescope}. The observations are associated with program \#13050. Support for program \#13050 was provided by NASA through a grant from the Space Telescope Science Institute, which is operated by the Association of Universities for Research in Astronomy, Inc., under NASA contract NAS 5-26555. Furthermore, this research made use of the NASA/IPAC Extragalactic Database (NED) which is operated by the Jet Propulsion Laboratory, California Institute of Technology, under contract with the National Aeronautics and Space Administration, and the SDSS DR8. Funding for the SDSS and SDSS-II has been provided by the Alfred P. Sloan Foundation, the Participating Institutions, the National Science Foundation, the U.S. Department of Energy, the National Aeronautics and Space Administration, the Japanese Monbukagakusho, the Max Planck Society, and the Higher Education Funding Council for England. The SDSS Web Site is http://www.sdss.org/.
The SDSS is managed by the Astrophysical Research Consortium for the Participating Institutions. The Participating Institutions are the American Museum of Natural History, Astrophysical Institute Potsdam, University of Basel, University of Cambridge, Case Western Reserve University, University of Chicago, Drexel University, Fermilab, the Institute for Advanced Study, the Japan Participation Group, Johns Hopkins University, the Joint Institute for Nuclear Astrophysics, the Kavli Institute for Particle Astrophysics and Cosmology, the Korean Scientist Group, the Chinese Academy of Sciences (LAMOST), Los Alamos National Laboratory, the Max-Planck-Institute for Astronomy (MPIA), the Max-Planck-Institute for Astrophysics (MPA), New Mexico State University, Ohio State University, University of Pittsburgh, University of Portsmouth, Princeton University, the United States Naval Observatory, and the University of Washington.

%=====================================================================

%=====================================================================
% FIGURES
%=====================================================================

%=====================================================================
% REFERENCES
%=====================================================================

\bibliographystyle{yahapj}
\bibliography{mn2e}

\begin{thebibliography}{210}
\providecommand\natexlab[1]{#1}
\providecommand\JournalTitle[1]{#1}

\bibitem[{{Auger} {et~al.}(2010){Auger}, {Treu}, {Bolton}, {Gavazzi},
  {Koopmans}, {Marshall}, {Moustakas}, \& {Burles}}]{2010ApJ...724..511A}
{Auger}, M.~W., {Treu}, T., {Bolton}, A.~S., {et~al.} 2010,
  \href{http://dx.doi.org/10.1088/0004-637X/724/1/511}{\JournalTitle{\apj},
  724, 511}

\bibitem[{{Baldry} {et~al.}(2004){Baldry}, {Glazebrook}, {Brinkmann},
  {Ivezi{\'c}}, {Lupton}, {Nichol}, \& {Szalay}}]{2004ApJ...600..681B}
{Baldry}, I.~K., {Glazebrook}, K., {Brinkmann}, J., {et~al.} 2004,
  \href{http://dx.doi.org/10.1086/380092}{\JournalTitle{\apj}, 600, 681}

\bibitem[{{Barnab{\`e}} {et~al.}(2011){Barnab{\`e}}, {Czoske}, {Koopmans},
  {Treu}, \& {Bolton}}]{2011MNRAS.415.2215B}
{Barnab{\`e}}, M., {Czoske}, O., {Koopmans}, L.~V.~E., {Treu}, T., \& {Bolton},
  A.~S. 2011,
  \href{http://dx.doi.org/10.1111/j.1365-2966.2011.18842.x}{\JournalTitle{\mnras},
  415, 2215}

\bibitem[{{Barnab{\`e}} {et~al.}(2009){Barnab{\`e}}, {Czoske}, {Koopmans},
  {Treu}, {Bolton}, \& {Gavazzi}}]{2009MNRAS.399...21B}
{Barnab{\`e}}, M., {Czoske}, O., {Koopmans}, L.~V.~E., {et~al.} 2009,
  \href{http://dx.doi.org/10.1111/j.1365-2966.2009.14941.x}{\JournalTitle{\mnras},
  399, 21}

\bibitem[{{Barnes}(1988)}]{1988ApJ...331..699B}
{Barnes}, J.~E. 1988,
  \href{http://dx.doi.org/10.1086/166593}{\JournalTitle{\apj}, 331, 699}

\bibitem[{{Barnes}(1989)}]{1989Natur.338..123B}
---. 1989, \href{http://dx.doi.org/10.1038/338123a0}{\JournalTitle{\nat}, 338,
  123}

\bibitem[{{Barnes} \& {Hernquist}(1996)}]{1996ApJ...471..115B}
{Barnes}, J.~E., \& {Hernquist}, L. 1996,
  \href{http://dx.doi.org/10.1086/177957}{\JournalTitle{\apj}, 471, 115}

\bibitem[{{Barro} {et~al.}(2013){Barro}, {Faber}, {P{\'e}rez-Gonz{\'a}lez},
  {Koo}, {Williams}, {Kocevski}, {Trump}, {Mozena}, {McGrath}, {van der Wel},
  {Wuyts}, {Bell}, {Croton}, {Ceverino}, {Dekel}, {Ashby}, {Cheung},
  {Ferguson}, {Fontana}, {Fang}, {Giavalisco}, {Grogin}, {Guo}, {Hathi},
  {Hopkins}, {Huang}, {Koekemoer}, {Kartaltepe}, {Lee}, {Newman}, {Porter},
  {Primack}, {Ryan}, {Rosario}, {Somerville}, {Salvato}, \&
  {Hsu}}]{2013ApJ...765..104B}
{Barro}, G., {Faber}, S.~M., {P{\'e}rez-Gonz{\'a}lez}, P.~G., {et~al.} 2013,
  \href{http://dx.doi.org/10.1088/0004-637X/765/2/104}{\JournalTitle{\apj},
  765, 104}

\bibitem[{{Barro} {et~al.}(2014){Barro}, {Faber}, {P{\'e}rez-Gonz{\'a}lez},
  {Pacifici}, {Trump}, {Koo}, {Wuyts}, {Guo}, {Bell}, {Dekel}, {Porter},
  {Primack}, {Ferguson}, {Ashby}, {Caputi}, {Ceverino}, {Croton}, {Fazio},
  {Giavalisco}, {Hsu}, {Kocevski}, {Koekemoer}, {Kurczynski}, {Kollipara},
  {Lee}, {McIntosh}, {McGrath}, {Moody}, {Somerville}, {Papovich}, {Salvato},
  {Santini}, {Tal}, {van der Wel}, {Williams}, {Willner}, \&
  {Zolotov}}]{2014ApJ...791...52B}
---. 2014,
  \href{http://dx.doi.org/10.1088/0004-637X/791/1/52}{\JournalTitle{\apj}, 791,
  52}

\bibitem[{{Barro} {et~al.}(2016){Barro}, {Kriek}, {P{\'e}rez-Gonz{\'a}lez},
  {Trump}, {Koo}, {Faber}, {Dekel}, {Primack}, {Guo}, {Kocevski},
  {Mu{\~n}oz-Mateos}, {Rujopakarn}, \& {Seth}}]{2016ApJ...827L..32B}
{Barro}, G., {Kriek}, M., {P{\'e}rez-Gonz{\'a}lez}, P.~G., {et~al.} 2016,
  \href{http://dx.doi.org/10.3847/2041-8205/827/2/L32}{\JournalTitle{\apjl},
  827, L32}

\bibitem[{{B{\'e}dorf} \& {Portegies Zwart}(2013)}]{2013MNRAS.431..767B}
{B{\'e}dorf}, J., \& {Portegies Zwart}, S. 2013,
  \href{http://dx.doi.org/10.1093/mnras/stt208}{\JournalTitle{\mnras}, 431,
  767}

\bibitem[{{Beifiori} {et~al.}(2014){Beifiori}, {Thomas}, {Maraston}, {Steele},
  {Masters}, {Pforr}, {Saglia}, {Bender}, {Tojeiro}, {Chen}, {Bolton},
  {Brownstein}, {Johansson}, {Leauthaud}, {Nichol}, {Schneider}, {Senger},
  {Skibba}, {Wake}, {Pan}, {Snedden}, {Bizyaev}, {Brewington}, {Malanushenko},
  {Malanushenko}, {Oravetz}, {Simmons}, {Shelden}, \&
  {Ebelke}}]{2014ApJ...789...92B}
{Beifiori}, A., {Thomas}, D., {Maraston}, C., {et~al.} 2014,
  \href{http://dx.doi.org/10.1088/0004-637X/789/2/92}{\JournalTitle{\apj}, 789,
  92}

\bibitem[{{Bell} \& {de Jong}(2001)}]{2001ApJ...550..212B}
{Bell}, E.~F., \& {de Jong}, R.~S. 2001,
  \href{http://dx.doi.org/10.1086/319728}{\JournalTitle{\apj}, 550, 212}

\bibitem[{{Bell} {et~al.}(2003){Bell}, {McIntosh}, {Katz}, \&
  {Weinberg}}]{2003ApJS..149..289B}
{Bell}, E.~F., {McIntosh}, D.~H., {Katz}, N., \& {Weinberg}, M.~D. 2003,
  \href{http://dx.doi.org/10.1086/378847}{\JournalTitle{\apjs}, 149, 289}

\bibitem[{{Bender} {et~al.}(1994){Bender}, {Saglia}, \&
  {Gerhard}}]{1994MNRAS.269..785B}
{Bender}, R., {Saglia}, R.~P., \& {Gerhard}, O.~E. 1994, \JournalTitle{\mnras},
  269, 785

\bibitem[{{Bezanson} {et~al.}(2013){Bezanson}, {van Dokkum}, {van de Sande},
  {Franx}, \& {Kriek}}]{2013ApJ...764L...8B}
{Bezanson}, R., {van Dokkum}, P., {van de Sande}, J., {Franx}, M., \& {Kriek},
  M. 2013,
  \href{http://dx.doi.org/10.1088/2041-8205/764/1/L8}{\JournalTitle{\apjl},
  764, L8}

\bibitem[{{Bezanson} {et~al.}(2009){Bezanson}, {van Dokkum}, {Tal},
  {Marchesini}, {Kriek}, {Franx}, \& {Coppi}}]{2009ApJ...697.1290B}
{Bezanson}, R., {van Dokkum}, P.~G., {Tal}, T., {et~al.} 2009,
  \href{http://dx.doi.org/10.1088/0004-637X/697/2/1290}{\JournalTitle{\apj},
  697, 1290}

\bibitem[{{Bezanson} {et~al.}(2011){Bezanson}, {van Dokkum}, {Franx},
  {Brammer}, {Brinchmann}, {Kriek}, {Labb{\'e}}, {Quadri}, {Rix}, {van de
  Sande}, {Whitaker}, \& {Williams}}]{2011ApJ...737L..31B}
{Bezanson}, R., {van Dokkum}, P.~G., {Franx}, M., {et~al.} 2011,
  \href{http://dx.doi.org/10.1088/2041-8205/737/2/L31}{\JournalTitle{\apjl},
  737, L31}

\bibitem[{{Binney}(1978)}]{1978MNRAS.183..501B}
{Binney}, J. 1978, \JournalTitle{\mnras}, 183, 501

\bibitem[{{Binney} \& {Merrifield}(1998)}]{1998gaas.book.....B}
{Binney}, J., \& {Merrifield}, M. 1998, {Galactic Astronomy}

\bibitem[{{Birnboim} \& {Dekel}(2003)}]{2003MNRAS.345..349B}
{Birnboim}, Y., \& {Dekel}, A. 2003,
  \href{http://dx.doi.org/10.1046/j.1365-8711.2003.06955.x}{\JournalTitle{\mnras},
  345, 349}

\bibitem[{{Boardman} {et~al.}(2016){Boardman}, {Weijmans}, {van den Bosch},
  {Zhu}, {Yildirim}, {van de Ven}, {Cappellari}, {de Zeeuw}, {Emsellem},
  {Krajnovi{\'c}}, \& {Naab}}]{2016MNRAS.460.3029B}
{Boardman}, N.~F., {Weijmans}, A.-M., {van den Bosch}, R., {et~al.} 2016,
  \href{http://dx.doi.org/10.1093/mnras/stw1187}{\JournalTitle{\mnras}, 460,
  3029}

\bibitem[{{Bolton} {et~al.}(2012){Bolton}, {Brownstein}, {Kochanek}, {Shu},
  {Schlegel}, {Eisenstein}, {Wake}, {Connolly}, {Maraston}, {Arneson}, \&
  {Weaver}}]{2012ApJ...757...82B}
{Bolton}, A.~S., {Brownstein}, J.~R., {Kochanek}, C.~S., {et~al.} 2012,
  \href{http://dx.doi.org/10.1088/0004-637X/757/1/82}{\JournalTitle{\apj}, 757,
  82}

\bibitem[{{Bower} {et~al.}(1992){Bower}, {Lucey}, \&
  {Ellis}}]{1992MNRAS.254..601B}
{Bower}, R.~G., {Lucey}, J.~R., \& {Ellis}, R.~S. 1992, \JournalTitle{\mnras},
  254, 601

\bibitem[{{Buitrago} {et~al.}(2008){Buitrago}, {Trujillo}, {Conselice},
  {Bouwens}, {Dickinson}, \& {Yan}}]{2008ApJ...687L..61B}
{Buitrago}, F., {Trujillo}, I., {Conselice}, C.~J., {et~al.} 2008,
  \href{http://dx.doi.org/10.1086/592836}{\JournalTitle{\apjl}, 687, L61}

\bibitem[{{Burkert} \& {Naab}(2005)}]{2005MNRAS.363..597B}
{Burkert}, A., \& {Naab}, T. 2005,
  \href{http://dx.doi.org/10.1111/j.1365-2966.2005.09482.x}{\JournalTitle{\mnras},
  363, 597}

\bibitem[{{Burkert} {et~al.}(2008){Burkert}, {Naab}, {Johansson}, \&
  {Jesseit}}]{2008ApJ...685..897B}
{Burkert}, A., {Naab}, T., {Johansson}, P.~H., \& {Jesseit}, R. 2008,
  \href{http://dx.doi.org/10.1086/591632}{\JournalTitle{\apj}, 685, 897}

\bibitem[{{Cappellari}(2002)}]{2002MNRAS.333..400C}
{Cappellari}, M. 2002,
  \href{http://dx.doi.org/10.1046/j.1365-8711.2002.05412.x}{\JournalTitle{\mnras},
  333, 400}

\bibitem[{{Cappellari}(2016)}]{2016ARA&A..54..597C}
---. 2016,
  \href{http://dx.doi.org/10.1146/annurev-astro-082214-122432}{\JournalTitle{\araa},
  54, 597}

\bibitem[{{Cappellari} \& {Copin}(2003)}]{2003MNRAS.342..345C}
{Cappellari}, M., \& {Copin}, Y. 2003,
  \href{http://dx.doi.org/10.1046/j.1365-8711.2003.06541.x}{\JournalTitle{\mnras},
  342, 345}

\bibitem[{{Cappellari} \& {Emsellem}(2004)}]{2004PASP..116..138C}
{Cappellari}, M., \& {Emsellem}, E. 2004,
  \href{http://dx.doi.org/10.1086/381875}{\JournalTitle{\pasp}, 116, 138}

\bibitem[{{Cappellari} {et~al.}(2006){Cappellari}, {Bacon}, {Bureau}, {Damen},
  {Davies}, {de Zeeuw}, {Emsellem}, {Falc{\'o}n-Barroso}, {Krajnovi{\'c}},
  {Kuntschner}, {McDermid}, {Peletier}, {Sarzi}, {van den Bosch}, \& {van de
  Ven}}]{2006MNRAS.366.1126C}
{Cappellari}, M., {Bacon}, R., {Bureau}, M., {et~al.} 2006,
  \href{http://dx.doi.org/10.1111/j.1365-2966.2005.09981.x}{\JournalTitle{\mnras},
  366, 1126}

\bibitem[{{Cappellari} {et~al.}(2007){Cappellari}, {Emsellem}, {Bacon},
  {Bureau}, {Davies}, {de Zeeuw}, {Falc{\'o}n-Barroso}, {Krajnovi{\'c}},
  {Kuntschner}, {McDermid}, {Peletier}, {Sarzi}, {van den Bosch}, \& {van de
  Ven}}]{2007MNRAS.379..418C}
{Cappellari}, M., {Emsellem}, E., {Bacon}, R., {et~al.} 2007,
  \href{http://dx.doi.org/10.1111/j.1365-2966.2007.11963.x}{\JournalTitle{\mnras},
  379, 418}

\bibitem[{{Cappellari} {et~al.}(2013{\natexlab{a}}){Cappellari}, {Scott},
  {Alatalo}, {Blitz}, {Bois}, {Bournaud}, {Bureau}, {Crocker}, {Davies},
  {Davis}, {de Zeeuw}, {Duc}, {Emsellem}, {Khochfar}, {Krajnovi{\'c}},
  {Kuntschner}, {McDermid}, {Morganti}, {Naab}, {Oosterloo}, {Sarzi}, {Serra},
  {Weijmans}, \& {Young}}]{2013MNRAS.432.1709C}
{Cappellari}, M., {Scott}, N., {Alatalo}, K., {et~al.} 2013{\natexlab{a}},
  \href{http://dx.doi.org/10.1093/mnras/stt562}{\JournalTitle{\mnras}, 432,
  1709}

\bibitem[{{Cappellari} {et~al.}(2013{\natexlab{b}}){Cappellari}, {McDermid},
  {Alatalo}, {Blitz}, {Bois}, {Bournaud}, {Bureau}, {Crocker}, {Davies},
  {Davis}, {de Zeeuw}, {Duc}, {Emsellem}, {Khochfar}, {Krajnovi{\'c}},
  {Kuntschner}, {Morganti}, {Naab}, {Oosterloo}, {Sarzi}, {Scott}, {Serra},
  {Weijmans}, \& {Young}}]{2013MNRAS.432.1862C}
{Cappellari}, M., {McDermid}, R.~M., {Alatalo}, K., {et~al.}
  2013{\natexlab{b}},
  \href{http://dx.doi.org/10.1093/mnras/stt644}{\JournalTitle{\mnras}, 432,
  1862}

\bibitem[{{Cappellari} {et~al.}(2015){Cappellari}, {Romanowsky}, {Brodie},
  {Forbes}, {Strader}, {Foster}, {Kartha}, {Pastorello}, {Pota}, {Spitler},
  {Usher}, \& {Arnold}}]{2015ApJ...804L..21C}
{Cappellari}, M., {Romanowsky}, A.~J., {Brodie}, J.~P., {et~al.} 2015,
  \href{http://dx.doi.org/10.1088/2041-8205/804/1/L21}{\JournalTitle{\apjl},
  804, L21}

\bibitem[{{Carilli} {et~al.}(2010){Carilli}, {Daddi}, {Riechers}, {Walter},
  {Weiss}, {Dannerbauer}, {Morrison}, {Wagg}, {Dav{\'e}}, {Elbaz}, {Stern},
  {Dickinson}, {Krips}, \& {Aravena}}]{2010ApJ...714.1407C}
{Carilli}, C.~L., {Daddi}, E., {Riechers}, D., {et~al.} 2010,
  \href{http://dx.doi.org/10.1088/0004-637X/714/2/1407}{\JournalTitle{\apj},
  714, 1407}

\bibitem[{{Carlberg}(1984)}]{1984ApJ...286..403C}
{Carlberg}, R.~G. 1984,
  \href{http://dx.doi.org/10.1086/162615}{\JournalTitle{\apj}, 286, 403}

\bibitem[{{Cenarro} \& {Trujillo}(2009)}]{2009ApJ...696L..43C}
{Cenarro}, A.~J., \& {Trujillo}, I. 2009,
  \href{http://dx.doi.org/10.1088/0004-637X/696/1/L43}{\JournalTitle{\apjl},
  696, L43}

\bibitem[{{Cervantes} \& {Vazdekis}(2009)}]{2009MNRAS.392..691C}
{Cervantes}, J.~L., \& {Vazdekis}, A. 2009,
  \href{http://dx.doi.org/10.1111/j.1365-2966.2008.14079.x}{\JournalTitle{\mnras},
  392, 691}

\bibitem[{{Chang} {et~al.}(2013){Chang}, {van der Wel}, {Rix}, {Wuyts},
  {Zibetti}, {Ramkumar}, \& {Holden}}]{2013ApJ...762...83C}
{Chang}, Y.-Y., {van der Wel}, A., {Rix}, H.-W., {et~al.} 2013,
  \href{http://dx.doi.org/10.1088/0004-637X/762/2/83}{\JournalTitle{\apj}, 762,
  83}

\bibitem[{{Chiosi} \& {Carraro}(2002)}]{2002MNRAS.335..335C}
{Chiosi}, C., \& {Carraro}, G. 2002,
  \href{http://dx.doi.org/10.1046/j.1365-8711.2002.05590.x}{\JournalTitle{\mnras},
  335, 335}

\bibitem[{{Cimatti} {et~al.}(2008){Cimatti}, {Cassata}, {Pozzetti}, {Kurk},
  {Mignoli}, {Renzini}, {Daddi}, {Bolzonella}, {Brusa}, {Rodighiero},
  {Dickinson}, {Franceschini}, {Zamorani}, {Berta}, {Rosati}, \&
  {Halliday}}]{2008A&A...482...21C}
{Cimatti}, A., {Cassata}, P., {Pozzetti}, L., {et~al.} 2008,
  \href{http://dx.doi.org/10.1051/0004-6361:20078739}{\JournalTitle{\aap}, 482,
  21}

\bibitem[{{Cole} {et~al.}(2001){Cole}, {Norberg}, {Baugh}, {Frenk},
  {Bland-Hawthorn}, {Bridges}, {Cannon}, {Colless}, {Collins}, {Couch},
  {Cross}, {Dalton}, {De Propris}, {Driver}, {Efstathiou}, {Ellis},
  {Glazebrook}, {Jackson}, {Lahav}, {Lewis}, {Lumsden}, {Maddox}, {Madgwick},
  {Peacock}, {Peterson}, {Sutherland}, \& {Taylor}}]{2001MNRAS.326..255C}
{Cole}, S., {Norberg}, P., {Baugh}, C.~M., {et~al.} 2001,
  \href{http://dx.doi.org/10.1046/j.1365-8711.2001.04591.x}{\JournalTitle{\mnras},
  326, 255}

\bibitem[{{Conroy} {et~al.}(2009){Conroy}, {Gunn}, \&
  {White}}]{2009ApJ...699..486C}
{Conroy}, C., {Gunn}, J.~E., \& {White}, M. 2009,
  \href{http://dx.doi.org/10.1088/0004-637X/699/1/486}{\JournalTitle{\apj},
  699, 486}

\bibitem[{{Cook} {et~al.}(2016){Cook}, {Conroy}, {Pillepich},
  {Rodriguez-Gomez}, \& {Hernquist}}]{2016ApJ...833..158C}
{Cook}, B.~A., {Conroy}, C., {Pillepich}, A., {Rodriguez-Gomez}, V., \&
  {Hernquist}, L. 2016,
  \href{http://dx.doi.org/10.3847/1538-4357/833/2/158}{\JournalTitle{\apj},
  833, 158}

\bibitem[{{Cox} {et~al.}(2006){Cox}, {Dutta}, {Di Matteo}, {Hernquist},
  {Hopkins}, {Robertson}, \& {Springel}}]{2006ApJ...650..791C}
{Cox}, T.~J., {Dutta}, S.~N., {Di Matteo}, T., {et~al.} 2006,
  \href{http://dx.doi.org/10.1086/507474}{\JournalTitle{\apj}, 650, 791}

\bibitem[{{Daddi} {et~al.}(2005){Daddi}, {Renzini}, {Pirzkal}, {Cimatti},
  {Malhotra}, {Stiavelli}, {Xu}, {Pasquali}, {Rhoads}, {Brusa}, {di Serego
  Alighieri}, {Ferguson}, {Koekemoer}, {Moustakas}, {Panagia}, \&
  {Windhorst}}]{2005ApJ...626..680D}
{Daddi}, E., {Renzini}, A., {Pirzkal}, N., {et~al.} 2005,
  \href{http://dx.doi.org/10.1086/430104}{\JournalTitle{\apj}, 626, 680}

\bibitem[{{Davies} {et~al.}(1983){Davies}, {Efstathiou}, {Fall}, {Illingworth},
  \& {Schechter}}]{1983ApJ...266...41D}
{Davies}, R.~L., {Efstathiou}, G., {Fall}, S.~M., {Illingworth}, G., \&
  {Schechter}, P.~L. 1983,
  \href{http://dx.doi.org/10.1086/160757}{\JournalTitle{\apj}, 266, 41}

\bibitem[{{de Francesco} {et~al.}(2006){de Francesco}, {Capetti}, \&
  {Marconi}}]{2006A&A...460..439D}
{de Francesco}, G., {Capetti}, A., \& {Marconi}, A. 2006,
  \href{http://dx.doi.org/10.1051/0004-6361:20065826}{\JournalTitle{\aap}, 460,
  439}

\bibitem[{{De Lucia} \& {Blaizot}(2007)}]{2007MNRAS.375....2D}
{De Lucia}, G., \& {Blaizot}, J. 2007,
  \href{http://dx.doi.org/10.1111/j.1365-2966.2006.11287.x}{\JournalTitle{\mnras},
  375, 2}

\bibitem[{{Dekel} \& {Burkert}(2014)}]{2014MNRAS.438.1870D}
{Dekel}, A., \& {Burkert}, A. 2014,
  \href{http://dx.doi.org/10.1093/mnras/stt2331}{\JournalTitle{\mnras}, 438,
  1870}

\bibitem[{{Dekel} {et~al.}(2009){Dekel}, {Birnboim}, {Engel}, {Freundlich},
  {Goerdt}, {Mumcuoglu}, {Neistein}, {Pichon}, {Teyssier}, \&
  {Zinger}}]{2009Natur.457..451D}
{Dekel}, A., {Birnboim}, Y., {Engel}, G., {et~al.} 2009,
  \href{http://dx.doi.org/10.1038/nature07648}{\JournalTitle{\nat}, 457, 451}

\bibitem[{{Di Matteo} {et~al.}(2009){Di Matteo}, {Pipino}, {Lehnert}, {Combes},
  \& {Semelin}}]{2009A&A...499..427D}
{Di Matteo}, P., {Pipino}, A., {Lehnert}, M.~D., {Combes}, F., \& {Semelin}, B.
  2009,
  \href{http://dx.doi.org/10.1051/0004-6361/200911715}{\JournalTitle{\aap},
  499, 427}

\bibitem[{{Dutton} {et~al.}(2013){Dutton}, {Macci{\`o}}, {Mendel}, \&
  {Simard}}]{2013MNRAS.432.2496D}
{Dutton}, A.~A., {Macci{\`o}}, A.~V., {Mendel}, J.~T., \& {Simard}, L. 2013,
  \href{http://dx.doi.org/10.1093/mnras/stt608}{\JournalTitle{\mnras}, 432,
  2496}

\bibitem[{{Eggen} {et~al.}(1962){Eggen}, {Lynden-Bell}, \&
  {Sandage}}]{1962ApJ...136..748E}
{Eggen}, O.~J., {Lynden-Bell}, D., \& {Sandage}, A.~R. 1962,
  \href{http://dx.doi.org/10.1086/147433}{\JournalTitle{\apj}, 136, 748}

\bibitem[{{Emsellem}(2013)}]{2013MNRAS.433.1862E}
{Emsellem}, E. 2013,
  \href{http://dx.doi.org/10.1093/mnras/stt840}{\JournalTitle{\mnras}, 433,
  1862}

\bibitem[{{Emsellem} {et~al.}(1994){Emsellem}, {Monnet}, \&
  {Bacon}}]{1994A&A...285..723E}
{Emsellem}, E., {Monnet}, G., \& {Bacon}, R. 1994, \JournalTitle{\aap}, 285,
  723

\bibitem[{{Emsellem} {et~al.}(2007){Emsellem}, {Cappellari}, {Krajnovi{\'c}},
  {van de Ven}, {Bacon}, {Bureau}, {Davies}, {de Zeeuw}, {Falc{\'o}n-Barroso},
  {Kuntschner}, {McDermid}, {Peletier}, \& {Sarzi}}]{2007MNRAS.379..401E}
{Emsellem}, E., {Cappellari}, M., {Krajnovi{\'c}}, D., {et~al.} 2007,
  \href{http://dx.doi.org/10.1111/j.1365-2966.2007.11752.x}{\JournalTitle{\mnras},
  379, 401}

\bibitem[{{Emsellem} {et~al.}(2011){Emsellem}, {Cappellari}, {Krajnovi{\'c}},
  {Alatalo}, {Blitz}, {Bois}, {Bournaud}, {Bureau}, {Davies}, {Davis}, {de
  Zeeuw}, {Khochfar}, {Kuntschner}, {Lablanche}, {McDermid}, {Morganti},
  {Naab}, {Oosterloo}, {Sarzi}, {Scott}, {Serra}, {van de Ven}, {Weijmans}, \&
  {Young}}]{2011MNRAS.414..888E}
---. 2011,
  \href{http://dx.doi.org/10.1111/j.1365-2966.2011.18496.x}{\JournalTitle{\mnras},
  414, 888}

\bibitem[{{Faber} {et~al.}(2007){Faber}, {Willmer}, {Wolf}, {Koo}, {Weiner},
  {Newman}, {Im}, {Coil}, {Conroy}, {Cooper}, {Davis}, {Finkbeiner}, {Gerke},
  {Gebhardt}, {Groth}, {Guhathakurta}, {Harker}, {Kaiser}, {Kassin},
  {Kleinheinrich}, {Konidaris}, {Kron}, {Lin}, {Luppino}, {Madgwick},
  {Meisenheimer}, {Noeske}, {Phillips}, {Sarajedini}, {Schiavon}, {Simard},
  {Szalay}, {Vogt}, \& {Yan}}]{2007ApJ...665..265F}
{Faber}, S.~M., {Willmer}, C.~N.~A., {Wolf}, C., {et~al.} 2007,
  \href{http://dx.doi.org/10.1086/519294}{\JournalTitle{\apj}, 665, 265}

\bibitem[{{Falc{\'o}n-Barroso} {et~al.}(2011){Falc{\'o}n-Barroso},
  {S{\'a}nchez-Bl{\'a}zquez}, {Vazdekis}, {Ricciardelli}, {Cardiel}, {Cenarro},
  {Gorgas}, \& {Peletier}}]{2011A&A...532A..95F}
{Falc{\'o}n-Barroso}, J., {S{\'a}nchez-Bl{\'a}zquez}, P., {Vazdekis}, A.,
  {et~al.} 2011,
  \href{http://dx.doi.org/10.1051/0004-6361/201116842}{\JournalTitle{\aap},
  532, A95}

\bibitem[{{Fan} {et~al.}(2010){Fan}, {Lapi}, {Bressan}, {Bernardi}, {De Zotti},
  \& {Danese}}]{2010ApJ...718.1460F}
{Fan}, L., {Lapi}, A., {Bressan}, A., {et~al.} 2010,
  \href{http://dx.doi.org/10.1088/0004-637X/718/2/1460}{\JournalTitle{\apj},
  718, 1460}

\bibitem[{{Fan} {et~al.}(2008){Fan}, {Lapi}, {De Zotti}, \&
  {Danese}}]{2008ApJ...689L.101F}
{Fan}, L., {Lapi}, A., {De Zotti}, G., \& {Danese}, L. 2008,
  \href{http://dx.doi.org/10.1086/595784}{\JournalTitle{\apjl}, 689, L101}

\bibitem[{{Ferr{\'e}-Mateu} {et~al.}(2015){Ferr{\'e}-Mateu}, {Mezcua},
  {Trujillo}, {Balcells}, \& {van den Bosch}}]{2015ApJ...808...79F}
{Ferr{\'e}-Mateu}, A., {Mezcua}, M., {Trujillo}, I., {Balcells}, M., \& {van
  den Bosch}, R.~C.~E. 2015,
  \href{http://dx.doi.org/10.1088/0004-637X/808/1/79}{\JournalTitle{\apj}, 808,
  79}

\bibitem[{{Ferr{\'e}-Mateu} {et~al.}(2017){Ferr{\'e}-Mateu}, {Trujillo},
  {Mart{\'{\i}}n-Navarro}, {Vazdekis}, {Mezcua}, {Balcells}, \&
  {Dom{\'{\i}}nguez}}]{2017arXiv170105197F}
{Ferr{\'e}-Mateu}, A., {Trujillo}, I., {Mart{\'{\i}}n-Navarro}, I., {et~al.}
  2017, \JournalTitle{ArXiv e-prints},
  \href{http://arxiv.org/abs/1701.05197}{{\sffamily arXiv:1701.05197}}

\bibitem[{{Franx} {et~al.}(2008){Franx}, {van Dokkum}, {Schreiber}, {Wuyts},
  {Labb{\'e}}, \& {Toft}}]{2008ApJ...688..770F}
{Franx}, M., {van Dokkum}, P.~G., {Schreiber}, N.~M.~F., {et~al.} 2008,
  \href{http://dx.doi.org/10.1086/592431}{\JournalTitle{\apj}, 688, 770}

\bibitem[{{Fukugita} {et~al.}(1998){Fukugita}, {Hogan}, \&
  {Peebles}}]{1998ApJ...503..518F}
{Fukugita}, M., {Hogan}, C.~J., \& {Peebles}, P.~J.~E. 1998,
  \href{http://dx.doi.org/10.1086/306025}{\JournalTitle{\apj}, 503, 518}

\bibitem[{{Gallazzi} {et~al.}(2006){Gallazzi}, {Charlot}, {Brinchmann}, \&
  {White}}]{2006MNRAS.370.1106G}
{Gallazzi}, A., {Charlot}, S., {Brinchmann}, J., \& {White}, S.~D.~M. 2006,
  \href{http://dx.doi.org/10.1111/j.1365-2966.2006.10548.x}{\JournalTitle{\mnras},
  370, 1106}

\bibitem[{{Gallazzi} {et~al.}(2005){Gallazzi}, {Charlot}, {Brinchmann},
  {White}, \& {Tremonti}}]{2005MNRAS.362...41G}
{Gallazzi}, A., {Charlot}, S., {Brinchmann}, J., {White}, S.~D.~M., \&
  {Tremonti}, C.~A. 2005,
  \href{http://dx.doi.org/10.1111/j.1365-2966.2005.09321.x}{\JournalTitle{\mnras},
  362, 41}

\bibitem[{{Gargiulo} {et~al.}(2012){Gargiulo}, {Saracco}, {Longhetti}, {La
  Barbera}, \& {Tamburri}}]{2012MNRAS.425.2698G}
{Gargiulo}, A., {Saracco}, P., {Longhetti}, M., {La Barbera}, F., \&
  {Tamburri}, S. 2012,
  \href{http://dx.doi.org/10.1111/j.1365-2966.2012.21670.x}{\JournalTitle{\mnras},
  425, 2698}

\bibitem[{{Gavazzi} {et~al.}(2007){Gavazzi}, {Treu}, {Rhodes}, {Koopmans},
  {Bolton}, {Burles}, {Massey}, \& {Moustakas}}]{2007ApJ...667..176G}
{Gavazzi}, R., {Treu}, T., {Rhodes}, J.~D., {et~al.} 2007,
  \href{http://dx.doi.org/10.1086/519237}{\JournalTitle{\apj}, 667, 176}

\bibitem[{{Girardi} {et~al.}(2000){Girardi}, {Bressan}, {Bertelli}, \&
  {Chiosi}}]{2000A&AS..141..371G}
{Girardi}, L., {Bressan}, A., {Bertelli}, G., \& {Chiosi}, C. 2000,
  \href{http://dx.doi.org/10.1051/aas:2000126}{\JournalTitle{\aaps}, 141, 371}

\bibitem[{{Gonzaga} {et~al.}(2012){Gonzaga}, {Hack}, {Fruchter}, \&
  {Mack}}]{2012drzp.book.....G}
{Gonzaga}, S., {Hack}, W., {Fruchter}, A., \& {Mack}, J. 2012, {The DrizzlePac
  Handbook}

\bibitem[{{Gonz{\'a}lez Delgado} {et~al.}(2014){Gonz{\'a}lez Delgado}, {Cid
  Fernandes}, {Garc{\'{\i}}a-Benito}, {P{\'e}rez}, {de Amorim},
  {Cortijo-Ferrero}, {Lacerda}, {L{\'o}pez Fern{\'a}ndez}, {S{\'a}nchez}, {Vale
  Asari}, {Alves}, {Bland-Hawthorn}, {Galbany}, {Gallazzi}, {Husemann},
  {Bekeraite}, {Jungwiert}, {L{\'o}pez-S{\'a}nchez}, {de Lorenzo-C{\'a}ceres},
  {Marino}, {Mast}, {Moll{\'a}}, {del Olmo}, {S{\'a}nchez-Bl{\'a}zquez}, {van
  de Ven}, {V{\'{\i}}lchez}, {Walcher}, {Wisotzki}, {Ziegler}, \&
  {Collaboration920}}]{2014ApJ...791L..16G}
{Gonz{\'a}lez Delgado}, R.~M., {Cid Fernandes}, R., {Garc{\'{\i}}a-Benito}, R.,
  {et~al.} 2014,
  \href{http://dx.doi.org/10.1088/2041-8205/791/1/L16}{\JournalTitle{\apjl},
  791, L16}

\bibitem[{{Graham} {et~al.}(2016){Graham}, {Durr{\'e}}, {Savorgnan}, {Medling},
  {Batcheldor}, {Scott}, {Watson}, \& {Marconi}}]{2016ApJ...819...43G}
{Graham}, A.~W., {Durr{\'e}}, M., {Savorgnan}, G.~A.~D., {et~al.} 2016,
  \href{http://dx.doi.org/10.3847/0004-637X/819/1/43}{\JournalTitle{\apj}, 819,
  43}

\bibitem[{{Graves} {et~al.}(2009{\natexlab{a}}){Graves}, {Faber}, \&
  {Schiavon}}]{2009ApJ...693..486G}
{Graves}, G.~J., {Faber}, S.~M., \& {Schiavon}, R.~P. 2009{\natexlab{a}},
  \href{http://dx.doi.org/10.1088/0004-637X/693/1/486}{\JournalTitle{\apj},
  693, 486}

\bibitem[{{Graves} {et~al.}(2009{\natexlab{b}}){Graves}, {Faber}, \&
  {Schiavon}}]{2009ApJ...698.1590G}
---. 2009{\natexlab{b}},
  \href{http://dx.doi.org/10.1088/0004-637X/698/2/1590}{\JournalTitle{\apj},
  698, 1590}

\bibitem[{{Guo} {et~al.}(2011){Guo}, {Giavalisco}, {Cassata}, {Ferguson},
  {Dickinson}, {Renzini}, {Koekemoer}, {Grogin}, {Papovich}, {Tundo},
  {Fontana}, {Lotz}, \& {Salimbeni}}]{2011ApJ...735...18G}
{Guo}, Y., {Giavalisco}, M., {Cassata}, P., {et~al.} 2011,
  \href{http://dx.doi.org/10.1088/0004-637X/735/1/18}{\JournalTitle{\apj}, 735,
  18}

\bibitem[{{Hernquist}(1992)}]{1992ApJ...400..460H}
{Hernquist}, L. 1992,
  \href{http://dx.doi.org/10.1086/172009}{\JournalTitle{\apj}, 400, 460}

\bibitem[{{Hernquist}(1993)}]{1993ApJ...409..548H}
---. 1993, \href{http://dx.doi.org/10.1086/172686}{\JournalTitle{\apj}, 409,
  548}

\bibitem[{{Hilz} {et~al.}(2013){Hilz}, {Naab}, \&
  {Ostriker}}]{2013MNRAS.429.2924H}
{Hilz}, M., {Naab}, T., \& {Ostriker}, J.~P. 2013,
  \href{http://dx.doi.org/10.1093/mnras/sts501}{\JournalTitle{\mnras}, 429,
  2924}

\bibitem[{{Hilz} {et~al.}(2012){Hilz}, {Naab}, {Ostriker}, {Thomas}, {Burkert},
  \& {Jesseit}}]{2012MNRAS.425.3119H}
{Hilz}, M., {Naab}, T., {Ostriker}, J.~P., {et~al.} 2012,
  \href{http://dx.doi.org/10.1111/j.1365-2966.2012.21541.x}{\JournalTitle{\mnras},
  425, 3119}

\bibitem[{{Hinshaw} {et~al.}(2009){Hinshaw}, {Weiland}, {Hill}, {Odegard},
  {Larson}, {Bennett}, {Dunkley}, {Gold}, {Greason}, {Jarosik}, {Komatsu},
  {Nolta}, {Page}, {Spergel}, {Wollack}, {Halpern}, {Kogut}, {Limon}, {Meyer},
  {Tucker}, \& {Wright}}]{2009ApJS..180..225H}
{Hinshaw}, G., {Weiland}, J.~L., {Hill}, R.~S., {et~al.} 2009,
  \href{http://dx.doi.org/10.1088/0067-0049/180/2/225}{\JournalTitle{\apjs},
  180, 225}

\bibitem[{{Hirschmann} {et~al.}(2015){Hirschmann}, {Naab}, {Ostriker},
  {Forbes}, {Duc}, {Dav{\'e}}, {Oser}, \& {Karabal}}]{2015MNRAS.449..528H}
{Hirschmann}, M., {Naab}, T., {Ostriker}, J.~P., {et~al.} 2015,
  \href{http://dx.doi.org/10.1093/mnras/stv274}{\JournalTitle{\mnras}, 449,
  528}

\bibitem[{{Hoffman} {et~al.}(2009){Hoffman}, {Cox}, {Dutta}, \&
  {Hernquist}}]{2009ApJ...705..920H}
{Hoffman}, L., {Cox}, T.~J., {Dutta}, S., \& {Hernquist}, L. 2009,
  \href{http://dx.doi.org/10.1088/0004-637X/705/1/920}{\JournalTitle{\apj},
  705, 920}

\bibitem[{{Hoffman} {et~al.}(2010){Hoffman}, {Cox}, {Dutta}, \&
  {Hernquist}}]{2010ApJ...723..818H}
---. 2010,
  \href{http://dx.doi.org/10.1088/0004-637X/723/1/818}{\JournalTitle{\apj},
  723, 818}

\bibitem[{{Hogg} {et~al.}(2002){Hogg}, {Blanton}, {Strateva}, {Bahcall},
  {Brinkmann}, {Csabai}, {Doi}, {Fukugita}, {Hennessy}, {Ivezi{\'c}}, {Knapp},
  {Lamb}, {Lupton}, {Munn}, {Nichol}, {Schlegel}, {Schneider}, \&
  {York}}]{2002AJ....124..646H}
{Hogg}, D.~W., {Blanton}, M., {Strateva}, I., {et~al.} 2002,
  \href{http://dx.doi.org/10.1086/341392}{\JournalTitle{\aj}, 124, 646}

\bibitem[{{Hopkins} {et~al.}(2010){Hopkins}, {Bundy}, {Hernquist}, {Wuyts}, \&
  {Cox}}]{2010MNRAS.401.1099H}
{Hopkins}, P.~F., {Bundy}, K., {Hernquist}, L., {Wuyts}, S., \& {Cox}, T.~J.
  2010,
  \href{http://dx.doi.org/10.1111/j.1365-2966.2009.15699.x}{\JournalTitle{\mnras},
  401, 1099}

\bibitem[{{Hopkins} {et~al.}(2009{\natexlab{a}}){Hopkins}, {Bundy}, {Murray},
  {Quataert}, {Lauer}, \& {Ma}}]{2009MNRAS.398..898H}
{Hopkins}, P.~F., {Bundy}, K., {Murray}, N., {et~al.} 2009{\natexlab{a}},
  \href{http://dx.doi.org/10.1111/j.1365-2966.2009.15062.x}{\JournalTitle{\mnras},
  398, 898}

\bibitem[{{Hopkins} {et~al.}(2009{\natexlab{b}}){Hopkins}, {Cox}, {Dutta},
  {Hernquist}, {Kormendy}, \& {Lauer}}]{2009ApJS..181..135H}
{Hopkins}, P.~F., {Cox}, T.~J., {Dutta}, S.~N., {et~al.} 2009{\natexlab{b}},
  \href{http://dx.doi.org/10.1088/0067-0049/181/1/135}{\JournalTitle{\apjs},
  181, 135}

\bibitem[{{Hopkins} {et~al.}(2009{\natexlab{c}}){Hopkins}, {Cox}, {Younger}, \&
  {Hernquist}}]{2009ApJ...691.1168H}
{Hopkins}, P.~F., {Cox}, T.~J., {Younger}, J.~D., \& {Hernquist}, L.
  2009{\natexlab{c}},
  \href{http://dx.doi.org/10.1088/0004-637X/691/2/1168}{\JournalTitle{\apj},
  691, 1168}

\bibitem[{{Husemann} {et~al.}(2012){Husemann}, {Kamann}, {Sandin},
  {S{\'a}nchez}, {Garc{\'{\i}}a-Benito}, \& {Mast}}]{2012A&A...545A.137H}
{Husemann}, B., {Kamann}, S., {Sandin}, C., {et~al.} 2012,
  \href{http://dx.doi.org/10.1051/0004-6361/201220102}{\JournalTitle{\aap},
  545, A137}

\bibitem[{{Husemann} {et~al.}(2013){Husemann}, {Jahnke}, {S{\'a}nchez},
  {Barrado}, {Bekerait\`{e}}, {Bomans}, {Castillo-Morales},
  {Catal{\'a}n-Torrecilla}, {Cid Fernandes}, {Falc{\'o}n-Barroso},
  {Garc{\'{\i}}a-Benito}, {Gonz{\'a}lez Delgado}, {Iglesias-P{\'a}ramo},
  {Johnson}, {Kupko}, {L{\'o}pez-Fernandez}, {Lyubenova}, {Marino}, {Mast},
  {Miskolczi}, {Monreal-Ibero}, {Gil de Paz}, {P{\'e}rez}, {P{\'e}rez},
  {Rosales-Ortega}, {Ruiz-Lara}, {Schilling}, {van de Ven}, {Walcher}, {Alves},
  {de Amorim}, {Backsmann}, {Barrera-Ballesteros}, {Bland-Hawthorn}, {Cortijo},
  {Dettmar}, {Demleitner}, {D{\'{\i}}az}, {Enke}, {Florido}, {Flores},
  {Galbany}, {Gallazzi}, {Garc{\'{\i}}a-Lorenzo}, {Gomes}, {Gruel}, {Haines},
  {Holmes}, {Jungwiert}, {Kalinova}, {Kehrig}, {Kennicutt}, {Klar}, {Lehnert},
  {L{\'o}pez-S{\'a}nchez}, {de Lorenzo-C{\'a}ceres}, {M{\'a}rmol-Queralt{\'o}},
  {M{\'a}rquez}, {Mendez-Abreu}, {Moll{\'a}}, {del Olmo}, {Meidt}, {Papaderos},
  {Puschnig}, {Quirrenbach}, {Roth}, {S{\'a}nchez-Bl{\'a}zquez}, {Spekkens},
  {Singh}, {Stanishev}, {Trager}, {Vilchez}, {Wild}, {Wisotzki}, {Zibetti}, \&
  {Ziegler}}]{2013A&A...549A..87H}
{Husemann}, B., {Jahnke}, K., {S{\'a}nchez}, S.~F., {et~al.} 2013,
  \href{http://dx.doi.org/10.1051/0004-6361/201220582}{\JournalTitle{\aap},
  549, A87}

\bibitem[{{Ishibashi} {et~al.}(2013){Ishibashi}, {Fabian}, \&
  {Canning}}]{2013MNRAS.431.2350I}
{Ishibashi}, W., {Fabian}, A.~C., \& {Canning}, R.~E.~A. 2013,
  \href{http://dx.doi.org/10.1093/mnras/stt333}{\JournalTitle{\mnras}, 431,
  2350}

\bibitem[{{Jesseit} {et~al.}(2005){Jesseit}, {Naab}, \&
  {Burkert}}]{2005MNRAS.360.1185J}
{Jesseit}, R., {Naab}, T., \& {Burkert}, A. 2005,
  \href{http://dx.doi.org/10.1111/j.1365-2966.2005.09129.x}{\JournalTitle{\mnras},
  360, 1185}

\bibitem[{{Jesseit} {et~al.}(2007){Jesseit}, {Naab}, {Peletier}, \&
  {Burkert}}]{2007MNRAS.376..997J}
{Jesseit}, R., {Naab}, T., {Peletier}, R.~F., \& {Burkert}, A. 2007,
  \href{http://dx.doi.org/10.1111/j.1365-2966.2007.11524.x}{\JournalTitle{\mnras},
  376, 997}

\bibitem[{{Jimenez} {et~al.}(2007){Jimenez}, {Bernardi}, {Haiman}, {Panter}, \&
  {Heavens}}]{2007ApJ...669..947J}
{Jimenez}, R., {Bernardi}, M., {Haiman}, Z., {Panter}, B., \& {Heavens}, A.~F.
  2007, \href{http://dx.doi.org/10.1086/521323}{\JournalTitle{\apj}, 669, 947}

\bibitem[{{Kauffmann} \& {Charlot}(1998)}]{1998MNRAS.297L..23K}
{Kauffmann}, G., \& {Charlot}, S. 1998,
  \href{http://dx.doi.org/10.1046/j.1365-8711.1998.01708.x}{\JournalTitle{\mnras},
  297, L23}

\bibitem[{{Kelz} {et~al.}(2006){Kelz}, {Verheijen}, {Roth}, {Bauer}, {Becker},
  {Paschke}, {Popow}, {S{\'a}nchez}, \& {Laux}}]{2006PASP..118..129K}
{Kelz}, A., {Verheijen}, M.~A.~W., {Roth}, M.~M., {et~al.} 2006,
  \href{http://dx.doi.org/10.1086/497455}{\JournalTitle{\pasp}, 118, 129}

\bibitem[{{Kere{\v s}} {et~al.}(2005){Kere{\v s}}, {Katz}, {Weinberg}, \&
  {Dav{\'e}}}]{2005MNRAS.363....2K}
{Kere{\v s}}, D., {Katz}, N., {Weinberg}, D.~H., \& {Dav{\'e}}, R. 2005,
  \href{http://dx.doi.org/10.1111/j.1365-2966.2005.09451.x}{\JournalTitle{\mnras},
  363, 2}

\bibitem[{{Kewley} {et~al.}(2010){Kewley}, {Rupke}, {Zahid}, {Geller}, \&
  {Barton}}]{2010ApJ...721L..48K}
{Kewley}, L.~J., {Rupke}, D., {Zahid}, H.~J., {Geller}, M.~J., \& {Barton},
  E.~J. 2010,
  \href{http://dx.doi.org/10.1088/2041-8205/721/1/L48}{\JournalTitle{\apjl},
  721, L48}

\bibitem[{{Kobayashi}(2004)}]{2004MNRAS.347..740K}
{Kobayashi}, C. 2004,
  \href{http://dx.doi.org/10.1111/j.1365-2966.2004.07258.x}{\JournalTitle{\mnras},
  347, 740}

\bibitem[{{Koopmans} {et~al.}(2006){Koopmans}, {Treu}, {Bolton}, {Burles}, \&
  {Moustakas}}]{2006ApJ...649..599K}
{Koopmans}, L.~V.~E., {Treu}, T., {Bolton}, A.~S., {Burles}, S., \&
  {Moustakas}, L.~A. 2006,
  \href{http://dx.doi.org/10.1086/505696}{\JournalTitle{\apj}, 649, 599}

\bibitem[{{Koopmans} {et~al.}(2009){Koopmans}, {Bolton}, {Treu}, {Czoske},
  {Auger}, {Barnab{\`e}}, {Vegetti}, {Gavazzi}, {Moustakas}, \&
  {Burles}}]{2009ApJ...703L..51K}
{Koopmans}, L.~V.~E., {Bolton}, A., {Treu}, T., {et~al.} 2009,
  \href{http://dx.doi.org/10.1088/0004-637X/703/1/L51}{\JournalTitle{\apjl},
  703, L51}

\bibitem[{{Kormendy} \& {Bender}(1996)}]{1996ApJ...464L.119K}
{Kormendy}, J., \& {Bender}, R. 1996,
  \href{http://dx.doi.org/10.1086/310095}{\JournalTitle{\apjl}, 464, L119}

\bibitem[{{Kormendy} {et~al.}(2009){Kormendy}, {Fisher}, {Cornell}, \&
  {Bender}}]{2009ApJS..182..216K}
{Kormendy}, J., {Fisher}, D.~B., {Cornell}, M.~E., \& {Bender}, R. 2009,
  \href{http://dx.doi.org/10.1088/0067-0049/182/1/216}{\JournalTitle{\apjs},
  182, 216}

\bibitem[{{Kormendy} \& {Ho}(2013)}]{2013ARA&A..51..511K}
{Kormendy}, J., \& {Ho}, L.~C. 2013,
  \href{http://dx.doi.org/10.1146/annurev-astro-082708-101811}{\JournalTitle{\araa},
  51, 511}

\bibitem[{{Krajnovi{\'c}} {et~al.}(2005){Krajnovi{\'c}}, {Cappellari},
  {Emsellem}, {McDermid}, \& {de Zeeuw}}]{2005MNRAS.357.1113K}
{Krajnovi{\'c}}, D., {Cappellari}, M., {Emsellem}, E., {McDermid}, R.~M., \&
  {de Zeeuw}, P.~T. 2005,
  \href{http://dx.doi.org/10.1111/j.1365-2966.2005.08715.x}{\JournalTitle{\mnras},
  357, 1113}

\bibitem[{{Kriek} {et~al.}(2008){Kriek}, {van der Wel}, {van Dokkum}, {Franx},
  \& {Illingworth}}]{2008ApJ...682..896K}
{Kriek}, M., {van der Wel}, A., {van Dokkum}, P.~G., {Franx}, M., \&
  {Illingworth}, G.~D. 2008,
  \href{http://dx.doi.org/10.1086/589677}{\JournalTitle{\apj}, 682, 896}

\bibitem[{{Kriek} {et~al.}(2009){Kriek}, {van Dokkum}, {Labb{\'e}}, {Franx},
  {Illingworth}, {Marchesini}, \& {Quadri}}]{2009ApJ...700..221K}
{Kriek}, M., {van Dokkum}, P.~G., {Labb{\'e}}, I., {et~al.} 2009,
  \href{http://dx.doi.org/10.1088/0004-637X/700/1/221}{\JournalTitle{\apj},
  700, 221}

\bibitem[{{Kriek} {et~al.}(2006){Kriek}, {van Dokkum}, {Franx}, {Quadri},
  {Gawiser}, {Herrera}, {Illingworth}, {Labb{\'e}}, {Lira}, {Marchesini},
  {Rix}, {Rudnick}, {Taylor}, {Toft}, {Urry}, \& {Wuyts}}]{2006ApJ...649L..71K}
{Kriek}, M., {van Dokkum}, P.~G., {Franx}, M., {et~al.} 2006,
  \href{http://dx.doi.org/10.1086/508371}{\JournalTitle{\apjl}, 649, L71}

\bibitem[{{Kuntschner} {et~al.}(2010){Kuntschner}, {Emsellem}, {Bacon},
  {Cappellari}, {Davies}, {de Zeeuw}, {Falc{\'o}n-Barroso}, {Krajnovi{\'c}},
  {McDermid}, {Peletier}, {Sarzi}, {Shapiro}, {van den Bosch}, \& {van de
  Ven}}]{2010MNRAS.408...97K}
{Kuntschner}, H., {Emsellem}, E., {Bacon}, R., {et~al.} 2010,
  \href{http://dx.doi.org/10.1111/j.1365-2966.2010.17161.x}{\JournalTitle{\mnras},
  408, 97}

\bibitem[{{La Barbera} {et~al.}(2013){La Barbera}, {Ferreras}, {Vazdekis}, {de
  la Rosa}, {de Carvalho}, {Trevisan}, {Falc{\'o}n-Barroso}, \&
  {Ricciardelli}}]{2013MNRAS.433.3017L}
{La Barbera}, F., {Ferreras}, I., {Vazdekis}, A., {et~al.} 2013,
  \href{http://dx.doi.org/10.1093/mnras/stt943}{\JournalTitle{\mnras}, 433,
  3017}

\bibitem[{{Lackner} {et~al.}(2012){Lackner}, {Cen}, {Ostriker}, \&
  {Joung}}]{2012MNRAS.425..641L}
{Lackner}, C.~N., {Cen}, R., {Ostriker}, J.~P., \& {Joung}, M.~R. 2012,
  \href{http://dx.doi.org/10.1111/j.1365-2966.2012.21525.x}{\JournalTitle{\mnras},
  425, 641}

\bibitem[{{Larson}(1974)}]{1974MNRAS.166..585L}
{Larson}, R.~B. 1974,
  \href{http://dx.doi.org/10.1093/mnras/166.3.585}{\JournalTitle{\mnras}, 166,
  585}

\bibitem[{{Lynden-Bell}(1967)}]{1967MNRAS.136..101L}
{Lynden-Bell}, D. 1967,
  \href{http://dx.doi.org/10.1093/mnras/136.1.101}{\JournalTitle{\mnras}, 136,
  101}

\bibitem[{{Lyubenova} {et~al.}(2016){Lyubenova}, {Mart{\'{\i}}n-Navarro}, {van
  de Ven}, {Falc{\'o}n-Barroso}, {Galbany}, {Gallazzi}, {Garc{\'{\i}}a-Benito},
  {Gonz{\'a}lez Delgado}, {Husemann}, {La Barbera}, {Marino}, {Mast},
  {Mendez-Abreu}, {Peletier}, {S{\'a}nchez-Bl{\'a}zquez}, {S{\'a}nchez},
  {Trager}, {van den Bosch}, {Vazdekis}, {Walcher}, {Zhu}, {Zibetti},
  {Ziegler}, {Bland-Hawthorn}, \& {CALIFA Collaboration}}]{2016MNRAS.463.3220L}
{Lyubenova}, M., {Mart{\'{\i}}n-Navarro}, I., {van de Ven}, G., {et~al.} 2016,
  \href{http://dx.doi.org/10.1093/mnras/stw2434}{\JournalTitle{\mnras}, 463,
  3220}

\bibitem[{{Martig} {et~al.}(2009){Martig}, {Bournaud}, {Teyssier}, \&
  {Dekel}}]{2009ApJ...707..250M}
{Martig}, M., {Bournaud}, F., {Teyssier}, R., \& {Dekel}, A. 2009,
  \href{http://dx.doi.org/10.1088/0004-637X/707/1/250}{\JournalTitle{\apj},
  707, 250}

\bibitem[{{Mart{\'{\i}}n-Navarro}
  {et~al.}(2015{\natexlab{a}}){Mart{\'{\i}}n-Navarro}, {Barbera}, {Vazdekis},
  {Falc{\'o}n-Barroso}, \& {Ferreras}}]{2015MNRAS.447.1033M}
{Mart{\'{\i}}n-Navarro}, I., {Barbera}, F.~L., {Vazdekis}, A.,
  {Falc{\'o}n-Barroso}, J., \& {Ferreras}, I. 2015{\natexlab{a}},
  \href{http://dx.doi.org/10.1093/mnras/stu2480}{\JournalTitle{\mnras}, 447,
  1033}

\bibitem[{{Mart{\'{\i}}n-Navarro} {et~al.}(2016){Mart{\'{\i}}n-Navarro},
  {Brodie}, {van den Bosch}, {Romanowsky}, \& {Forbes}}]{2016ApJ...832L..11M}
{Mart{\'{\i}}n-Navarro}, I., {Brodie}, J.~P., {van den Bosch}, R.~C.~E.,
  {Romanowsky}, A.~J., \& {Forbes}, D.~A. 2016,
  \href{http://dx.doi.org/10.3847/2041-8205/832/1/L11}{\JournalTitle{\apjl},
  832, L11}

\bibitem[{{Mart{\'{\i}}n-Navarro}
  {et~al.}(2015{\natexlab{b}}){Mart{\'{\i}}n-Navarro}, {La Barbera},
  {Vazdekis}, {Ferr{\'e}-Mateu}, {Trujillo}, \&
  {Beasley}}]{2015MNRAS.451.1081M}
{Mart{\'{\i}}n-Navarro}, I., {La Barbera}, F., {Vazdekis}, A., {et~al.}
  2015{\natexlab{b}},
  \href{http://dx.doi.org/10.1093/mnras/stv1022}{\JournalTitle{\mnras}, 451,
  1081}

\bibitem[{{McLure} {et~al.}(2013){McLure}, {Pearce}, {Dunlop}, {Cirasuolo},
  {Curtis-Lake}, {Bruce}, {Caputi}, {Almaini}, {Bonfield}, {Bradshaw},
  {Buitrago}, {Chuter}, {Foucaud}, {Hartley}, \&
  {Jarvis}}]{2013MNRAS.428.1088M}
{McLure}, R.~J., {Pearce}, H.~J., {Dunlop}, J.~S., {et~al.} 2013,
  \href{http://dx.doi.org/10.1093/mnras/sts092}{\JournalTitle{\mnras}, 428,
  1088}

\bibitem[{{Monnet} {et~al.}(1992){Monnet}, {Bacon}, \&
  {Emsellem}}]{1992A&A...253..366M}
{Monnet}, G., {Bacon}, R., \& {Emsellem}, E. 1992, \JournalTitle{\aap}, 253,
  366

\bibitem[{{Naab} \& {Burkert}(2003)}]{2003ApJ...597..893N}
{Naab}, T., \& {Burkert}, A. 2003,
  \href{http://dx.doi.org/10.1086/378581}{\JournalTitle{\apj}, 597, 893}

\bibitem[{{Naab} {et~al.}(2006){Naab}, {Jesseit}, \&
  {Burkert}}]{2006MNRAS.372..839N}
{Naab}, T., {Jesseit}, R., \& {Burkert}, A. 2006,
  \href{http://dx.doi.org/10.1111/j.1365-2966.2006.10902.x}{\JournalTitle{\mnras},
  372, 839}

\bibitem[{{Naab} {et~al.}(2009){Naab}, {Johansson}, \&
  {Ostriker}}]{2009ApJ...699L.178N}
{Naab}, T., {Johansson}, P.~H., \& {Ostriker}, J.~P. 2009,
  \href{http://dx.doi.org/10.1088/0004-637X/699/2/L178}{\JournalTitle{\apjl},
  699, L178}

\bibitem[{{Naab} {et~al.}(2007){Naab}, {Johansson}, {Ostriker}, \&
  {Efstathiou}}]{2007ApJ...658..710N}
{Naab}, T., {Johansson}, P.~H., {Ostriker}, J.~P., \& {Efstathiou}, G. 2007,
  \href{http://dx.doi.org/10.1086/510841}{\JournalTitle{\apj}, 658, 710}

\bibitem[{{Naab} \& {Ostriker}(2009)}]{2009ApJ...690.1452N}
{Naab}, T., \& {Ostriker}, J.~P. 2009,
  \href{http://dx.doi.org/10.1088/0004-637X/690/2/1452}{\JournalTitle{\apj},
  690, 1452}

\bibitem[{{Naab} {et~al.}(2014){Naab}, {Oser}, {Emsellem}, {Cappellari},
  {Krajnovi{\'c}}, {McDermid}, {Alatalo}, {Bayet}, {Blitz}, {Bois}, {Bournaud},
  {Bureau}, {Crocker}, {Davies}, {Davis}, {de Zeeuw}, {Duc}, {Hirschmann},
  {Johansson}, {Khochfar}, {Kuntschner}, {Morganti}, {Oosterloo}, {Sarzi},
  {Scott}, {Serra}, {Ven}, {Weijmans}, \& {Young}}]{2014MNRAS.444.3357N}
{Naab}, T., {Oser}, L., {Emsellem}, E., {et~al.} 2014,
  \href{http://dx.doi.org/10.1093/mnras/stt1919}{\JournalTitle{\mnras}, 444,
  3357}

\bibitem[{{Navarro} {et~al.}(1996){Navarro}, {Frenk}, \&
  {White}}]{1996ApJ...462..563N}
{Navarro}, J.~F., {Frenk}, C.~S., \& {White}, S.~D.~M. 1996,
  \href{http://dx.doi.org/10.1086/177173}{\JournalTitle{\apj}, 462, 563}

\bibitem[{{Navarro} {et~al.}(1997){Navarro}, {Frenk}, \&
  {White}}]{1997ApJ...490..493N}
---. 1997, \href{http://dx.doi.org/10.1086/304888}{\JournalTitle{\apj}, 490,
  493}

\bibitem[{{Newman} {et~al.}(2015){Newman}, {Belli}, \&
  {Ellis}}]{2015ApJ...813L...7N}
{Newman}, A.~B., {Belli}, S., \& {Ellis}, R.~S. 2015,
  \href{http://dx.doi.org/10.1088/2041-8205/813/1/L7}{\JournalTitle{\apjl},
  813, L7}

\bibitem[{{Newman} {et~al.}(2012){Newman}, {Ellis}, {Bundy}, \&
  {Treu}}]{2012ApJ...746..162N}
{Newman}, A.~B., {Ellis}, R.~S., {Bundy}, K., \& {Treu}, T. 2012,
  \href{http://dx.doi.org/10.1088/0004-637X/746/2/162}{\JournalTitle{\apj},
  746, 162}

\bibitem[{{Nipoti} {et~al.}(2012){Nipoti}, {Treu}, {Leauthaud}, {Bundy},
  {Newman}, \& {Auger}}]{2012MNRAS.422.1714N}
{Nipoti}, C., {Treu}, T., {Leauthaud}, A., {et~al.} 2012,
  \href{http://dx.doi.org/10.1111/j.1365-2966.2012.20749.x}{\JournalTitle{\mnras},
  422, 1714}

\bibitem[{{Oldham} {et~al.}(2017){Oldham}, {Auger}, {Fassnacht}, {Treu},
  {Brewer}, {Koopmans}, {Lagattuta}, {Marshall}, {McKean}, \&
  {Vegetti}}]{2017MNRAS.465.3185O}
{Oldham}, L., {Auger}, M.~W., {Fassnacht}, C.~D., {et~al.} 2017,
  \href{http://dx.doi.org/10.1093/mnras/stw2832}{\JournalTitle{\mnras}, 465,
  3185}

\bibitem[{{Oser} {et~al.}(2012){Oser}, {Naab}, {Ostriker}, \&
  {Johansson}}]{2012ApJ...744...63O}
{Oser}, L., {Naab}, T., {Ostriker}, J.~P., \& {Johansson}, P.~H. 2012,
  \href{http://dx.doi.org/10.1088/0004-637X/744/1/63}{\JournalTitle{\apj}, 744,
  63}

\bibitem[{{Oser} {et~al.}(2010){Oser}, {Ostriker}, {Naab}, {Johansson}, \&
  {Burkert}}]{2010ApJ...725.2312O}
{Oser}, L., {Ostriker}, J.~P., {Naab}, T., {Johansson}, P.~H., \& {Burkert}, A.
  2010,
  \href{http://dx.doi.org/10.1088/0004-637X/725/2/2312}{\JournalTitle{\apj},
  725, 2312}

\bibitem[{{Panter} {et~al.}(2008){Panter}, {Jimenez}, {Heavens}, \&
  {Charlot}}]{2008MNRAS.391.1117P}
{Panter}, B., {Jimenez}, R., {Heavens}, A.~F., \& {Charlot}, S. 2008,
  \href{http://dx.doi.org/10.1111/j.1365-2966.2008.13981.x}{\JournalTitle{\mnras},
  391, 1117}

\bibitem[{{Patel} {et~al.}(2013){Patel}, {van Dokkum}, {Franx}, {Quadri},
  {Muzzin}, {Marchesini}, {Williams}, {Holden}, \&
  {Stefanon}}]{2013ApJ...766...15P}
{Patel}, S.~G., {van Dokkum}, P.~G., {Franx}, M., {et~al.} 2013,
  \href{http://dx.doi.org/10.1088/0004-637X/766/1/15}{\JournalTitle{\apj}, 766,
  15}

\bibitem[{{P{\'e}rez} {et~al.}(2013){P{\'e}rez}, {Cid Fernandes}, {Gonz{\'a}lez
  Delgado}, {Garc{\'{\i}}a-Benito}, {S{\'a}nchez}, {Husemann}, {Mast},
  {Rod{\'o}n}, {Kupko}, {Backsmann}, {de Amorim}, {van de Ven}, {Walcher},
  {Wisotzki}, {Cortijo-Ferrero}, \& {CALIFA
  Collaboration}}]{2013ApJ...764L...1P}
{P{\'e}rez}, E., {Cid Fernandes}, R., {Gonz{\'a}lez Delgado}, R.~M., {et~al.}
  2013,
  \href{http://dx.doi.org/10.1088/2041-8205/764/1/L1}{\JournalTitle{\apjl},
  764, L1}

\bibitem[{{Pietrinferni} {et~al.}(2004){Pietrinferni}, {Cassisi}, {Salaris}, \&
  {Castelli}}]{2004ApJ...612..168P}
{Pietrinferni}, A., {Cassisi}, S., {Salaris}, M., \& {Castelli}, F. 2004,
  \href{http://dx.doi.org/10.1086/422498}{\JournalTitle{\apj}, 612, 168}

\bibitem[{{Pietrinferni} {et~al.}(2006){Pietrinferni}, {Cassisi}, {Salaris}, \&
  {Castelli}}]{2006ApJ...642..797P}
---. 2006, \href{http://dx.doi.org/10.1086/501344}{\JournalTitle{\apj}, 642,
  797}

\bibitem[{{Pipino} {et~al.}(2010){Pipino}, {D'Ercole}, {Chiappini}, \&
  {Matteucci}}]{2010MNRAS.407.1347P}
{Pipino}, A., {D'Ercole}, A., {Chiappini}, C., \& {Matteucci}, F. 2010,
  \href{http://dx.doi.org/10.1111/j.1365-2966.2010.17007.x}{\JournalTitle{\mnras},
  407, 1347}

\bibitem[{{Quilis} \& {Trujillo}(2013)}]{2013ApJ...773L...8Q}
{Quilis}, V., \& {Trujillo}, I. 2013,
  \href{http://dx.doi.org/10.1088/2041-8205/773/1/L8}{\JournalTitle{\apjl},
  773, L8}

\bibitem[{{Remus} {et~al.}(2013){Remus}, {Burkert}, {Dolag}, {Johansson},
  {Naab}, {Oser}, \& {Thomas}}]{2013ApJ...766...71R}
{Remus}, R.-S., {Burkert}, A., {Dolag}, K., {et~al.} 2013,
  \href{http://dx.doi.org/10.1088/0004-637X/766/2/71}{\JournalTitle{\apj}, 766,
  71}

\bibitem[{{Remus} {et~al.}(2017){Remus}, {Dolag}, {Naab}, {Burkert},
  {Hirschmann}, {Hoffmann}, \& {Johansson}}]{2017MNRAS.464.3742R}
{Remus}, R.-S., {Dolag}, K., {Naab}, T., {et~al.} 2017,
  \href{http://dx.doi.org/10.1093/mnras/stw2594}{\JournalTitle{\mnras}, 464,
  3742}

\bibitem[{{Riechers} {et~al.}(2011){Riechers}, {Hodge}, {Walter}, {Carilli}, \&
  {Bertoldi}}]{2011ApJ...739L..31R}
{Riechers}, D.~A., {Hodge}, J., {Walter}, F., {Carilli}, C.~L., \& {Bertoldi},
  F. 2011,
  \href{http://dx.doi.org/10.1088/2041-8205/739/1/L31}{\JournalTitle{\apjl},
  739, L31}

\bibitem[{{Robertson} {et~al.}(2006){Robertson}, {Cox}, {Hernquist}, {Franx},
  {Hopkins}, {Martini}, \& {Springel}}]{2006ApJ...641...21R}
{Robertson}, B., {Cox}, T.~J., {Hernquist}, L., {et~al.} 2006,
  \href{http://dx.doi.org/10.1086/500360}{\JournalTitle{\apj}, 641, 21}

\bibitem[{{Roth} {et~al.}(2005){Roth}, {Kelz}, {Fechner}, {Hahn}, {Bauer},
  {Becker}, {B{\"o}hm}, {Christensen}, {Dionies}, {Paschke}, {Popow}, {Wolter},
  {Schmoll}, {Laux}, \& {Altmann}}]{2005PASP..117..620R}
{Roth}, M.~M., {Kelz}, A., {Fechner}, T., {et~al.} 2005,
  \href{http://dx.doi.org/10.1086/429877}{\JournalTitle{\pasp}, 117, 620}

\bibitem[{{Ruff} {et~al.}(2011){Ruff}, {Gavazzi}, {Marshall}, {Treu}, {Auger},
  \& {Brault}}]{2011ApJ...727...96R}
{Ruff}, A.~J., {Gavazzi}, R., {Marshall}, P.~J., {et~al.} 2011,
  \href{http://dx.doi.org/10.1088/0004-637X/727/2/96}{\JournalTitle{\apj}, 727,
  96}

\bibitem[{{Rybicki}(1987)}]{1987IAUS..127..397R}
{Rybicki}, G.~B. 1987, in IAU Symposium, Vol. 127, Structure and Dynamics of
  Elliptical Galaxies, ed. P.~T. {de Zeeuw}, 397

\bibitem[{{S{\'a}nchez} {et~al.}(2012){S{\'a}nchez}, {Kennicutt}, {Gil de Paz},
  {van de Ven}, {V{\'{\i}}lchez}, {Wisotzki}, {Walcher}, {Mast}, {Aguerri},
  {Albiol-P{\'e}rez}, {Alonso-Herrero}, {Alves}, {Bakos}, {Bart{\'a}kov{\'a}},
  {Bland-Hawthorn}, {Boselli}, {Bomans}, {Castillo-Morales}, {Cortijo-Ferrero},
  {de Lorenzo-C{\'a}ceres}, {Del Olmo}, {Dettmar}, {D{\'{\i}}az}, {Ellis},
  {Falc{\'o}n-Barroso}, {Flores}, {Gallazzi}, {Garc{\'{\i}}a-Lorenzo},
  {Gonz{\'a}lez Delgado}, {Gruel}, {Haines}, {Hao}, {Husemann},
  {Igl{\'e}sias-P{\'a}ramo}, {Jahnke}, {Johnson}, {Jungwiert}, {Kalinova},
  {Kehrig}, {Kupko}, {L{\'o}pez-S{\'a}nchez}, {Lyubenova}, {Marino},
  {M{\'a}rmol-Queralt{\'o}}, {M{\'a}rquez}, {Masegosa}, {Meidt},
  {Mendez-Abreu}, {Monreal-Ibero}, {Montijo}, {Mour{\~a}o}, {Palacios-Navarro},
  {Papaderos}, {Pasquali}, {Peletier}, {P{\'e}rez}, {P{\'e}rez}, {Quirrenbach},
  {Rela{\~n}o}, {Rosales-Ortega}, {Roth}, {Ruiz-Lara},
  {S{\'a}nchez-Bl{\'a}zquez}, {Sengupta}, {Singh}, {Stanishev}, {Trager},
  {Vazdekis}, {Viironen}, {Wild}, {Zibetti}, \&
  {Ziegler}}]{2012A&A...538A...8S}
{S{\'a}nchez}, S.~F., {Kennicutt}, R.~C., {Gil de Paz}, A., {et~al.} 2012,
  \href{http://dx.doi.org/10.1051/0004-6361/201117353}{\JournalTitle{\aap},
  538, A8}

\bibitem[{{S{\'a}nchez-Bl{\'a}zquez} {et~al.}(2007){S{\'a}nchez-Bl{\'a}zquez},
  {Forbes}, {Strader}, {Brodie}, \& {Proctor}}]{2007MNRAS.377..759S}
{S{\'a}nchez-Bl{\'a}zquez}, P., {Forbes}, D.~A., {Strader}, J., {Brodie}, J.,
  \& {Proctor}, R. 2007,
  \href{http://dx.doi.org/10.1111/j.1365-2966.2007.11647.x}{\JournalTitle{\mnras},
  377, 759}

\bibitem[{{S{\'a}nchez-Bl{\'a}zquez} {et~al.}(2006){S{\'a}nchez-Bl{\'a}zquez},
  {Peletier}, {Jim{\'e}nez-Vicente}, {Cardiel}, {Cenarro},
  {Falc{\'o}n-Barroso}, {Gorgas}, {Selam}, \& {Vazdekis}}]{2006MNRAS.371..703S}
{S{\'a}nchez-Bl{\'a}zquez}, P., {Peletier}, R.~F., {Jim{\'e}nez-Vicente}, J.,
  {et~al.} 2006,
  \href{http://dx.doi.org/10.1111/j.1365-2966.2006.10699.x}{\JournalTitle{\mnras},
  371, 703}

\bibitem[{{Saulder} {et~al.}(2015){Saulder}, {van den Bosch}, \&
  {Mieske}}]{2015A&A...578A.134S}
{Saulder}, C., {van den Bosch}, R.~C.~E., \& {Mieske}, S. 2015,
  \href{http://dx.doi.org/10.1051/0004-6361/201425472}{\JournalTitle{\aap},
  578, A134}

\bibitem[{{Schlafly} \& {Finkbeiner}(2011)}]{2011ApJ...737..103S}
{Schlafly}, E.~F., \& {Finkbeiner}, D.~P. 2011,
  \href{http://dx.doi.org/10.1088/0004-637X/737/2/103}{\JournalTitle{\apj},
  737, 103}

\bibitem[{{Spolaor} {et~al.}(2010){Spolaor}, {Kobayashi}, {Forbes}, {Couch}, \&
  {Hau}}]{2010MNRAS.408..272S}
{Spolaor}, M., {Kobayashi}, C., {Forbes}, D.~A., {Couch}, W.~J., \& {Hau},
  G.~K.~T. 2010,
  \href{http://dx.doi.org/10.1111/j.1365-2966.2010.17080.x}{\JournalTitle{\mnras},
  408, 272}

\bibitem[{{Szomoru} {et~al.}(2012){Szomoru}, {Franx}, \& {van
  Dokkum}}]{2012ApJ...749..121S}
{Szomoru}, D., {Franx}, M., \& {van Dokkum}, P.~G. 2012,
  \href{http://dx.doi.org/10.1088/0004-637X/749/2/121}{\JournalTitle{\apj},
  749, 121}

\bibitem[{{Szomoru} {et~al.}(2013){Szomoru}, {Franx}, {van Dokkum}, {Trenti},
  {Illingworth}, {Labb{\'e}}, \& {Oesch}}]{2013ApJ...763...73S}
{Szomoru}, D., {Franx}, M., {van Dokkum}, P.~G., {et~al.} 2013,
  \href{http://dx.doi.org/10.1088/0004-637X/763/2/73}{\JournalTitle{\apj}, 763,
  73}

\bibitem[{{Szomoru} {et~al.}(2010){Szomoru}, {Franx}, {van Dokkum}, {Trenti},
  {Illingworth}, {Labb{\'e}}, {Bouwens}, {Oesch}, \&
  {Carollo}}]{2010ApJ...714L.244S}
---. 2010,
  \href{http://dx.doi.org/10.1088/2041-8205/714/2/L244}{\JournalTitle{\apjl},
  714, L244}

\bibitem[{{Taylor} {et~al.}(2010){Taylor}, {Franx}, {Glazebrook}, {Brinchmann},
  {van der Wel}, \& {van Dokkum}}]{2010ApJ...720..723T}
{Taylor}, E.~N., {Franx}, M., {Glazebrook}, K., {et~al.} 2010,
  \href{http://dx.doi.org/10.1088/0004-637X/720/1/723}{\JournalTitle{\apj},
  720, 723}

\bibitem[{{Thomas} {et~al.}(2003){Thomas}, {Maraston}, \&
  {Bender}}]{2003MNRAS.339..897T}
{Thomas}, D., {Maraston}, C., \& {Bender}, R. 2003,
  \href{http://dx.doi.org/10.1046/j.1365-8711.2003.06248.x}{\JournalTitle{\mnras},
  339, 897}

\bibitem[{{Thomas} {et~al.}(2005){Thomas}, {Maraston}, {Bender}, \& {Mendes de
  Oliveira}}]{2005ApJ...621..673T}
{Thomas}, D., {Maraston}, C., {Bender}, R., \& {Mendes de Oliveira}, C. 2005,
  \href{http://dx.doi.org/10.1086/426932}{\JournalTitle{\apj}, 621, 673}

\bibitem[{{Toft} {et~al.}(2012){Toft}, {Gallazzi}, {Zirm}, {Wold}, {Zibetti},
  {Grillo}, \& {Man}}]{2012ApJ...754....3T}
{Toft}, S., {Gallazzi}, A., {Zirm}, A., {et~al.} 2012,
  \href{http://dx.doi.org/10.1088/0004-637X/754/1/3}{\JournalTitle{\apj}, 754,
  3}

\bibitem[{{Toft} {et~al.}(2005){Toft}, {van Dokkum}, {Franx}, {Thompson},
  {Illingworth}, {Bouwens}, \& {Kriek}}]{2005ApJ...624L...9T}
{Toft}, S., {van Dokkum}, P., {Franx}, M., {et~al.} 2005,
  \href{http://dx.doi.org/10.1086/430346}{\JournalTitle{\apjl}, 624, L9}

\bibitem[{{Toft} {et~al.}(2007){Toft}, {van Dokkum}, {Franx}, {Labbe},
  {F{\"o}rster Schreiber}, {Wuyts}, {Webb}, {Rudnick}, {Zirm}, {Kriek}, {van
  der Werf}, {Blakeslee}, {Illingworth}, {Rix}, {Papovich}, \&
  {Moorwood}}]{2007ApJ...671..285T}
---. 2007, \href{http://dx.doi.org/10.1086/521810}{\JournalTitle{\apj}, 671,
  285}

\bibitem[{{Toft} {et~al.}(2014){Toft}, {Smol{\v c}i{\'c}}, {Magnelli}, {Karim},
  {Zirm}, {Michalowski}, {Capak}, {Sheth}, {Schawinski}, {Krogager}, {Wuyts},
  {Sanders}, {Man}, {Lutz}, {Staguhn}, {Berta}, {Mccracken}, {Krpan}, \&
  {Riechers}}]{2014ApJ...782...68T}
{Toft}, S., {Smol{\v c}i{\'c}}, V., {Magnelli}, B., {et~al.} 2014,
  \href{http://dx.doi.org/10.1088/0004-637X/782/2/68}{\JournalTitle{\apj}, 782,
  68}

\bibitem[{{Toomre}(1977)}]{1977egsp.conf..401T}
{Toomre}, A. 1977, in Evolution of Galaxies and Stellar Populations, ed. B.~M.
  {Tinsley} \& R.~B.~G. {Larson}, D.~Campbell, 401

\bibitem[{{Toomre} \& {Toomre}(1972)}]{1972ApJ...178..623T}
{Toomre}, A., \& {Toomre}, J. 1972,
  \href{http://dx.doi.org/10.1086/151823}{\JournalTitle{\apj}, 178, 623}

\bibitem[{{Tortora} {et~al.}(2016){Tortora}, {La Barbera}, {Napolitano}, {Roy},
  {Radovich}, {Cavuoti}, {Brescia}, {Longo}, {Getman}, {Capaccioli}, {Grado},
  {Kuijken}, {de Jong}, {McFarland}, \& {Puddu}}]{2016MNRAS.457.2845T}
{Tortora}, C., {La Barbera}, F., {Napolitano}, N.~R., {et~al.} 2016,
  \href{http://dx.doi.org/10.1093/mnras/stw184}{\JournalTitle{\mnras}, 457,
  2845}

\bibitem[{{Trujillo} {et~al.}(2009){Trujillo}, {Cenarro}, {de
  Lorenzo-C{\'a}ceres}, {Vazdekis}, {de la Rosa}, \&
  {Cava}}]{2009ApJ...692L.118T}
{Trujillo}, I., {Cenarro}, A.~J., {de Lorenzo-C{\'a}ceres}, A., {et~al.} 2009,
  \href{http://dx.doi.org/10.1088/0004-637X/692/2/L118}{\JournalTitle{\apjl},
  692, L118}

\bibitem[{{Trujillo} {et~al.}(2014){Trujillo}, {Ferr{\'e}-Mateu}, {Balcells},
  {Vazdekis}, \& {S{\'a}nchez-Bl{\'a}zquez}}]{2014ApJ...780L..20T}
{Trujillo}, I., {Ferr{\'e}-Mateu}, A., {Balcells}, M., {Vazdekis}, A., \&
  {S{\'a}nchez-Bl{\'a}zquez}, P. 2014,
  \href{http://dx.doi.org/10.1088/2041-8205/780/2/L20}{\JournalTitle{\apjl},
  780, L20}

\bibitem[{{Trujillo} {et~al.}(2006){Trujillo}, {F{\"o}rster Schreiber},
  {Rudnick}, {Barden}, {Franx}, {Rix}, {Caldwell}, {McIntosh}, {Toft},
  {H{\"a}ussler}, {Zirm}, {van Dokkum}, {Labb{\'e}}, {Moorwood},
  {R{\"o}ttgering}, {van der Wel}, {van der Werf}, \& {van
  Starkenburg}}]{2006ApJ...650...18T}
{Trujillo}, I., {F{\"o}rster Schreiber}, N.~M., {Rudnick}, G., {et~al.} 2006,
  \href{http://dx.doi.org/10.1086/506464}{\JournalTitle{\apj}, 650, 18}

\bibitem[{{van de Sande} {et~al.}(2011){van de Sande}, {Kriek}, {Franx}, {van
  Dokkum}, {Bezanson}, {Whitaker}, {Brammer}, {Labb{\'e}}, {Groot}, \&
  {Kaper}}]{2011ApJ...736L...9V}
{van de Sande}, J., {Kriek}, M., {Franx}, M., {et~al.} 2011,
  \href{http://dx.doi.org/10.1088/2041-8205/736/1/L9}{\JournalTitle{\apjl},
  736, L9}

\bibitem[{{van de Sande} {et~al.}(2013){van de Sande}, {Kriek}, {Franx}, {van
  Dokkum}, {Bezanson}, {Bouwens}, {Quadri}, {Rix}, \&
  {Skelton}}]{2013ApJ...771...85V}
---. 2013,
  \href{http://dx.doi.org/10.1088/0004-637X/771/2/85}{\JournalTitle{\apj}, 771,
  85}

\bibitem[{{van den Bosch}(2016)}]{2016ApJ...831..134V}
{van den Bosch}, R.~C.~E. 2016,
  \href{http://dx.doi.org/10.3847/0004-637X/831/2/134}{\JournalTitle{\apj},
  831, 134}

\bibitem[{{van den Bosch} {et~al.}(2012){van den Bosch}, {Gebhardt},
  {G{\"u}ltekin}, {van de Ven}, {van der Wel}, \&
  {Walsh}}]{2012Natur.491..729V}
{van den Bosch}, R.~C.~E., {Gebhardt}, K., {G{\"u}ltekin}, K., {et~al.} 2012,
  \href{http://dx.doi.org/10.1038/nature11592}{\JournalTitle{\nat}, 491, 729}

\bibitem[{{van den Bosch} {et~al.}(2015){van den Bosch}, {Gebhardt},
  {G{\"u}ltekin}, {Y{\i}ld{\i}r{\i}m}, \& {Walsh}}]{2015ApJS..218...10V}
{van den Bosch}, R.~C.~E., {Gebhardt}, K., {G{\"u}ltekin}, K.,
  {Y{\i}ld{\i}r{\i}m}, A., \& {Walsh}, J.~L. 2015,
  \href{http://dx.doi.org/10.1088/0067-0049/218/1/10}{\JournalTitle{\apjs},
  218, 10}

\bibitem[{{van den Bosch} \& {van de Ven}(2009)}]{2009MNRAS.398.1117V}
{van den Bosch}, R.~C.~E., \& {van de Ven}, G. 2009,
  \href{http://dx.doi.org/10.1111/j.1365-2966.2009.15177.x}{\JournalTitle{\mnras},
  398, 1117}

\bibitem[{{van den Bosch} {et~al.}(2008){van den Bosch}, {van de Ven},
  {Verolme}, {Cappellari}, \& {de Zeeuw}}]{2008MNRAS.385..647V}
{van den Bosch}, R.~C.~E., {van de Ven}, G., {Verolme}, E.~K., {Cappellari},
  M., \& {de Zeeuw}, P.~T. 2008,
  \href{http://dx.doi.org/10.1111/j.1365-2966.2008.12874.x}{\JournalTitle{\mnras},
  385, 647}

\bibitem[{{van der Marel}(1991)}]{1991MNRAS.253..710V}
{van der Marel}, R.~P. 1991, \JournalTitle{\mnras}, 253, 710

\bibitem[{{van der Marel} \& {Franx}(1993)}]{1993ApJ...407..525V}
{van der Marel}, R.~P., \& {Franx}, M. 1993,
  \href{http://dx.doi.org/10.1086/172534}{\JournalTitle{\apj}, 407, 525}

\bibitem[{{van der Wel} {et~al.}(2008){van der Wel}, {Holden}, {Zirm}, {Franx},
  {Rettura}, {Illingworth}, \& {Ford}}]{2008ApJ...688...48V}
{van der Wel}, A., {Holden}, B.~P., {Zirm}, A.~W., {et~al.} 2008,
  \href{http://dx.doi.org/10.1086/592267}{\JournalTitle{\apj}, 688, 48}

\bibitem[{{van der Wel} {et~al.}(2011){van der Wel}, {Rix}, {Wuyts}, {McGrath},
  {Koekemoer}, {Bell}, {Holden}, {Robaina}, \&
  {McIntosh}}]{2011ApJ...730...38V}
{van der Wel}, A., {Rix}, H.-W., {Wuyts}, S., {et~al.} 2011,
  \href{http://dx.doi.org/10.1088/0004-637X/730/1/38}{\JournalTitle{\apj}, 730,
  38}

\bibitem[{{van der Wel} {et~al.}(2012){van der Wel}, {Bell}, {H{\"a}ussler},
  {McGrath}, {Chang}, {Guo}, {McIntosh}, {Rix}, {Barden}, {Cheung}, {Faber},
  {Ferguson}, {Galametz}, {Grogin}, {Hartley}, {Kartaltepe}, {Kocevski},
  {Koekemoer}, {Lotz}, {Mozena}, {Peth}, \& {Peng}}]{2012ApJS..203...24V}
{van der Wel}, A., {Bell}, E.~F., {H{\"a}ussler}, B., {et~al.} 2012,
  \href{http://dx.doi.org/10.1088/0067-0049/203/2/24}{\JournalTitle{\apjs},
  203, 24}

\bibitem[{{van der Wel} {et~al.}(2014){van der Wel}, {Franx}, {van Dokkum},
  {Skelton}, {Momcheva}, {Whitaker}, {Brammer}, {Bell}, {Rix}, {Wuyts},
  {Ferguson}, {Holden}, {Barro}, {Koekemoer}, {Chang}, {McGrath},
  {H{\"a}ussler}, {Dekel}, {Behroozi}, {Fumagalli}, {Leja}, {Lundgren},
  {Maseda}, {Nelson}, {Wake}, {Patel}, {Labb{\'e}}, {Faber}, {Grogin}, \&
  {Kocevski}}]{2014ApJ...788...28V}
{van der Wel}, A., {Franx}, M., {van Dokkum}, P.~G., {et~al.} 2014,
  \href{http://dx.doi.org/10.1088/0004-637X/788/1/28}{\JournalTitle{\apj}, 788,
  28}

\bibitem[{{van Dokkum} {et~al.}(2009){van Dokkum}, {Kriek}, \&
  {Franx}}]{2009Natur.460..717V}
{van Dokkum}, P.~G., {Kriek}, M., \& {Franx}, M. 2009,
  \href{http://dx.doi.org/10.1038/nature08220}{\JournalTitle{\nat}, 460, 717}

\bibitem[{{van Dokkum} {et~al.}(2008){van Dokkum}, {Franx}, {Kriek}, {Holden},
  {Illingworth}, {Magee}, {Bouwens}, {Marchesini}, {Quadri}, {Rudnick},
  {Taylor}, \& {Toft}}]{2008ApJ...677L...5V}
{van Dokkum}, P.~G., {Franx}, M., {Kriek}, M., {et~al.} 2008,
  \href{http://dx.doi.org/10.1086/587874}{\JournalTitle{\apjl}, 677, L5}

\bibitem[{{van Dokkum} {et~al.}(2010){van Dokkum}, {Whitaker}, {Brammer},
  {Franx}, {Kriek}, {Labb{\'e}}, {Marchesini}, {Quadri}, {Bezanson},
  {Illingworth}, {Muzzin}, {Rudnick}, {Tal}, \& {Wake}}]{2010ApJ...709.1018V}
{van Dokkum}, P.~G., {Whitaker}, K.~E., {Brammer}, G., {et~al.} 2010,
  \href{http://dx.doi.org/10.1088/0004-637X/709/2/1018}{\JournalTitle{\apj},
  709, 1018}

\bibitem[{{van Dokkum} {et~al.}(2015){van Dokkum}, {Nelson}, {Franx}, {Oesch},
  {Momcheva}, {Brammer}, {F{\"o}rster Schreiber}, {Skelton}, {Whitaker}, {van
  der Wel}, {Bezanson}, {Fumagalli}, {Illingworth}, {Kriek}, {Leja}, \&
  {Wuyts}}]{2015ApJ...813...23V}
{van Dokkum}, P.~G., {Nelson}, E.~J., {Franx}, M., {et~al.} 2015,
  \href{http://dx.doi.org/10.1088/0004-637X/813/1/23}{\JournalTitle{\apj}, 813,
  23}

\bibitem[{{Vazdekis} {et~al.}(2010){Vazdekis}, {S{\'a}nchez-Bl{\'a}zquez},
  {Falc{\'o}n-Barroso}, {Cenarro}, {Beasley}, {Cardiel}, {Gorgas}, \&
  {Peletier}}]{2010MNRAS.404.1639V}
{Vazdekis}, A., {S{\'a}nchez-Bl{\'a}zquez}, P., {Falc{\'o}n-Barroso}, J.,
  {et~al.} 2010,
  \href{http://dx.doi.org/10.1111/j.1365-2966.2010.16407.x}{\JournalTitle{\mnras},
  404, 1639}

\bibitem[{{Vazdekis} {et~al.}(2015){Vazdekis}, {Coelho}, {Cassisi},
  {Ricciardelli}, {Falc{\'o}n-Barroso}, {S{\'a}nchez-Bl{\'a}zquez}, {Barbera},
  {Beasley}, \& {Pietrinferni}}]{2015MNRAS.449.1177V}
{Vazdekis}, A., {Coelho}, P., {Cassisi}, S., {et~al.} 2015,
  \href{http://dx.doi.org/10.1093/mnras/stv151}{\JournalTitle{\mnras}, 449,
  1177}

\bibitem[{{Verheijen} {et~al.}(2004){Verheijen}, {Bershady}, {Andersen},
  {Swaters}, {Westfall}, {Kelz}, \& {Roth}}]{2004AN....325..151V}
{Verheijen}, M.~A.~W., {Bershady}, M.~A., {Andersen}, D.~R., {et~al.} 2004,
  \href{http://dx.doi.org/10.1002/asna.200310197}{\JournalTitle{Astronomische
  Nachrichten}, 325, 151}

\bibitem[{{Vogelsberger} {et~al.}(2014){Vogelsberger}, {Genel}, {Springel},
  {Torrey}, {Sijacki}, {Xu}, {Snyder}, {Nelson}, \&
  {Hernquist}}]{2014MNRAS.444.1518V}
{Vogelsberger}, M., {Genel}, S., {Springel}, V., {et~al.} 2014,
  \href{http://dx.doi.org/10.1093/mnras/stu1536}{\JournalTitle{\mnras}, 444,
  1518}

\bibitem[{{Walcher} {et~al.}(2015){Walcher}, {Coelho}, {Gallazzi}, {Bruzual},
  {Charlot}, \& {Chiappini}}]{2015A&A...582A..46W}
{Walcher}, C.~J., {Coelho}, P.~R.~T., {Gallazzi}, A., {et~al.} 2015,
  \href{http://dx.doi.org/10.1051/0004-6361/201525924}{\JournalTitle{\aap},
  582, A46}

\bibitem[{{Walsh} {et~al.}(2012){Walsh}, {van den Bosch}, {Barth}, \&
  {Sarzi}}]{2012ApJ...753...79W}
{Walsh}, J.~L., {van den Bosch}, R.~C.~E., {Barth}, A.~J., \& {Sarzi}, M. 2012,
  \href{http://dx.doi.org/10.1088/0004-637X/753/1/79}{\JournalTitle{\apj}, 753,
  79}

\bibitem[{{Walsh} {et~al.}(2015){Walsh}, {van den Bosch}, {Gebhardt},
  {Yildirim}, {G{\"u}ltekin}, {Husemann}, \& {Richstone}}]{2015ApJ...808..183W}
{Walsh}, J.~L., {van den Bosch}, R.~C.~E., {Gebhardt}, K., {et~al.} 2015,
  \href{http://dx.doi.org/10.1088/0004-637X/808/2/183}{\JournalTitle{\apj},
  808, 183}

\bibitem[{{Walsh} {et~al.}(2017){Walsh}, {van den Bosch}, {Gebhardt},
  {Y{\i}ld{\i}r{\i}m}, {G{\"u}ltekin}, {Husemann}, \&
  {Richstone}}]{2017ApJ...835..208W}
---. 2017,
  \href{http://dx.doi.org/10.3847/1538-4357/835/2/208}{\JournalTitle{\apj},
  835, 208}

\bibitem[{{Walsh} {et~al.}(2016){Walsh}, {van den Bosch}, {Gebhardt},
  {Y{\i}ld{\i}r{\i}m}, {Richstone}, {G{\"u}ltekin}, \&
  {Husemann}}]{2016ApJ...817....2W}
---. 2016,
  \href{http://dx.doi.org/10.3847/0004-637X/817/1/2}{\JournalTitle{\apj}, 817,
  2}

\bibitem[{{Weijmans} {et~al.}(2014){Weijmans}, {de Zeeuw}, {Emsellem},
  {Krajnovi{\'c}}, {Lablanche}, {Alatalo}, {Blitz}, {Bois}, {Bournaud},
  {Bureau}, {Cappellari}, {Crocker}, {Davies}, {Davis}, {Duc}, {Khochfar},
  {Kuntschner}, {McDermid}, {Morganti}, {Naab}, {Oosterloo}, {Sarzi}, {Scott},
  {Serra}, {Verdoes Kleijn}, \& {Young}}]{2014MNRAS.444.3340W}
{Weijmans}, A.-M., {de Zeeuw}, P.~T., {Emsellem}, E., {et~al.} 2014,
  \href{http://dx.doi.org/10.1093/mnras/stu1603}{\JournalTitle{\mnras}, 444,
  3340}

\bibitem[{{Wellons} {et~al.}(2015){Wellons}, {Torrey}, {Ma}, {Rodriguez-Gomez},
  {Vogelsberger}, {Kriek}, {van Dokkum}, {Nelson}, {Genel}, {Pillepich},
  {Springel}, {Sijacki}, {Snyder}, {Nelson}, {Sales}, \&
  {Hernquist}}]{2015MNRAS.449..361W}
{Wellons}, S., {Torrey}, P., {Ma}, C.-P., {et~al.} 2015,
  \href{http://dx.doi.org/10.1093/mnras/stv303}{\JournalTitle{\mnras}, 449,
  361}

\bibitem[{{Wellons} {et~al.}(2016){Wellons}, {Torrey}, {Ma}, {Rodriguez-Gomez},
  {Pillepich}, {Nelson}, {Genel}, {Vogelsberger}, \&
  {Hernquist}}]{2016MNRAS.456.1030W}
---. 2016,
  \href{http://dx.doi.org/10.1093/mnras/stv2738}{\JournalTitle{\mnras}, 456,
  1030}

\bibitem[{{White}(1980)}]{1980MNRAS.191P...1W}
{White}, S.~D.~M. 1980,
  \href{http://dx.doi.org/10.1093/mnras/191.1.1P}{\JournalTitle{\mnras}, 191,
  1P}

\bibitem[{{White} \& {Rees}(1978)}]{1978MNRAS.183..341W}
{White}, S.~D.~M., \& {Rees}, M.~J. 1978, \JournalTitle{\mnras}, 183, 341

\bibitem[{{Williams} {et~al.}(2009){Williams}, {Quadri}, {Franx}, {van Dokkum},
  \& {Labb{\'e}}}]{2009ApJ...691.1879W}
{Williams}, R.~J., {Quadri}, R.~F., {Franx}, M., {van Dokkum}, P., \&
  {Labb{\'e}}, I. 2009,
  \href{http://dx.doi.org/10.1088/0004-637X/691/2/1879}{\JournalTitle{\apj},
  691, 1879}

\bibitem[{{Wuyts} {et~al.}(2010){Wuyts}, {Cox}, {Hayward}, {Franx},
  {Hernquist}, {Hopkins}, {Jonsson}, \& {van Dokkum}}]{2010ApJ...722.1666W}
{Wuyts}, S., {Cox}, T.~J., {Hayward}, C.~C., {et~al.} 2010,
  \href{http://dx.doi.org/10.1088/0004-637X/722/2/1666}{\JournalTitle{\apj},
  722, 1666}

\bibitem[{{Y{\i}ld{\i}r{\i}m} {et~al.}(2015){Y{\i}ld{\i}r{\i}m}, {van den
  Bosch}, {van de Ven}, {Husemann}, {Lyubenova}, {Walsh}, {Gebhardt}, \&
  {G{\"u}ltekin}}]{2015MNRAS.452.1792Y}
{Y{\i}ld{\i}r{\i}m}, A., {van den Bosch}, R.~C.~E., {van de Ven}, G., {et~al.}
  2015, \href{http://dx.doi.org/10.1093/mnras/stv1381}{\JournalTitle{\mnras},
  452, 1792}

\bibitem[{{Y{\i}ld{\i}r{\i}m} {et~al.}(2016){Y{\i}ld{\i}r{\i}m}, {van den
  Bosch}, {van de Ven}, {Dutton}, {L{\"a}sker}, {Husemann}, {Walsh},
  {Gebhardt}, {G{\"u}ltekin}, \& {Mart{\'{\i}}n-Navarro}}]{2016MNRAS.456..538Y}
---. 2016,
  \href{http://dx.doi.org/10.1093/mnras/stv2665}{\JournalTitle{\mnras}, 456,
  538}

\bibitem[{{Zirm} {et~al.}(2007){Zirm}, {van der Wel}, {Franx}, {Labb{\'e}},
  {Trujillo}, {van Dokkum}, {Toft}, {Daddi}, {Rudnick}, {Rix},
  {R{\"o}ttgering}, \& {van der Werf}}]{2007ApJ...656...66Z}
{Zirm}, A.~W., {van der Wel}, A., {Franx}, M., {et~al.} 2007,
  \href{http://dx.doi.org/10.1086/510713}{\JournalTitle{\apj}, 656, 66}

\end{thebibliography}

%%=====================================================================
%% APPENDICES
\appendix
\section{Multi-Gaussian Expansion}
\label{sec:appendix_mge}
%%=====================================================================

\clearpage

\begin{table}
	\caption{Multi-Gaussian-Expansion of MRK\,1216's \textit{HST} (F160W) \textit{H}-band image at a fixed PA of 70.15\,\degree (measured counter-clockwise with the image aligned N.-E., i.e. north is up and east is left), adopting an extinction correction of 0.017\,mag and an \textit{H}-band absolute magnitude for the sun of 3.32. The columns display the number of each Gaussian (1), its surface density (2), dispersion (3), corresponding flattening (4) and offset from the PA, measured counter-clockwise (5). All tables are available in their entirety as Supporting Information online. A portion is shown here for guidance.}
	\begin{center}
	\begin{tabular}{ c  c  c  c  c }
		\hline
		\# & I [$L_{\odot}\,pc^{-2}$] & $\sigma$ [arcsec] & $q$ & $\Psi_{PA}$ \\
		\hline
           1& 89721.781&      0.0940199&      0.7200000&      0.0000000\\
           2& 65370.125&      0.2298265&      0.7700000&      0.0000000\\
           -& -&      -&      -&      -\\
          10&    17.127&     29.7630386&      0.9900000&      0.0000000\\
    	\hline
	\end{tabular}
	\vspace{2ex}
	\label{tab:mrk1216_mge}
	\end{center}
\end{table}

%%=====================================================================
%% APPENDICES
%\appendix
\section{Data and dynamical model predictions}
\label{sec:data_model}
%%=====================================================================

\begin{figure*}
		\begin{center}
		\includegraphics[width=.95\textwidth]{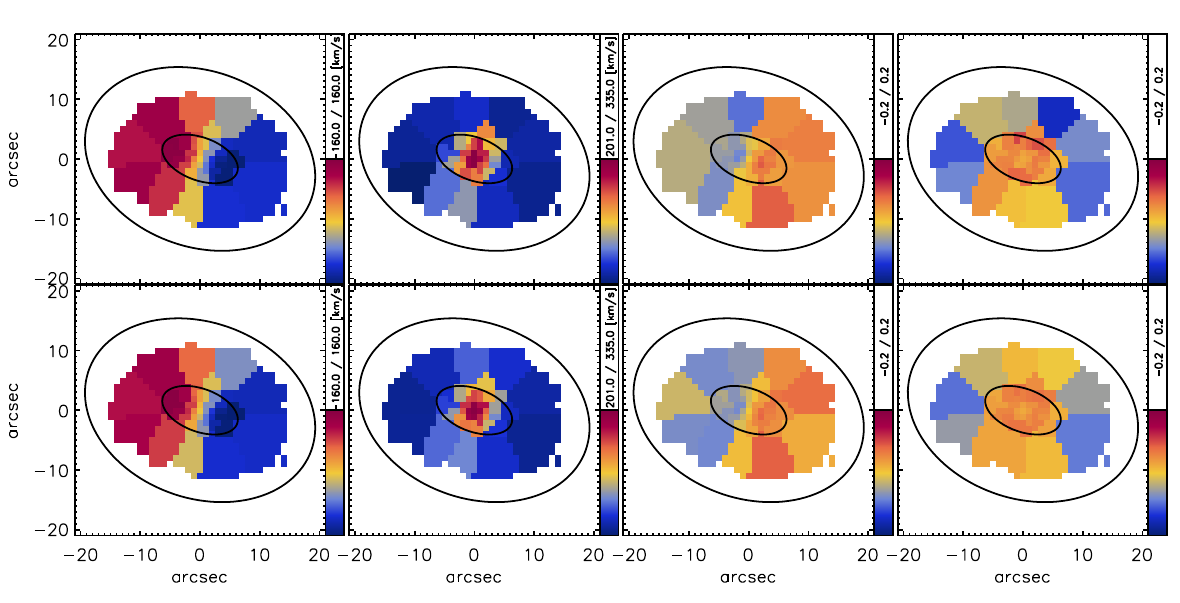}
		\end{center}
	\caption{Top: Bi-symmetrised \ppak\ IFU stellar kinematic maps of MRK\,1216, showing the mean stellar line-of-sight velocity $v$, velocity dispersion $\sigma$ and higher order Gauss-Hermite moments $h_3$ and $h_4$. Overplotted are contours of constant SB at 1 and 3\,\Reff\, to illustrate the extent of the kinematic data. The maps show fast and regular rotation around the short axis of 160\,\kms and a central velocity dispersion of 335\,\kms. Bottom: Best-fitting Schwarzschild model predictions. All maps are oriented north-east, i.e. north is up and east is left.}
	\label{fig:mrk1216_data_model}
\end{figure*}

\begin{figure*}
		\begin{center}
		\includegraphics[width=.95\textwidth]{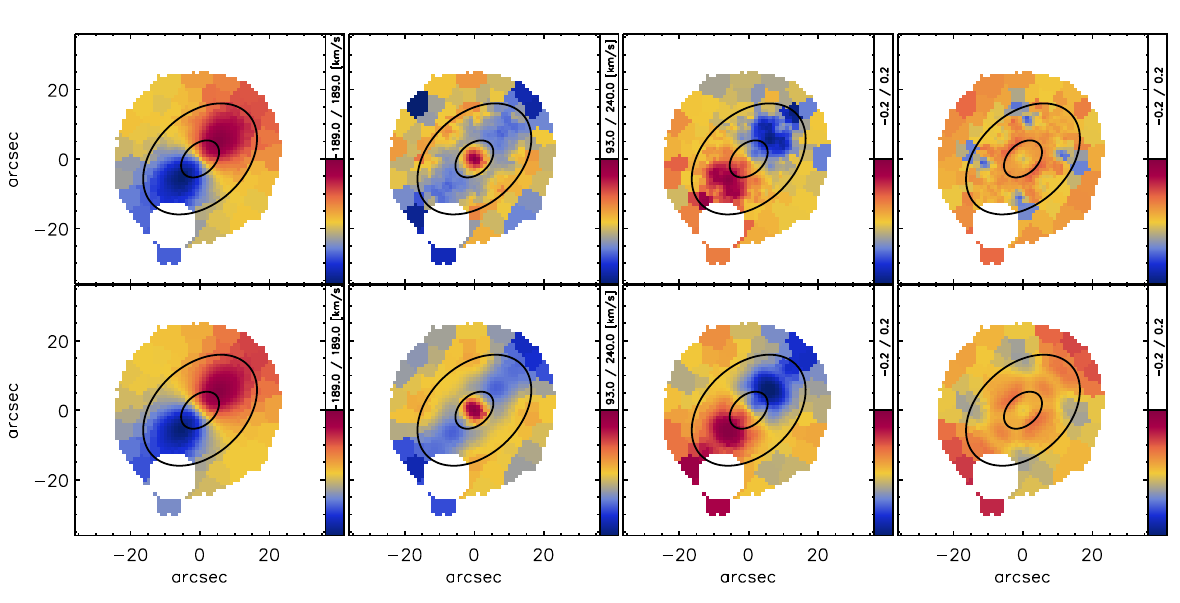}
		\end{center}
	\caption{Top: Bi-symmetrised \ppak\ IFU stellar kinematic maps of NGC\,0384, showing the mean stellar line-of-sight velocity $v$, velocity dispersion $\sigma$ and higher order Gauss-Hermite moments $h_3$ and $h_4$. Overplotted are contours of constant SB at 1 and 3\,\Reff\, to illustrate the extent of the kinematic data. The maps show fast and regular rotation around the short axis of 189\,\kms and a central velocity dispersion of 240\,\kms. Bottom: Best-fitting Schwarzschild model predictions. All maps are oriented north-east, i.e. north is up and east is left.}
	\label{fig:ngc0384_data_model}
\end{figure*}

\begin{figure*}
		\begin{center}
		\includegraphics[width=.95\textwidth]{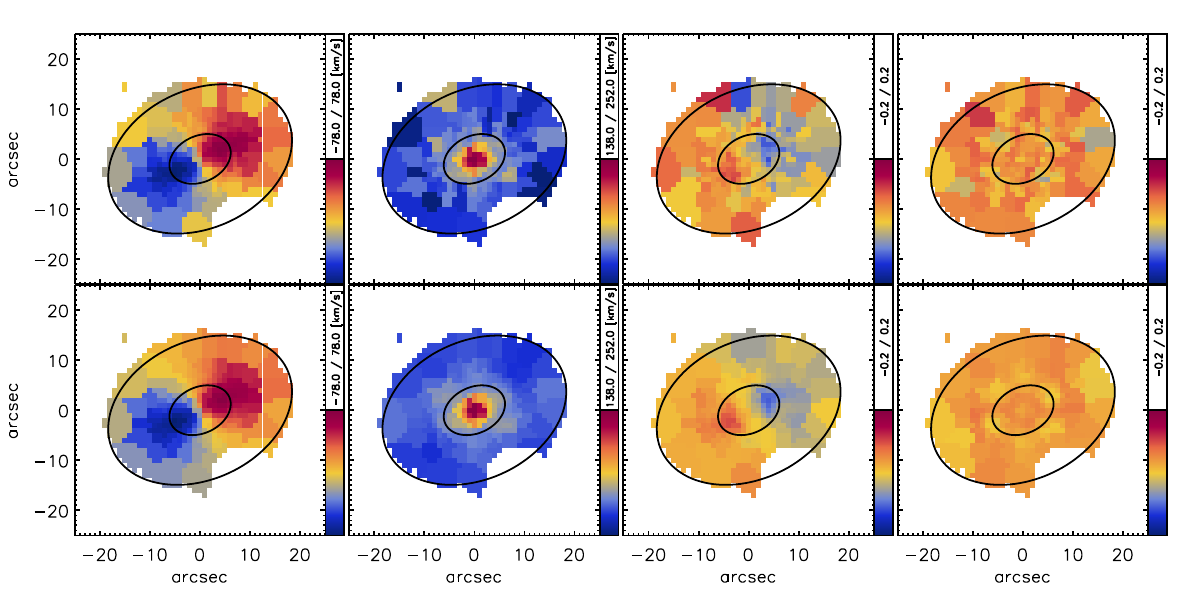}
		\end{center}
	\caption{Top: Point-symmetrised \ppak\ IFU stellar kinematic maps of NGC\,0472, showing the mean stellar line-of-sight velocity $v$, velocity dispersion $\sigma$ and higher order Gauss-Hermite moments $h_3$ and $h_4$. Overplotted are contours of constant SB at 1 and 3\,\Reff\, to illustrate the extent of the kinematic data. The maps show regular rotation around the short axis of 78\,\kms and a central velocity dispersion of 252\,\kms. Bottom: Best-fitting Schwarzschild model predictions. All maps are oriented north-east, i.e. north is up and east is left.}
	\label{fig:ngc0472_data_model}
\end{figure*}

\begin{figure*}
		\begin{center}
		\includegraphics[width=.95\textwidth]{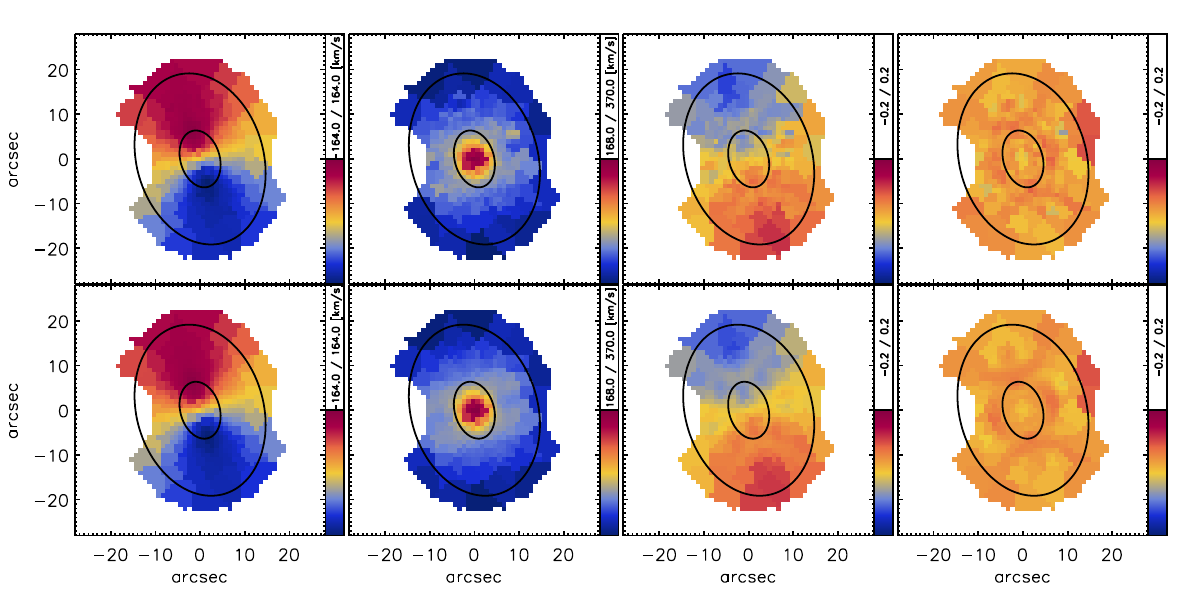}
		\end{center}
	\caption{Top: Bi-symmetrised \ppak\ IFU stellar kinematic maps of NGC\,1270, showing the mean stellar line-of-sight velocity $v$, velocity dispersion $\sigma$ and higher order Gauss-Hermite moments $h_3$ and $h_4$. Overplotted are contours of constant SB at 1 and 3\,\Reff\, to illustrate the extent of the kinematic data. The maps show fast and regular rotation around the short axis of 164\,\kms and a central velocity dispersion of 370\,\kms. Bottom: Best-fitting Schwarzschild model predictions. All maps are oriented north-east, i.e. north is up and east is left.}
	\label{fig:ngc1270_data_model}
\end{figure*}

\begin{figure*}
		\begin{center}
		\includegraphics[width=.95\textwidth]{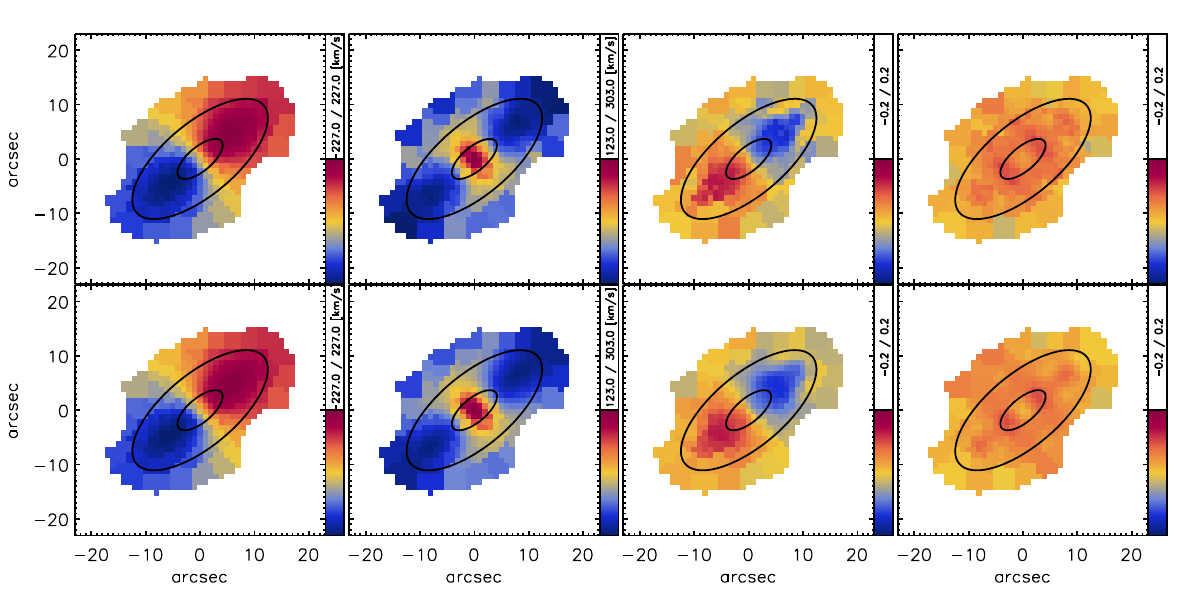}
		\end{center}
	\caption{Top: Bi-symmetrised \ppak\ IFU stellar kinematic maps of NGC\,1271, showing the mean stellar line-of-sight velocity $v$, velocity dispersion $\sigma$ and higher order Gauss-Hermite moments $h_3$ and $h_4$. Overplotted are contours of constant SB at 1 and 3\,\Reff\, to illustrate the extent of the kinematic data. The maps show fast and regular rotation around the short axis of 227\,\kms and a central velocity dispersion of 303\,\kms. Bottom: Best-fitting Schwarzschild model predictions. All maps are oriented north-east, i.e. north is up and east is left.}
	\label{fig:ngc1271_data_model}
\end{figure*}

\begin{figure*}
		\begin{center}
		\includegraphics[width=.95\textwidth]{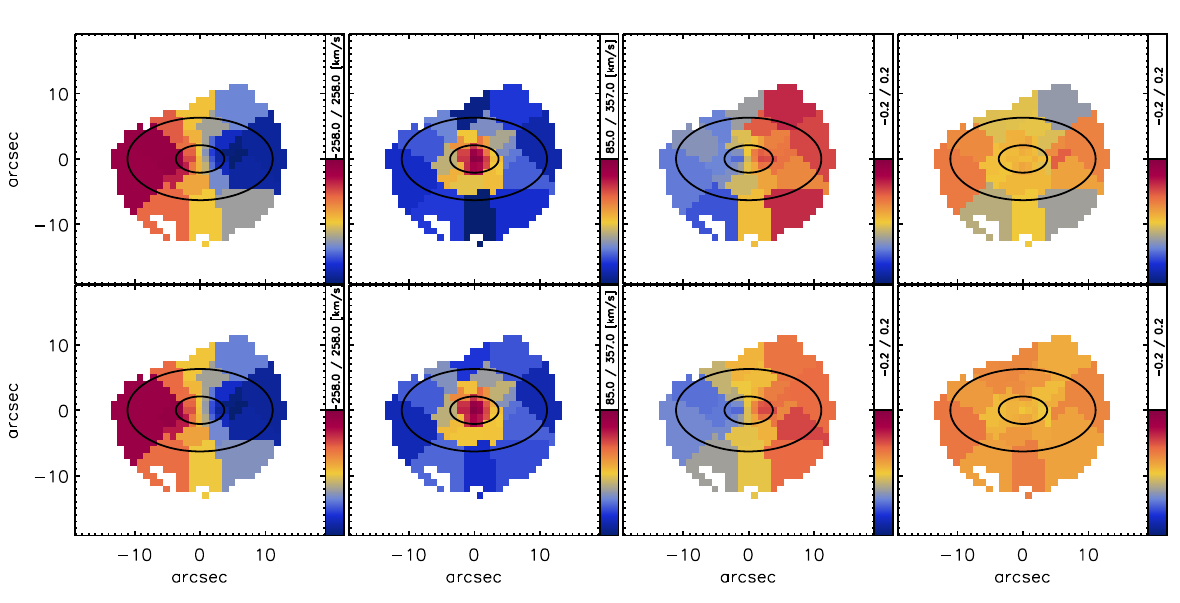}
		\end{center}
	\caption{Top: Bi-symmetrised \ppak\ IFU stellar kinematic maps of NGC\,1277, showing the mean stellar line-of-sight velocity $v$, velocity dispersion $\sigma$ and higher order Gauss-Hermite moments $h_3$ and $h_4$. Overplotted are contours of constant SB at 1 and 3\,\Reff\, to illustrate the extent of the kinematic data. The maps show fast and regular rotation around the short axis of 258\,\kms and a central velocity dispersion of 357\,\kms. Bottom: Best-fitting Schwarzschild model predictions. All maps are oriented north-east, i.e. north is up and east is left.}
	\label{fig:ngc1277_data_model}
\end{figure*}

\begin{figure*}
		\begin{center}
		\includegraphics[width=.95\textwidth]{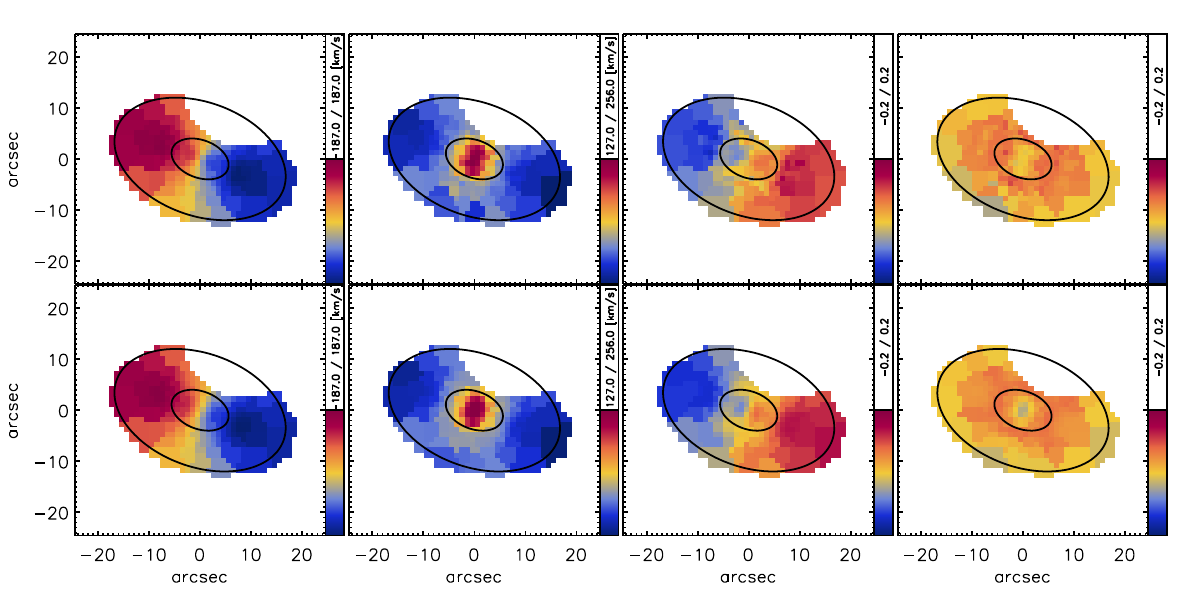}
		\end{center}
	\caption{Top: Bi-symmetrised \ppak\ IFU stellar kinematic maps of NGC\,1281, showing the mean stellar line-of-sight velocity $v$, velocity dispersion $\sigma$ and higher order Gauss-Hermite moments $h_3$ and $h_4$. Overplotted are contours of constant SB at 1 and 3\,\Reff\, to illustrate the extent of the kinematic data. The maps show fast and regular rotation around the short axis of 187\,\kms and a central velocity dispersion of 256\,\kms. Bottom: Best-fitting Schwarzschild model predictions. All maps are oriented north-east, i.e. north is up and east is left.}
	\label{fig:ngc1281_data_model}
\end{figure*}

\begin{figure*}
		\begin{center}
		\includegraphics[width=.95\textwidth]{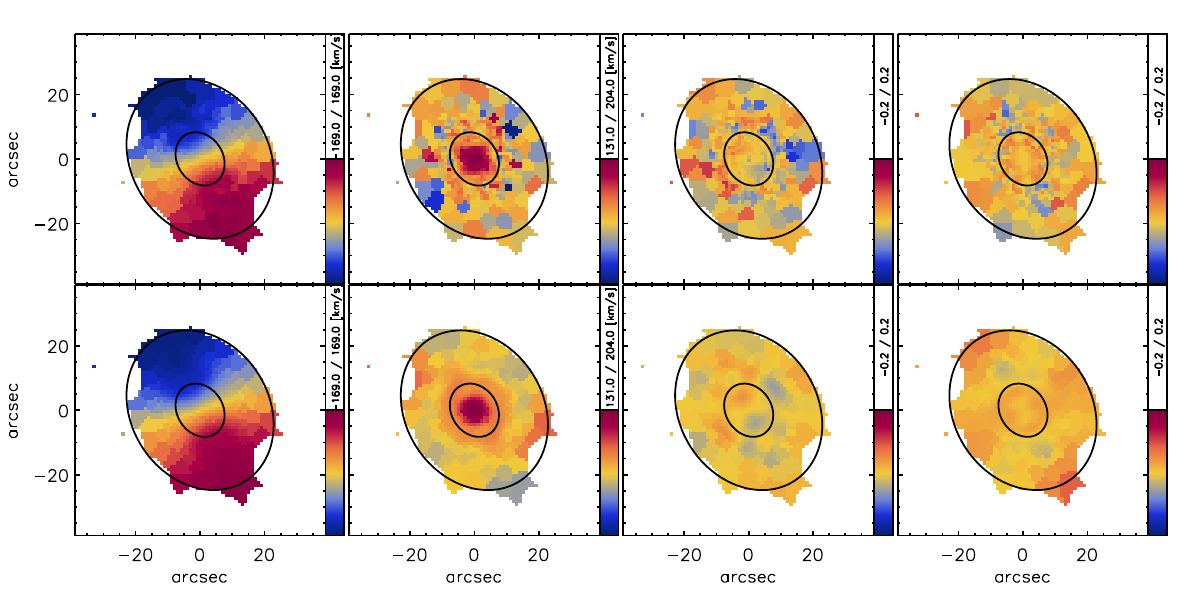}
		\end{center}
	\caption{Top: Point-symmetrised \ppak\ IFU stellar kinematic maps of NGC\,1282, showing the mean stellar line-of-sight velocity $v$, velocity dispersion $\sigma$ and higher order Gauss-Hermite moments $h_3$ and $h_4$. Overplotted are contours of constant SB at 1 and 3\,\Reff\, to illustrate the extent of the kinematic data. The maps show fast and regular rotation around the short axis of 169\,\kms and a central velocity dispersion of 204\,\kms. Bottom: Best-fitting Schwarzschild model predictions. All maps are oriented north-east, i.e. north is up and east is left.}
	\label{fig:ngc1282_data_model}
\end{figure*}

\begin{figure*}
		\begin{center}
		\includegraphics[width=.95\textwidth]{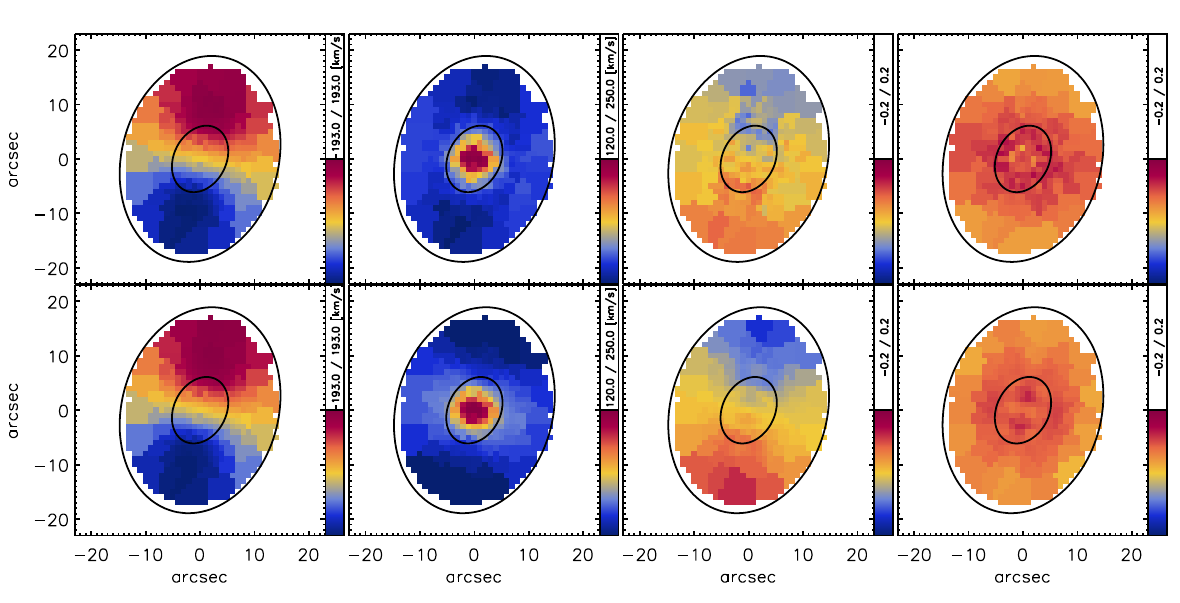}
		\end{center}
	\caption{Top: Bi-symmetrised \ppak\ IFU stellar kinematic maps of NGC\,2767, showing the mean stellar line-of-sight velocity $v$, velocity dispersion $\sigma$ and higher order Gauss-Hermite moments $h_3$ and $h_4$. Overplotted are contours of constant SB at 1 and 3\,\Reff\, to illustrate the extent of the kinematic data. The maps show fast and regular rotation around the short axis of 193\,\kms and a central velocity dispersion of 250\,\kms. Bottom: Best-fitting Schwarzschild model predictions. All maps are oriented north-east, i.e. north is up and east is left.}
	\label{fig:ngc2767_data_model}
\end{figure*}

\begin{figure*}
		\begin{center}
		\includegraphics[width=.95\textwidth]{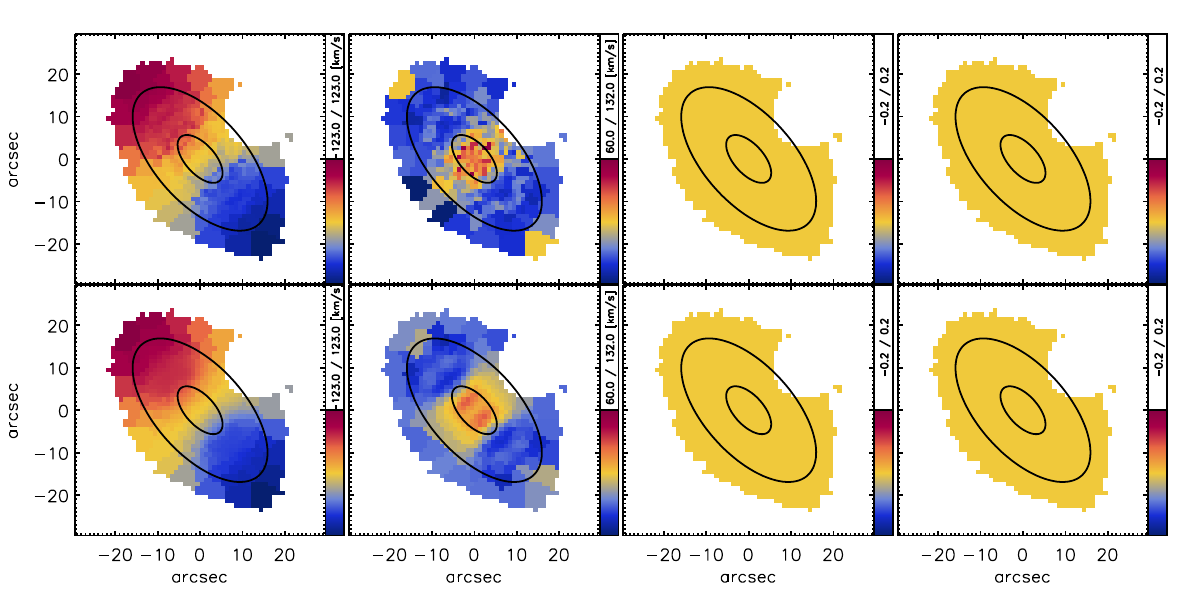}
		\end{center}
	\caption{Top: Bi-symmetrised \ppak\ IFU stellar kinematic maps of NGC\,3390, showing the mean stellar line-of-sight velocity $v$, velocity dispersion $\sigma$. The higher order Gauss-Hermite moments $h_3$ and $h_4$ have been omitted from the fits, as the velocity dispersion measurements are generally below the \ppak\ instrumental resolution. Overplotted are contours of constant SB at 1 and 3\,\Reff\, to illustrate the extent of the kinematic data. The maps show fast and regular rotation around the short axis of 123\,\kms and a central velocity dispersion of 107\,\kms. Bottom: Best-fitting Schwarzschild model predictions. All maps are oriented north-east, i.e. north is up and east is left.}
	\label{fig:ng3990_data_model}
\end{figure*}

\begin{figure*}
		\begin{center}
		\includegraphics[width=.95\textwidth]{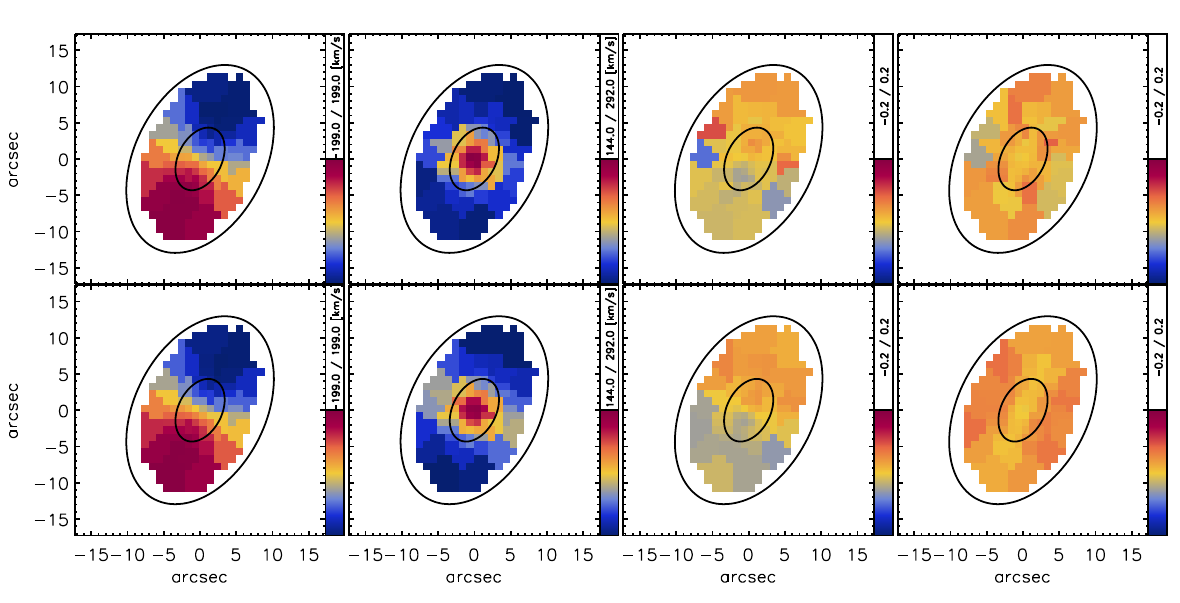}
		\end{center}
	\caption{Top: Bi-symmetrised \ppak\ IFU stellar kinematic maps of PGC\,11179, showing the mean stellar line-of-sight velocity $v$, velocity dispersion $\sigma$ and higher order Gauss-Hermite moments $h_3$ and $h_4$. Overplotted are contours of constant SB at 1 and 3\,\Reff\, to illustrate the extent of the kinematic data. The maps show fast and regular rotation around the short axis of 204\,\kms and a central velocity dispersion of 292\,\kms. Bottom: Best-fitting Schwarzschild model predictions. All maps are oriented north-east, i.e. north is up and east is left.}
	\label{fig:pgc11179_data_model}
\end{figure*}

\begin{figure*}
		\begin{center}
		\includegraphics[width=.95\textwidth]{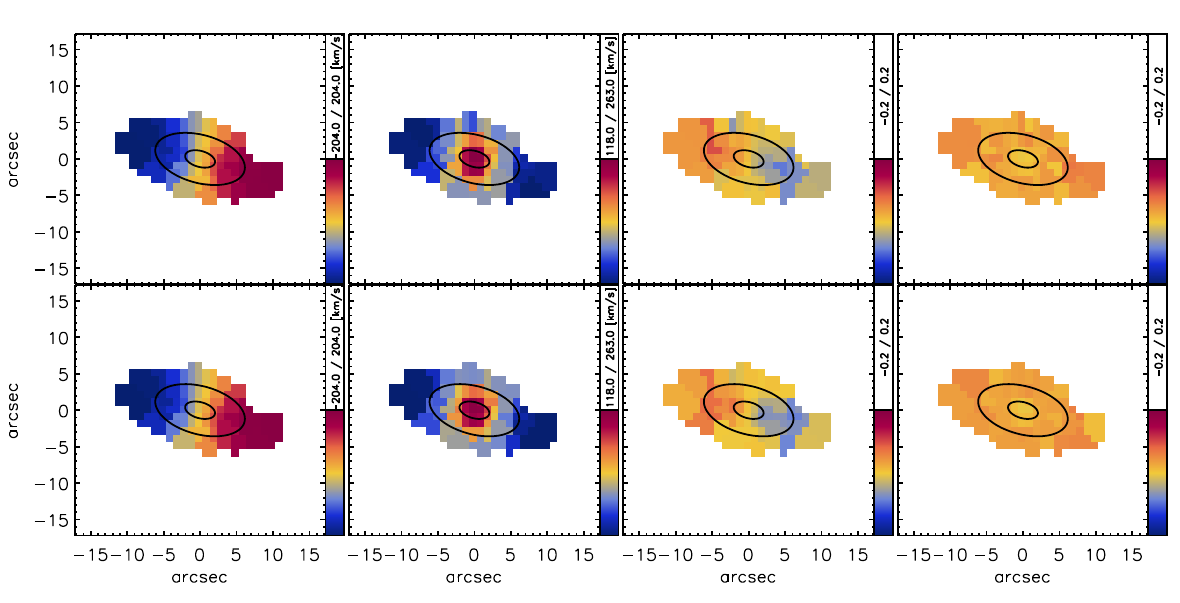}
		\end{center}
	\caption{Top: Bi-symmetrised \ppak\ IFU stellar kinematic maps of PGC\,12562, showing the mean stellar line-of-sight velocity $v$, velocity dispersion $\sigma$ and higher order Gauss-Hermite moments $h_3$ and $h_4$. Overplotted are contours of constant SB at 1 and 3\,\Reff\, to illustrate the extent of the kinematic data. The maps show fast and regular rotation around the short axis of 280\,\kms and a central velocity dispersion of 263\,\kms. Bottom: Best-fitting Schwarzschild model predictions. All maps are oriented north-east, i.e. north is up and east is left.}
	\label{fig:pgc12562_data_model}
\end{figure*}

\begin{figure*}
		\begin{center}
		\includegraphics[width=.95\textwidth]{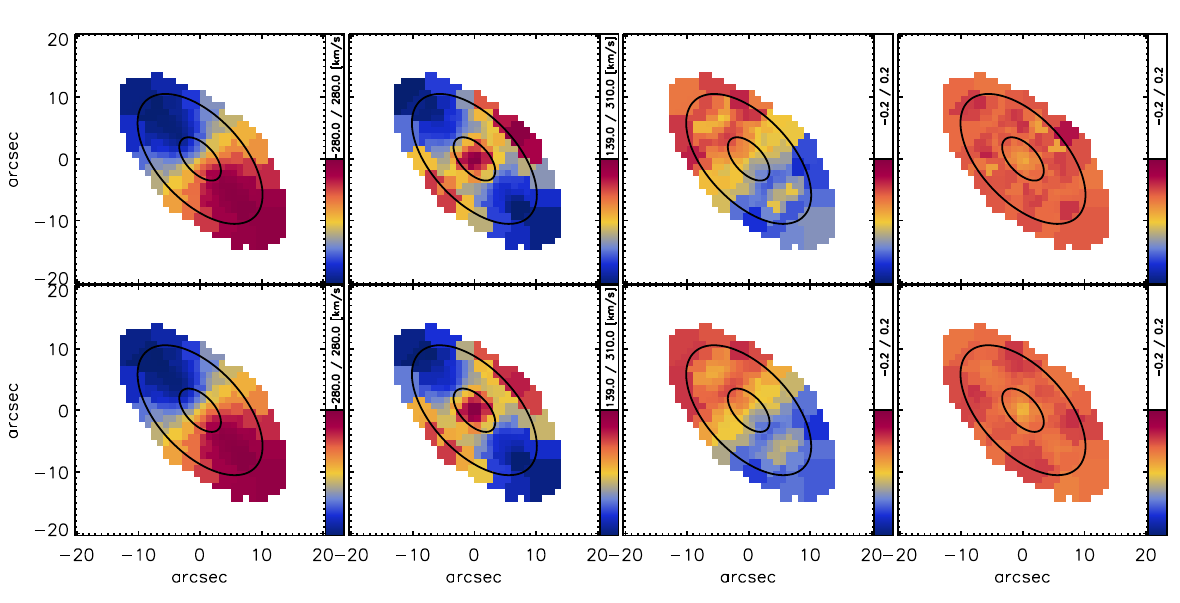}
		\end{center}
	\caption{Top: Bi-symmetrised \ppak\ IFU stellar kinematic maps of PGC\,32873, showing the mean stellar line-of-sight velocity $v$, velocity dispersion $\sigma$ and higher order Gauss-Hermite moments $h_3$ and $h_4$. Overplotted are contours of constant SB at 1 and 3\,\Reff\, to illustrate the extent of the kinematic data. The maps show fast and regular rotation around the short axis of 193\,\kms and a central velocity dispersion of 310\,\kms. Bottom: Best-fitting Schwarzschild model predictions. All maps are oriented north-east, i.e. north is up and east is left.}
	\label{fig:pgc32873_data_model}
\end{figure*}

\begin{figure*}
		\begin{center}
		\includegraphics[width=.95\textwidth]{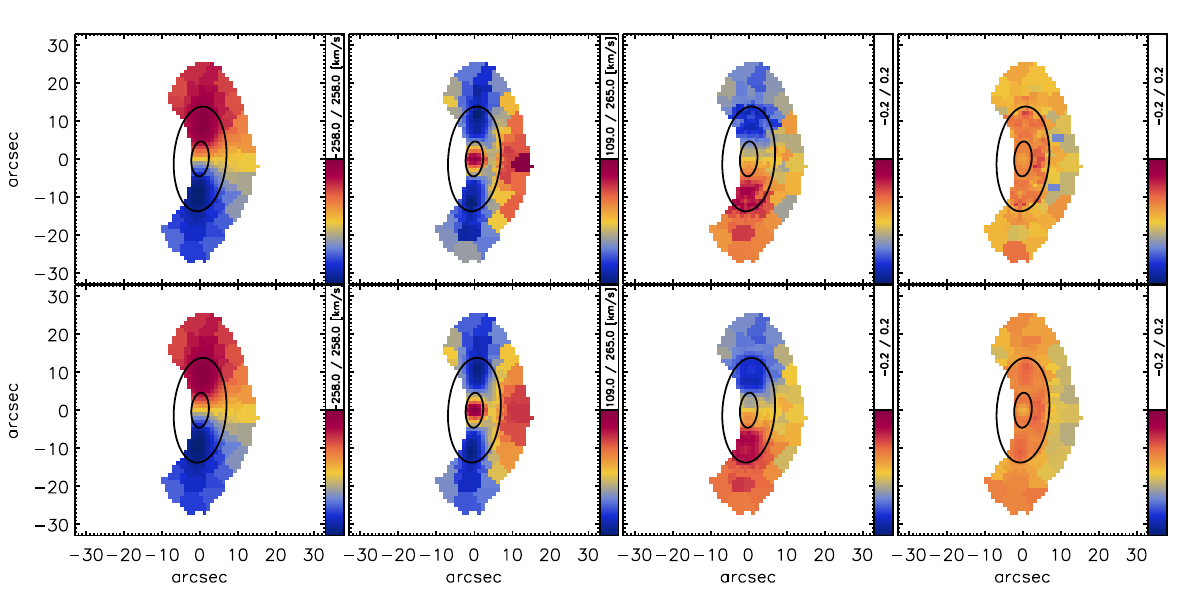}
		\end{center}
	\caption{Top: Bi-symmetrised \ppak\ IFU stellar kinematic maps of PGC\,70520, showing the mean stellar line-of-sight velocity $v$, velocity dispersion $\sigma$ and higher order Gauss-Hermite moments $h_3$ and $h_4$. Overplotted are contours of constant SB at 1 and 3\,\Reff\, to illustrate the extent of the kinematic data. The maps show fast and regular rotation around the short axis of 258\,\kms and a central velocity dispersion of 265\,\kms. Bottom: Best-fitting Schwarzschild model predictions. All maps are oriented north-east, i.e. north is up and east is left.}
	\label{fig:pgc70520_data_model}
\end{figure*}

\begin{figure*}
		\begin{center}
		\includegraphics[width=.95\textwidth]{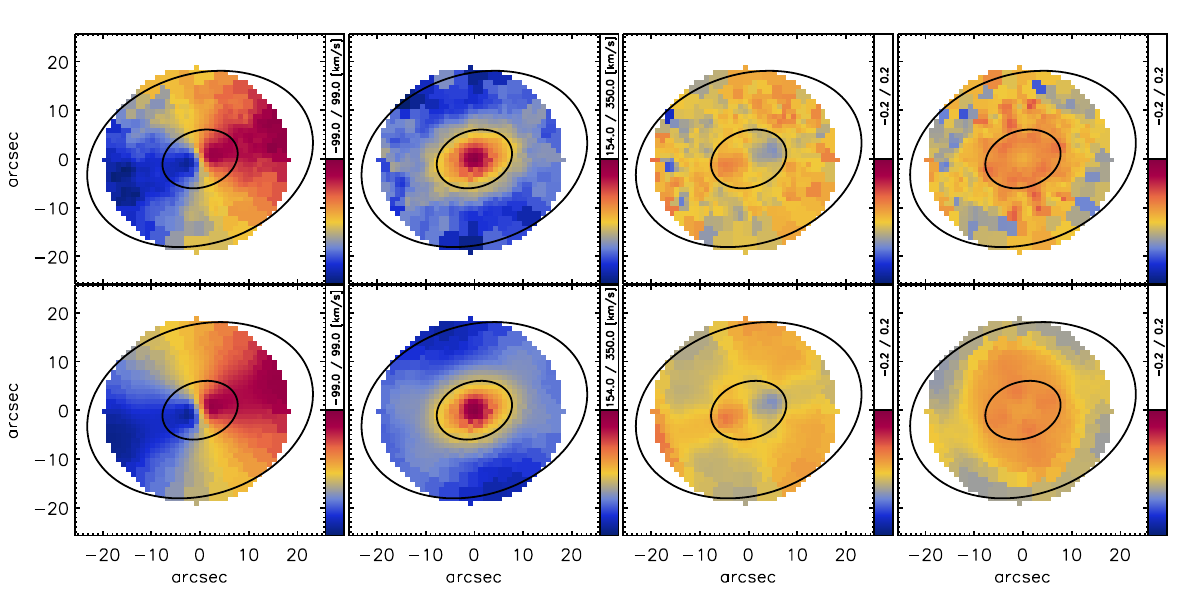}
		\end{center}
	\caption{Top: Bi-symmetrised \ppak\ IFU stellar kinematic maps of UGC\,2698, showing the mean stellar line-of-sight velocity $v$, velocity dispersion $\sigma$ and higher order Gauss-Hermite moments $h_3$ and $h_4$. Overplotted are contours of constant SB at 1 and 3\,\Reff\, to illustrate the extent of the kinematic data. The maps show fast and regular rotation around the short axis of 99\,\kms and a central velocity dispersion of 350\,\kms. Bottom: Best-fitting Schwarzschild model predictions. All maps are oriented north-east, i.e. north is up and east is left.}
	\label{fig:ugc2698_data_model}
\end{figure*}

\begin{figure*}
		\begin{center}
		\includegraphics[width=.95\textwidth]{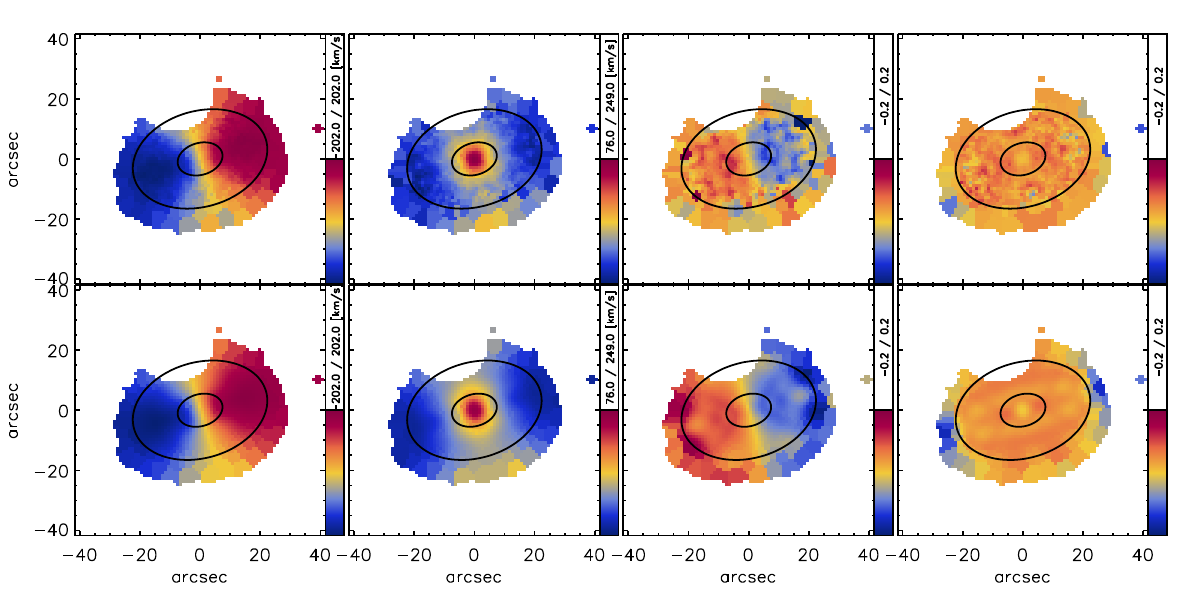}
		\end{center}
	\caption{Top: Bi-symmetrised \ppak\ IFU stellar kinematic maps of UGC\,3816, showing the mean stellar line-of-sight velocity $v$, velocity dispersion $\sigma$ and higher order Gauss-Hermite moments $h_3$ and $h_4$. Overplotted are contours of constant SB at 1 and 3\,\Reff\, to illustrate the extent of the kinematic data. The maps show fast and regular rotation around the short axis of 202\,\kms and a central velocity dispersion of 249\,\kms. Bottom: Best-fitting Schwarzschild model predictions. All maps are oriented north-east, i.e. north is up and east is left.}
	\label{fig:ugc2698_data_model}
\end{figure*}

%\bsp

\label{lastpage}

\end{document}